\documentclass[manuscript]{acmart}
\usepackage{graphicx}
\usepackage{subfigure}
\usepackage{xcolor}
\usepackage{enumitem}
\usepackage{url}
\usepackage{makecell}
\usepackage{multirow}
\usepackage{epstopdf}
\usepackage{hyperref}

\AtBeginDocument{%
  }

\begin{document}

\title{Network Edge Inference for Large Language Models: Principles, Techniques, and Opportunities}

\author{Zhixiong Chen}
\email{zhixiong.chen@qmul.ac.uk}
\affiliation{%
  \institution{Queen Mary University of London}
  \city{London}
  \country{United Kingdom}
}

\author{Bingjie Zhu}
\email{bjzhu1@stu.xidian.edu.cn}
\affiliation{%
  \institution{Xidian University}
  \city{Xian}
  \country{China}
}

\author{Jiangzhou Wang}
\email{j.z.wang@seu.edu.cn}
\affiliation{%
  \institution{Southeast University}
  \city{Nanjing}
  \country{China}
}

\author{Hyundong Shin}
\email{hshin@khu.ac.kr}
\affiliation{%
  \institution{Kyung Hee University}
  \city{Yongin-si}
  \country{Republic of Korea}
}

\author{Arumugam Nallanathan}
\email{a.nallanathan@qmul.ac.uk}
\affiliation{%
  \institution{Queen Mary University of London}
  \city{London}
  \country{United Kingdom}
}

\author{Dusit Niyato}
\email{dniyato@ntu.edu.sg}
\affiliation{%
  \institution{Nanyang Technological University}
  \city{Nanyang Avenue}
  \country{Singapore}
}

\renewcommand{\shortauthors}{Chen et al.}

\begin{abstract}
Large language models (LLMs) have advanced rapidly, emerging as versatile tools across fields thanks to their exceptional language understanding, generation, and reasoning capabilities. However, performing LLM inference at the network edge remains challenging due to their large memory and compute demands. This survey outlines the challenges specific to LLM edge inference and provides a comprehensive overview of recent progress, covering system architectures, model optimization and deployment, and resource management and scheduling. By synthesizing state-of-the-art techniques and mapping future directions, this survey aims to unlock the potential of LLMs in resource-constrained edge environments.
\end{abstract}



\maketitle

\section{Introduction}

\subsection{Background}
In recent years, large language models (LLMs) have become a cornerstone of artificial intelligence (AI) progress, demonstrating exceptional performance across a wide range of language-related tasks. Building on foundational advances, a vibrant ecosystem of LLMs has emerged, including the GPT-family \cite{NEURIPS2020_1457c0d6}, LLaMA-family \cite{touvron2023llama}, and open models such as OPT \cite{zhang2022opt} and Mistral \cite{jiang2023mistral7b}, along with specialized variants fine-tuned for coding, reasoning, and multimodal understanding. From machine translation and sentiment analysis to question answering and text generation, these models exhibit outstanding capabilities in natural language comprehension, synthesis, reasoning, and generation. This progress has reshaped both research and industry, enabling the development of widely used applications such as ChatGPT \cite{achiam2023gpt} and DeepSeek \cite{liu2024deepseek}. There is a growing consensus regarding this trajectory as a substantial step toward artificial general intelligence (AGI), though its timeline and scope remain active topics of debate \cite{Debates_Melanie}.

Driven by scaling laws \cite{NEURIPS2024_b6341525, kaplan2020scaling}, the development of LLMs has trended toward ever-increasing parameter counts. Nowadays, LLMs with tens and even hundreds of billions of parameters are becoming increasingly common. The sheer size of LLMs imposes heavy demands on memory, computation, and energy resources, creating deployment and inference bottlenecks not only on resource-constrained devices but also on advanced accelerators. For example, a LLaMA model \cite{touvron2023llama} with 70B parameter requires roughly 140 GB just to hold 16-bit floating-point (FP16) weights, excluding the additional headroom needed for the key-value (KV) cache and activations during inference. Such requirements exceed the capacity of most edge devices and even data center GPUs, such as NVIDIA's A100 80 GB or H100 80 GB.
Moreover, once the model is loaded, long-context prompts and multi-user concurrency can quickly exhaust the available memory.

Owing to their substantial computational and memory demands, mainstream commercial LLMs, e.g., ChatGPT \cite{achiam2023gpt} and DeepSeek \cite{liu2024deepseek},  are deployed in cloud data centers to provide inference services for users. However, cloud-centric inference has inherent limitations in interactive, latency-critical, privacy-sensitive, and bandwidth-constrained scenarios.
First, because each request must traverse the wide-area network to reach the cloud, end-to-end latency is susceptible to network delay and jitter, resulting in longer response times and unstable service quality. This worsens under peak load or on congested links, such as mobile networks during commutes, stadium events, or campus rush hours.
Second, transmitting prompts and context to the cloud raises data-residency, privacy, and compliance risks. This is especially problematic in privacy-sensitive sectors such as healthcare and finance, where clinical notes, imaging summaries, and transaction data are governed by laws (e.g., GDPR \cite{protection2018general}) and strict internal policies.
Transmitting them off device risks noncompliance. 
Third, bandwidth costs rise with long contexts and multimodal inputs (e.g., video clips for vision-language analysis) and are further amplified when many users are served concurrently. Finally, reliance on remote cloud infrastructure undermines availability in weak-connectivity environments such as rural areas and disaster zones. These realities motivate techniques that compress, place, and schedule LLMs to run on consumer-grade devices while preserving user experience.

To deploy LLMs on consumer-grade devices, a new wave of lightweight yet capable models and toolchains has emerged, co-optimizing model size, runtime, and hardware utilization. On the model side, Google's Gemini Nano runs locally inside Android's AICore on supported phones such as Pixel 8 Pro, with 1.8B and 3.25B variants designed for low-latency on-device tasks \cite{google_gemini_nano_2025}. Apple's Apple Intelligence ships on iPhone, iPad, and Mac, executing models on device when possible and selectively using private cloud for heavier requests \cite{apple_pcc_2024}. Chip vendors are broadening coverage, e.g., Qualcomm and Meta announced support for running LLaMA models on Snapdragon-powered phones and PCs, and LLaMA 3.2 introduced mobile-optimized 1B and 3B variants aimed at edge and handset use \cite{qualcomm_llama3_snapdragon_2024}.
Toolchains such as ONNX Runtime Mobile \cite{onnx_runtime_mobile} further streamline deployment of quantized models on iOS and Android with low memory overhead.

\subsection{Motivation and Challenges}

According to the above discussion, cloud-centric inference has been crucial for bringing LLMs to scale, but bandwidth, latency, privacy, and coverage constraints make it difficult to meet practical application requirements. On-device LLMs minimize latency and keep data local, yet lightweight models typically underperform relative to state-of-the-art cloud models. Even when deployment succeeds, many devices cannot sustain long prompts, multi-turn sessions, or concurrent requests. These limitations motivate shifting toward edge serving that blends device, near-edge, and cloud resources.

Edge serving avoids the extremes of cloud-centric and on-device inference by deploying LLM services close to user equipment at the network edge (for example, base stations and on-premises edge servers) while retaining on-demand access to near-edge and cloud resources.
Compared with cloud-centric inference, edge serving reduces dependence on the wide-area network, shortens network paths, and lowers end-to-end latency.
Compared with on-device inference, it provides more compute and memory headroom, enabling larger models, longer contexts, and greater concurrency.
Consequently, moving from cloud-centric or on-device architectures to edge serving reframes the objective from fitting everything on a single tier to co-designing model, memory, computation, and communication resources across tiers. The result is a practical pathway to sustaining quality of experience (QoE) with short response times, high throughput, and robust operation under real-world networks, without constant cloud backhaul or unrealistic on-device headroom.

However, deploying LLM inference at the edge remains a complex challenge at the intersection of LLM serving, mobile computing, and wireless communications.
Although cloud-centric and purely on-device LLM inference, as well as edge inference for conventional deep neural networks (DNNs), are well studied, their techniques do not scale or transfer directly to edge LLM serving due to LLM- and edge-specific constraints.
Relative to cloud and on-device deployments, LLM edge inference faces distinctive challenges: 1) \emph{Device Heterogeneity}: cloud-centric and on-device techniques often assume relatively uniform hardware and stable interconnects, whereas edge devices are heterogeneous across communication, computation, and memory. 2) \emph{Geo-Distributed Resources}: edge devices are geographically dispersed and offer fragmented resources, making coordination and scheduling harder than in centralized clouds or single-device setups.
3) \emph{Time-Varying Wireless Channels}: because LLM edge inference depends on the radio access network, effects like fading, interference, mobility, and handovers cause time-varying channel quality, resulting in bursty latency, uneven throughput, and packet loss.
Relative to conventional DNN inference, LLMs introduce extra challenges: 1) \emph{Stateful Inference Process}: unlike traditional DNNs that run a single forward pass per request, LLMs generate token by token, requiring multiple forward passes and needing to maintain persistent states (mainly KV cache that grows and shrinks dynamically) across multiple invocations. 2) \emph{Uncertain Resource Demand}: conventional DNNs often assume fixed-size inputs and outputs, whereas LLM prompts vary widely and output length is unpredictable, memory usage and processing strategies must adapt on the fly. 3) \emph{Large-Scale Model Parameters}: even small LLMs occupy gigabytes, stressing device storage, bandwidth, and computation more than DNNs. 4) \emph{Two-Phase Asymmetry}: the inference of LLMs involves two asymmetric phases, i.e., compute-bounded prefill and memory-bounded decoding, which requires phase-aware system designs.

In view of these challenges, enabling LLM inference at the wireless network edge requires transformative cross-layer designs spanning models, system architectures, and algorithms for managing and scheduling communication, computation, and memory resources.
To this end, this survey focuses on a comprehensive review of emerging techniques in this area and provides practical guidance for research and deployment.

\subsection{Comparisons with Prior Surveys and Our Contributions}
Prior surveys on LLM inference, such as \cite{10938426, park2025survey, 10.1145/3754448, khoshnoodi2024comprehensive, li2024survey, xia2024unlocking}, primarily focus on cloud data centers and synthesize advances in model compression, memory management, and cluster orchestration under relatively homogeneous GPU-rich infrastructure. These works overlook the resource constraints, device heterogeneity, and variability of wireless communications.
Moreover, surveys on on-device LLMs, e.g., \cite{10.1145/3719664, xu2024device}, focus on single-device deployments for phones and personal computers. They emphasize compression and intra-device serving, but rarely address multi-user edge environments.
In addition, existing surveys on LLM edge inference, e.g., \cite{10835069, 10398474}, remain high level, offering architectures or application taxonomies but lacking a detailed treatment of LLM-specific challenges, cross-layer co-design, and reproducible evaluation.
This survey fills these gaps by adopting a cross-layer perspective that integrates model optimization and deployment, system architecture design, and resource management to enable communication-, compute-, and memory-aware LLM inference at the network edge.
The main contributions of this paper are summarized as follows:
\begin{itemize}
  \item We provide a comprehensive overview of LLM inference and the unique challenges at the wireless network edge. Building on this foundation, we present reference architectures for LLM edge serving that span the device, near-edge, and cloud tiers, covering distributed deployment and inference frameworks.

  \item We provide a comprehensive review of optimization techniques for LLM inference at the network edge, covering model optimization and deployment, and resource management across communications, computation, and memory.
  We also curate the metrics, models, datasets, and platforms needed for rigorous, reproducible evaluation and to support real-world deployment.

  \item We identify several crucial research directions for LLM edge inference, including scalable serving for multi-model and multimodal LLMs, secure and privacy-preserving execution, and green inference.
\end{itemize}

This survey is organized as follows.
Section \ref{sec:fundamentals} illustrates the fundamentals of LLM edge inference.
Section \ref{sec:architectures} presents system architectures for LLM edge inference and their associated challenges, which define the execution topology and the resulting communication and state-transfer patterns.
Building on this, Section \ref{sec:model_optimization} reviews model optimization and deployment techniques that make each architecture practical and efficient under edge constraints.
Section \ref{sec:resource_scheduling} then covers joint optimization of communication, computation, and memory for improved efficiency and user-perceived service quality.
Section \ref{sec:evaluation_method} details evaluation methodology.
Section \ref{sec:future_research} outlines future research opportunities.
The conclusion is drawn in Section \ref{sec:conclusion}.

\section{Fundamentals of LLM Edge Inference}\label{sec:fundamentals}

\subsection{Generative LLM Architecture}
LLMs are typically Transformer-based neural networks with massive parameters.
Depending on the Transformer components used and the attention-masking scheme, LLMs typically have three primary structures:
\begin{itemize}
\item \textbf{Encoder-only LLMs}: These models consist of a stack of Transformer encoder blocks \cite{NIPS2017_3f5ee243} that map input text to contextual vector representations, without an explicit decoding phase for free-form generation. They are typically pretrained using a masked language modeling paradigm, i.e., predicting masked words in a sentence using bidirectional self-attention (no causal mask) conditioning on both left and right context. Canonical examples include BERT \cite{devlin-etal-2019-bert}, RoBERTa \cite{liu2019roberta}, and ALBERT \cite{lan2019albert}.

\item \textbf{Encoder-decoder LLMs}: These models pair a bidirectional encoder with an autoregressive decoder, linked via cross-attention from the decoder to the encoder.
The encoder converts the input text into context-rich embeddings, while the decoder generates the target sequence left-to-right using causal self-attention plus cross-attention to the encoded representations output by the encoder. A canonical example is T5 \cite{JMLR:v21:20-074}, with closely related variants such as mT5 \cite{xue-etal-2021-mt5} and FLAN-T5 \cite{wei2022finetuned}.

\item \textbf{Decoder-only LLMs}: These models are composed of a stack of Transformer decoder blocks and are pretrained using an autoregressive language modeling paradigm with causal self-attention to generate the next token given the preceding tokens. For each token, the attention mechanism makes only the preceding tokens visible, enforcing causality consistent with natural human speech. This makes them natural generators for tasks such as open-ended text generation, dialogue, summarization, code generation, and question answering. They also support in-context learning (ICL) \cite{dong2022survey} and are often adapted with instruction tuning \cite{pmlr-v202-longpre23a} for general-purpose assistance. Representative models under this architecture include GPT-3 \cite{NEURIPS2020_1457c0d6}, OPT \cite{zhang2022opt}, and LLaMA \cite{touvron2023llama}.
\end{itemize}

Early in the development of LLMs, decoder-only models were less popular than encoder-only and encoder-decoder models.
However, the release of the game-changing GPT-3 \cite{NEURIPS2020_1457c0d6} shifted the landscape, spurring a rapid rise of decoder-only architectures, which now dominate modern LLM design.
\begin{figure}[ht]
\centering
\subfigure[Architecture of decoder-only LLMs]{\includegraphics[width=0.46\textwidth]{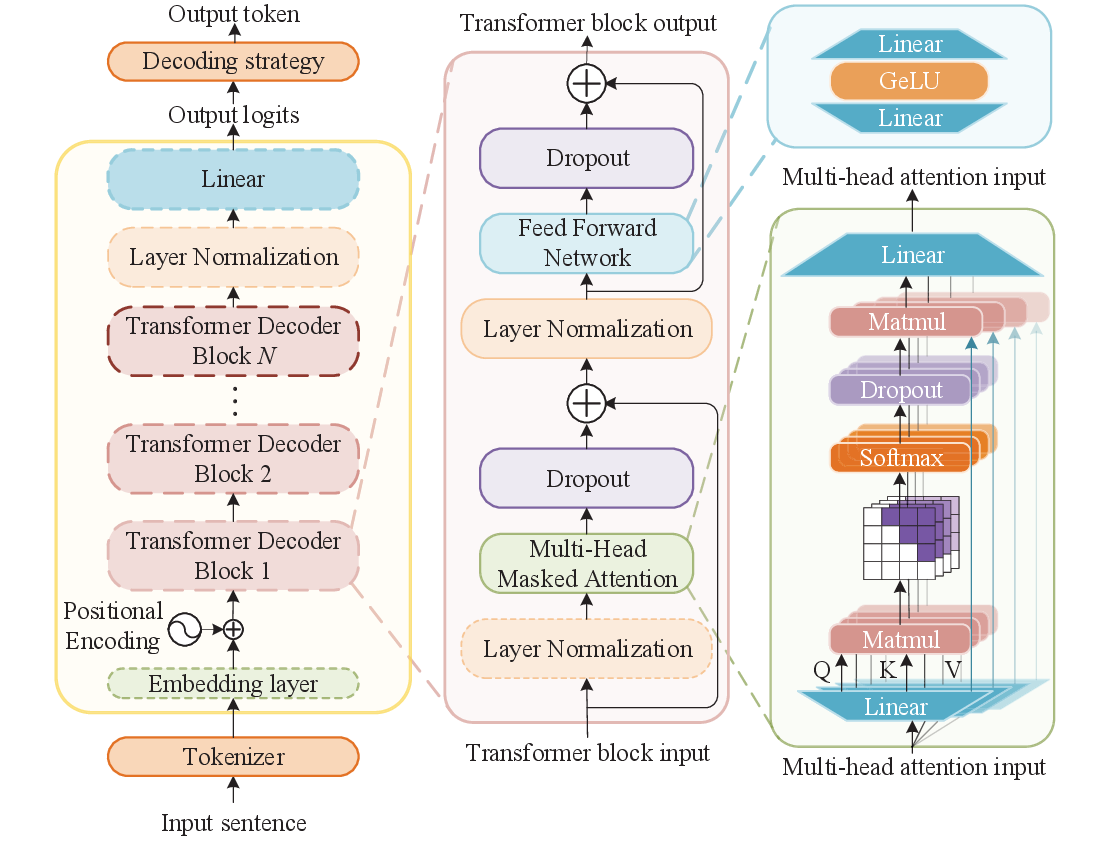}\label{fig:LLM_decoder_only}}
\hspace{0.2cm}
\subfigure[Inference process of decoder-only LLMs]{\includegraphics[width=0.48\textwidth]{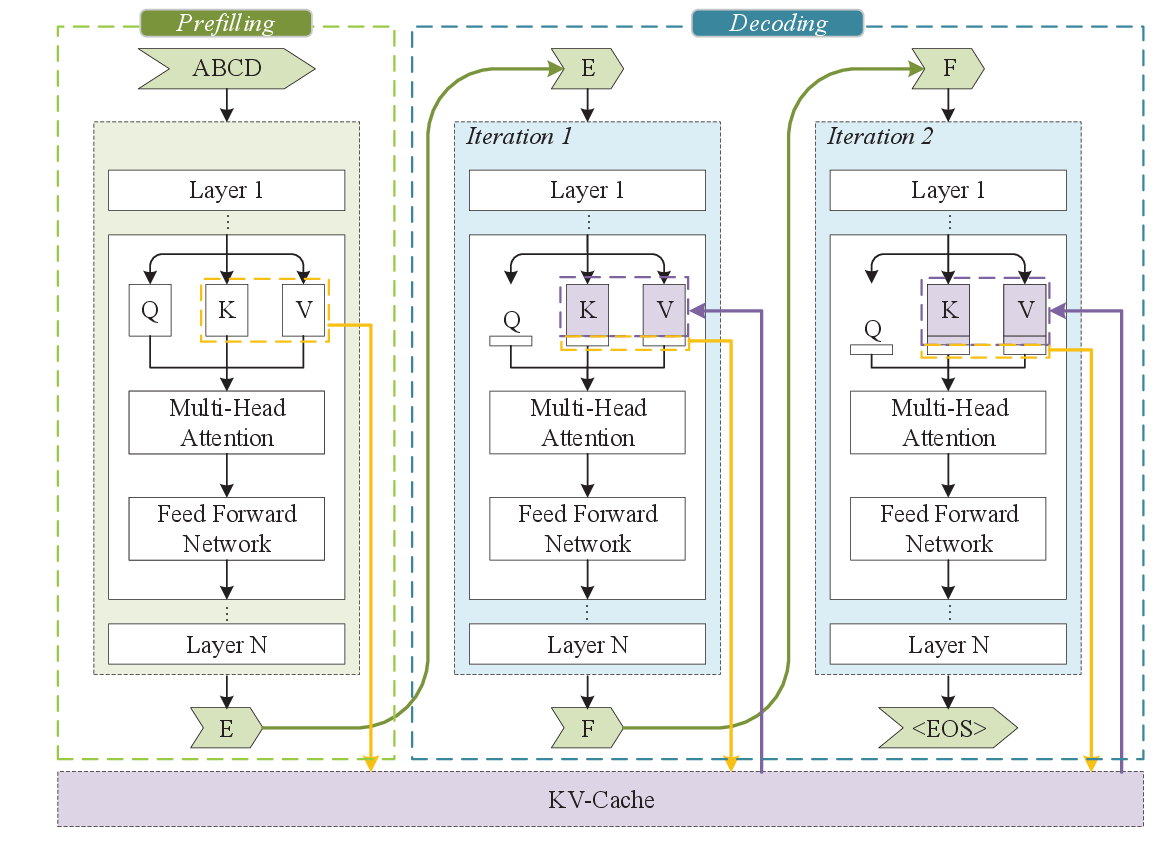}\label{fig:LLM_inference}}
\caption{Architectures and inference process of LLMs.}
\end{figure}

\subsection{Inference Process of LLMs}
Given the prevalence of decoder-only LLMs, we provide a concise overview of their architecture and inference process. In fact, encoder-only inference can be viewed as a special case of the decoder-only pipeline, consisting only of the prefill stage and processing the input once to produce contextual representations. Encoder-decoder models add an explicit encoding step, first run a single pass through the encoder to produce fixed representations, then perform left-to-right decoding with causal self-attention while cross-attending to the static encoder memory at each step. Other model variants, such as vision-language models (VLMs), extend the text-only pipeline with a modality front-end. In VLMs like Qwen2-VL \cite{team2024qwen2} and LLaVA-1.5 \cite{Liu_2024_CVPR}, a vision encoder converts images or video frames into feature sequences, projects them into the LLM embedding space, and combines them with text tokens (or uses them as cross-attention memory). The LLM then follows the same prefill and autoregressive decoding as in text-only inference, while the modality encoder, typically Transformer-based, resembles an encoder-only model at inference time.

The architecture of decoder-only LLMs is illustrated in Fig. \ref{fig:LLM_decoder_only}, which consists of a tokenizer, an embedding layer, a positional encoding module, multiple stacked Transformer decoder blocks, and a decoding head composed of a linear projection followed by a softmax operator. Each Transformer decoder block follows the design of the decoder component in the original Transformer model \cite{NIPS2017_3f5ee243} (or one of its modern variants) and typically contains three core submodules: a multi-head self-attention (MHSA) mechanism with causal masking, a FFN, and layer normalization (LN). Residual connections are applied around each submodule to stabilize training and preserve information flow. The blocks are stacked sequentially, where the output of one block serves as the input to the next.

Decoder-only LLMs rely on an autoregressive generation strategy, producing sequences in a strictly sequential manner by generating one token at a time, as shown in Fig. \ref{fig:LLM_inference}. At each generation step, the model processes the entire sequence observed so far, including both the original input tokens and the tokens it has already generated, to predict the next token. While conceptually straightforward, this process becomes increasingly expensive as sequence length grows, since naively recomputing self-attention over all prior tokens at every step leads to quadratic growth in computation and memory. To address this challenge, modern LLMs adopt the KV caching technique \cite{NEURIPS2024_028fcbcf}, in which the key (K) and value (V) vectors produced within the MHSA blocks are stored after their first computation and then reused in subsequent decoding steps. Building on this, the inference process of decoder-only LLMs can be conceptually divided into two stages:
\begin{itemize}
  \item \textbf{Prefill Stage}: In this stage, the LLM processes the entire initial input (the prompt) to compute hidden states and build the KV cache in each self-attention layer, as shown in Fig. \ref{fig:LLM_inference}. Concretely, the prompt is tokenized (text to token IDs), mapped to embeddings, combined with positional encodings, and passed through the stacked Transformer decoder blocks. The output of the last Transformer block is fed into the LLM head to produce output logits. A decoding method (e.g., greedy, top-k/top-p sampling with temperature, beam search, and contrastive search \cite{bai2024beyond}) then selects the first output token based on the logits.

  \item \textbf{Decoding Stage}: After prefill, the LLM enters an iterative loop that generates one token at a time, as shown in Fig. \ref{fig:LLM_inference}. At each step, the previously generated token is fed directly into the embedding layer without tokenization, since it is already a token rather than raw text. The resulting embedding is combined with positional encodings and passed through the Transformer stack and the LM head to produce the next token. During this process, the LLM reuses the existing KV cache from prior steps and appends new keys and values for the freshly generated token, avoiding recomputation over the entire history. The procedure continues until a stop condition is met, namely when an end-of-sequence token is generated or the configured maximum sequence length is reached. Note that without a KV cache, each step would require reprocessing all past tokens with the new token appended.
\end{itemize}

\subsection{Edge Inference}
Edge inference executes trained AI models at or near the data source, with on-demand access to nearby edge and cloud resources, rather than relying solely on distant cloud infrastructure or underpowered local devices. It offers:
(1) low latency and high responsiveness by avoiding wide-area round trips and backhaul jitter, (2) stronger privacy by keeping sensitive data (e.g., voice, medical text, personal documents) local, and (3) offline capability in disconnected or low-connectivity settings such as rural areas, underground transit systems, and disaster zones.
These advantages make edge inference a key enabler for a broad spectrum of AI applications including on-device metaverse perception and rendering, autonomous driving with millisecond-scale perception and planning loops, and real-time smart-city video analytics.

\begin{figure}[ht]
\centering
\subfigure[]{\includegraphics[width=0.084\textwidth]{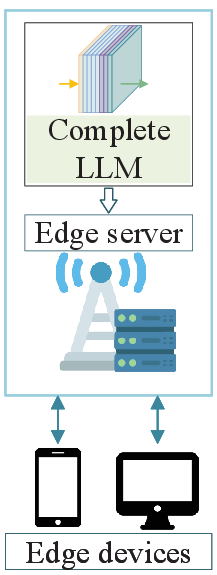}\label{fig:architecture_singleNode}}
\hspace{0.2cm}
\subfigure[]{\includegraphics[width=0.48\textwidth]{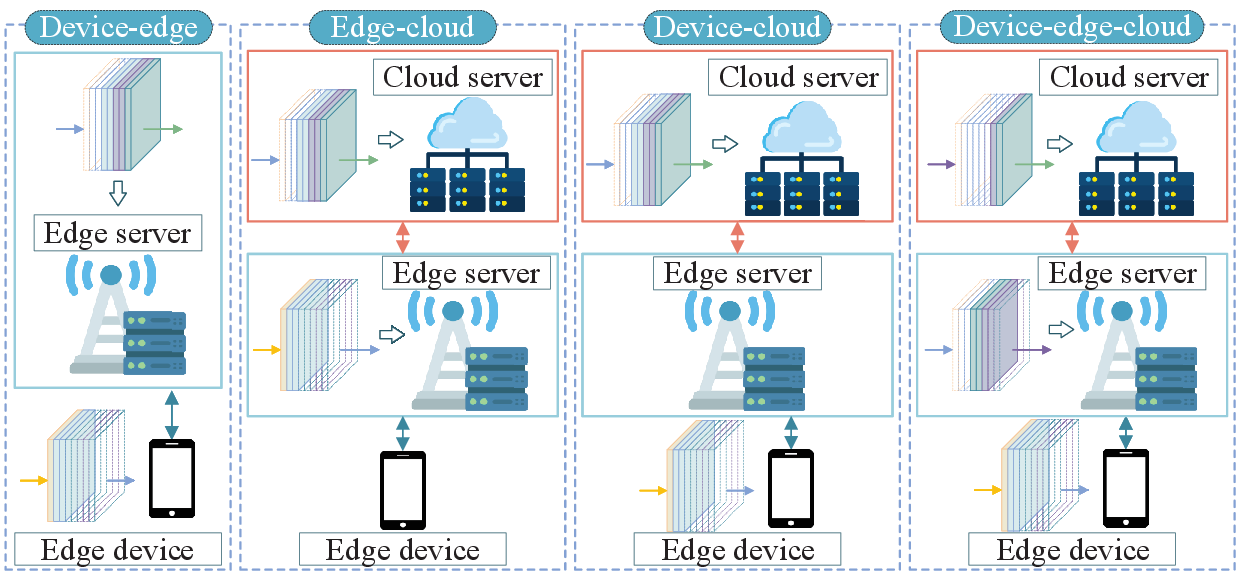}\label{fig:architecture_vertical}}
\hspace{0.2cm}
\subfigure[]{\includegraphics[width=0.2\textwidth]{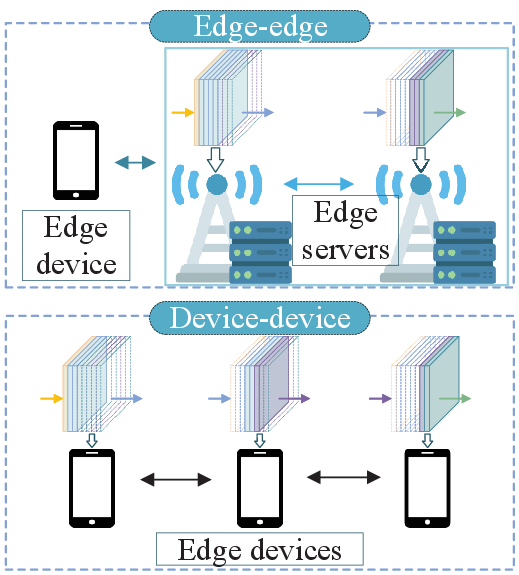}\label{fig:architecture_horizontal}}
\hspace{0.2cm}
\subfigure[]{\includegraphics[width=0.14\textwidth]{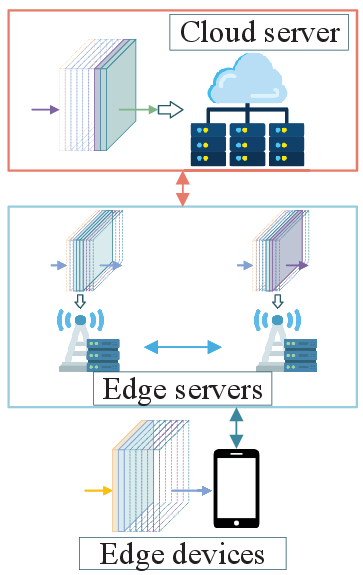}\label{fig:architecture_hybrid}}
\vspace{-0.1cm}
\caption{Architectures of LLM edge inference: (a) single-edge-node inference, (b) vertical collaborative inference, (c) horizontal collaborative inference, (d) hybrid collaborative inference.}
\label{fig:architectures}
\end{figure}

\section{System Architectures and Challenges for LLM Edge Inference}\label{sec:architectures}
This section introduces system architectures for LLM edge inference and highlights the associated research challenges. As shown in Fig. \ref{fig:architectures}, we categorize LLM edge inference into four architectural patterns. We describe each architecture and its typical characteristics below, with an overview in Table \ref{tab:sys_Architectures}.

\subsection{\textbf{Single-Edge-Node Inference}}
This architecture deploys an LLM on a single edge server to perform inference computations, serving multiple users in its coverage area and, when needed, nearby regions via low-latency metro or backhaul links, as shown in Fig. \ref{fig:architecture_singleNode}. Unlike on-device inference that serves requests on a single device, single-edge-node inference runs on a shared access or metro node that receives user prompts, performs LLM inference, and returns the results to their devices. In line with this architecture, \cite{10759588} deploys a quantized LLM on an edge server and adopts batching to increase inference throughput while meeting latency targets. Recognizing that users may require different LLMs for diverse tasks, \cite{11161094} and \cite{11154975} deploy multiple LLMs on each edge server and optimize their placement to match server resource constraints. \cite{10591707} further optimizes LLM task offloading and resource allocation via active inference. Overall, single-edge-node inference offers low coordination overhead, reduces backhaul traffic, and strengthens privacy, but it is constrained by the server's memory, compute, and power budget. It works well when quantized or distilled LLMs fit with sufficient KV-cache headroom, while very large models often exceed a single node's capacity.

\subsection{\textbf{Vertical Collaborative Inference}}
This architecture partitions an LLM across devices, edge servers, and the cloud to meet latency, throughput, and privacy goals within each tier's resource budget, enabling support for very large LLMs and bursty traffic. It is typically realized as device-edge, edge-cloud, device-cloud, or device-edge-cloud collaboration, as shown in Fig. \ref{fig:architecture_vertical}, with split points adjusted at runtime based on link quality, queueing delay, and compute load. This flexibility improves perceived latency and elasticity while preserving privacy. In line with this architecture, prior work uses DRL \cite{chen2025adaptive, 11160773} or dynamic programming \cite{mudvari2024splitllm} to select split points for deploying an LLM across the user device and the edge server. In \cite{xie2025novel}, the LLM is partitioned into three submodels, with the input and output submodels running on the device and the middle submodel, which contains most decoder layers, hosted in the cloud. By pooling resources across devices, edge servers, and the cloud, vertical collaborative inference can lower perceived latency, provide elastic capacity, and strengthen privacy. However, it increases orchestration complexity and can be bandwidth-sensitive, especially for long prompts, due to repeated activation transfers across tiers.

\subsection{\textbf{Horizontal Collaborative Inference}}
This architecture adopts multiple peer edge devices in the same tier to cooperate on LLM inference by splitting work across devices, as shown in Fig. \ref{fig:architecture_horizontal}. It can employ tensor parallelism to shard weights and run layers in lockstep, pipeline parallelism to place consecutive layer stages on different devices, and data-parallel replication to smooth bursts and raise throughput. Towards this architecture, \cite{zhao-etal-2024-lingualinked} partitions the LLM across end devices to enable collaborative inference and optimizes model assignment policy via linear programming. In \cite{pmlr-v235-jiang24f} and \cite{10818760}, the LLM is partitioned across heterogeneous devices or edge servers, and dynamic programming is employed to optimize the pipeline layout for maximizing throughput. \cite{10.1145/3627535.3638480} combines model quantization with phase-aware partitioning across multiple edge devices to further improve throughput. This architecture scales beyond a single device's memory and compute limits, improving elasticity and availability. However, it introduces orchestration complexity, increases sensitivity to inter-device link quality, and raises state-management overhead. It is most useful when an LLM cannot fit on one device and when nearby edge nodes offer sufficiently fast links.

\subsection{\textbf{Hybrid Collaborative Inference}}
This architecture combines vertical collaboration across the device, edge, and cloud with horizontal collaboration among peer nodes within each tier, as depicted in Fig. \ref{fig:architecture_hybrid}. Compared with using only vertical or only horizontal collaboration, it can serve larger models, absorb bursts, and reduce perceived latency by anchoring decoding near users while borrowing cloud capacity for heavy prefill or long-context retrieval.
For example, \cite{NEURIPS2023_28bf1419} aggregates idle computation resources from multiple research groups and volunteers to execute LLM inference, effectively accelerating inference. The trade-off is greater orchestration and state-management complexity, since KV caches and session state may need to be partitioned, compressed, and migrated across tiers and peers without violating latency targets.

\begin{table}[ht]\small
\caption{An Overview of Architectures for LLM Edge Inference}
\centering
\label{tab:sys_Architectures}
\begin{tabular}{|m{1.4cm}<{\centering}|m{1.4cm}<{\centering}|m{1.4cm}<{\centering}|m{1.4cm}<{\centering}|m{4.8cm}<{\centering}|m{2.2cm}<{\centering}|}
\hline
Arch. & Placement & Strength & Limits & Enabling techniques &  Rep. Examples  \\
\hline
Single-Edge-Node & Single edge server & Simple & Lightweight LLMs only & Compression (\ref{subsec:model_compress}), Decoding acceleration (\ref{subsec:decoding_strategy}), Caching/selection (\ref{subsubsec:LLM_caching}), Resource scheduling (\ref{sec:resource_scheduling}) & \cite{10759588, 11161094, 11154975, 10591707}\\
\hline
Vertical Collaborative & Combination between device, edge, cloud & Strong and elastic capacity & Backhaul bandwidth & Compression (\ref{subsec:model_compress}), Decoding acceleration (\ref{subsec:decoding_strategy}), Partition \& placement (\ref{subsubsec:LLM_partition}), Disaggregated prefill/decoding (\ref{subsubsec:disaggregrate_pd}), Resource scheduling (\ref{sec:resource_scheduling}) & \cite{chen2025adaptive, 11160773, mudvari2024splitllm, xie2025novel}\\
\hline
Horizontal Collaborative & Peer edge nodes & Higher throughput & Inter-device bandwidth & Compression (\ref{subsec:model_compress}), Decoding acceleration (\ref{subsec:decoding_strategy}), Partition \& placement (\ref{subsubsec:LLM_partition}), Resource scheduling (\ref{sec:resource_scheduling}) & \cite{zhao-etal-2024-lingualinked, pmlr-v235-jiang24f, 10818760, 10.1145/3627535.3638480} \\
\hline
Hybrid Collaborative & Vertical + horizontal & Support very large LLMs & Orchestration complexity & Compression (\ref{subsec:model_compress}), Decoding acceleration (\ref{subsec:decoding_strategy}), LLM deployment strategies (\ref{subsec:LLM_deployment}), Resource scheduling (\ref{sec:resource_scheduling}) & \cite{NEURIPS2023_28bf1419}\\
\hline
\end{tabular}
\end{table}

\subsection{Challenges for LLM Edge Inference}\label{subsec:LLM_edge_challenges}
As discussed in Section \ref{sec:fundamentals}, LLM inference differs from conventional DNN inference in both computational and systems behavior. On resource-constrained edge platforms, this divergence introduces distinct challenges:

\subsubsection{\textbf{Stateful Inference Process}}
Unlike stateless DNN inference (a single forward pass with a relatively fixed memory footprint), LLM inference is autoregressive and performs many forward passes, one per generated token. At each step the system maintains a per-request KV cache at every layer to avoid recomputation, and this cache grows with context length and is released only when the request completes \cite{park2025survey}. This statefulness complicates edge serving, since schedulers must anticipate per-request memory growth, reserve headroom to prevent out-of-memory (OOM) errors, and manage costly preemption or migration when KV caches must be transferred or rebuilt.

\subsubsection{\textbf{Two-Phase Asymmetry}}
LLM inference proceeds in two phases, i.e., prefill and decoding. In prefill, the model processes the entire prompt, builds the KV cache, and computes full self-attention. The workload is dominated by large dense matrix multiplications (Q/K/V projections and FFNs) and quadratic attention score products. Because weights are reused across all prompt tokens, arithmetic intensity is high with strong on-chip reuse, making prefill largely compute-bound and primarily limited by compute throughput \cite{10938426}. In contrast, the decoding phase generates one token per step, repeatedly reading the full KV cache and appending new token's KV entries, with most computation reduced to matrix-vector operations. Arithmetic intensity drops and memory traffic dominates, so decoding depends more on memory bandwidth than compute throughput and is typically memory-bound \cite{bai2024beyond}.
This asymmetry complicates edge serving, requiring compute-focused optimizations for prefill and memory bandwidth-centric optimizations for decoding.

\subsubsection{\textbf{Large-Scale Parameter Counts}}
Mainstream LLMs contain billions or even trillions of parameters. For instance, the LLaMA-70B model \cite{touvron2023llama} has 70 billion parameters, while the GPT-3 model \cite{NEURIPS2020_1457c0d6} scales up to 175 billion parameters. This scale drives up computation demand, memory capacity requirements, and memory bandwidth pressure during inference. Even with quantization, weight footprints can exceed on-device memory, increasing reliance on model partitioning, weight offloading, or aggressive compression.

\subsubsection{\textbf{Device Heterogeneity}}
Unlike cloud data centers with relatively uniform hardware and stable interconnects, edge devices are often resource limited and heterogeneous in computing, communication, memory capabilities \cite{10089235}. The scale of modern LLMs frequently exceeds the capacity of a single device, and inference typically requires spanning multiple edge devices or cooperating with the cloud. This makes deployment, request scheduling, resource management, and memory allocation substantially more complex.

\subsubsection{\textbf{Data Heterogeneity}}
Unlike the conventional DNN inference with fixed-size inputs, LLM input prompts vary widely in length and structure \cite{10.1145/3719664}. Moreover, the inference requests come from different users and have different difficulty, leading to uneven prefill cost, decoding time, and KV-cache growth. This variability complicates batching and scheduling, which must balance throughput against latency, pick batch windows that do not penalize short prompts, and mitigate head-of-line blocking from long prompts.

\subsubsection{\textbf{Uncertain Resource Demand}}
Unlike conventional DNN inference with fixed-size outputs,
LLM inference has stochastic generation lengths, making workloads variable and unpredictable \cite{park2025survey, 10.1145/3719664}. As a result, the computation, memory, and communication demands are not known in advance, complicating edge scheduling and capacity planning. Systems need to provision headroom for KV-cache growth and adapt resources dynamically during a request. In edge-cloud settings, they should also adjust offloading to network and queueing conditions and prefetch or migrate KV state to avoid OOM.

\subsubsection{\textbf{Time-Varying Wireless Channels}}
Most cloud-centric and on-device LLM inference studies (e.g., \cite{10812936, 10609649}) model device-cloud and inter-device links as bit pipes with constant or stationary random rates, which improves tractability and is often reasonable when core-network latency or high-speed on-device links dominate. In edge deployments, however, the radio access network induces time-varying rate, loss, and jitter due to fading and interference \cite{goldsmith2005wireless}. This variability matters because LLM edge inference is communication-coupled, with prompts or context and intermediate activations traversing wireless links, which increases end-to-end and tail latency, reduces token throughput, and can stall pipelines or batches. In streaming settings, it can also make token delivery bursty and trigger SLO-driven adaptations such as prompt or context truncation, switching to smaller or cached models, repartitioning, or more aggressive compression or early-exit, thereby trading quality for latency. Channel dynamics similarly affect distributed training and fine-tuning across edge nodes, where intermittent uplinks create stragglers and stale updates, forcing synchronization waits or participant dropout and slowing or destabilizing convergence.

\subsubsection{\textbf{Mobility and Intermittent Connectivity}}
In wireless edge environments, user mobility can trigger frequent handovers across base stations or serving edge nodes, while deep fading and coverage holes may cause transient disconnections \cite{goldsmith2005wireless}. These dynamics are particularly problematic because LLM inference is stateful, and maintaining session continuity may require migrating runtime states (e.g., KV caches and conversation context), re-establishing transport connections, and re-optimizing placement and scheduling under new radio conditions. If not handled carefully, handovers and outages can interrupt service, increase tail latency, and waste compute and radio resources due to retries or restarted decoding. This challenge is prominent in high-mobility and unreliable-link scenarios such as high-speed rail, unmanned aerial vehicle (UAV) swarms, and rural connectivity,  motivating handover-aware state management and graceful degradation mechanisms.

\section{Model Optimization and Deployment}\label{sec:model_optimization}
This section reviews model optimization and deployment techniques for resource-efficient, low-latency LLM inference at the network edge, including model compression, decoding mechanisms, and deployment strategies. These techniques are tightly coupled with the collaborative architectures in Section \ref{sec:architectures}. In particular, the architecture determines where computation and state reside and what must be transferred across the network, while the techniques in this section provide the concrete levers to make each architecture feasible and efficient.

\subsection{Model Compression}\label{subsec:model_compress}
The formidable size and compute demands of LLMs make practical deployment challenging in resource-constrained wireless edge environments. A natural and widely adopted solution is model compression, which reduces the storage and computational footprint of LLMs while preserving accuracy as much as possible and complying with edge resource constraints. Although compression techniques are broadly applicable to conventional DNNs, their design and effectiveness for LLM edge inference are influenced by LLM-specific properties, such as pronounced weight/activation outliers and prompt-dependent activation ranges. This subsection systematically reviews compression methods for LLMs, including quantization, pruning, knowledge distillation, and low-rank factorization. Table \ref{tab:Compression_compare} summarizes their key characteristics.

\begin{table}[ht]\small
\caption{Comparison of LLM Compression Techniques}
\centering
\label{tab:Compression_compare}
\begin{tabular}{|m{1.5cm}<{\centering}|m{2.8cm}<{\centering}|m{2.8cm}<{\centering}|m{3.6cm}<{\centering}|m{2.3cm}<{\centering}|}
\hline
Technique & Key Idea & Advantages & Limitations & Rep. Examples \\
\hline
Quantization  & Reduce weight/activation precision  & Reduces model size; supported by many accelerators & Accuracy degradation at low precision; may require retraining  & AWQ \cite{MLSYS2024_42a452cb}, Q-BERT \cite{Shen_Dong_Ye_Ma_Yao_Gholami_Mahoney_Keutzer_2020}, LLM.int8() \cite{NEURIPS2022_c3ba4962}, PACT \cite{choi2018pact}   \\
\hline
Pruning    & Remove redundant weights, neurons, heads, or layers  & Produces sparse models; reduces computation and memory footprint & Unstructured pruning needs special hardware; structured pruning harm accuracy & SparseGPT \cite{pmlr-v202-frantar23a}, Wanda \cite{sun2023simple}, LLM-Pruner \cite{NEURIPS2023_44956951} \\
\hline
Knowledge Distillation     & Train a smaller ``student'' to mimic a larger ``teacher''  & Produces compact dense models with good accuracy retention & Requires costly teacher training; student may underfit complex reasoning tasks & MiniLLM \cite{gu2024minillm}, FUSELLM \cite{wan2024knowledge}, PromptMix \cite{sahu-etal-2023-promptmix}\\
\hline
Low-rank Factorization        & Approximate large weight matrix with low-rank matrices  & Significant parameter reduction & Compression limited by intrinsic rank; accuracy loss if rank too small  & ASVD \cite{yuan2023asvd}, LASER \cite{sharma2023truth}, DSFormer \cite{chand2023dsformer} \\
\hline
\end{tabular}
\end{table}

\subsubsection{\textbf{Quantization}}
Quantization is a key LLM compression technique that reduces model size with only modest degradation in inference performance. Depending on when it is applied, quantization is categorized as post-training quantization (PTQ), which quantizes a pretrained model without retraining, and quantization-aware training (QAT), which simulates quantization effects during training or fine-tuning to improve robustness at low bit-widths. Furthermore, quantization can be grouped by the target component, i.e., weight-only quantization, and weight-activation quantization.
\begin{enumerate}[fullwidth, itemindent=1em, label={\arabic*})]
\item \emph{Weight-only Quantization}:
This category compresses LLMs by representing model parameters at reduced numerical precision while typically keeping activations in higher precision. It is particularly effective for edge LLM inference because weights dominate the memory footprint, lowering weight precision reduces memory traffic and can significantly improve token throughput. However, compared with conventional DNNs, LLMs often exhibit pronounced outlier weights that disproportionately contribute to quantization error, motivating outlier-aware PTQ designs. For example, AWQ \cite{MLSYS2024_42a452cb} preserves only 1\% of salient weights identified via activation statistics to mitigate quantization error, achieving more than $3\times$ speedup over FP16 implementation on both desktop and mobile GPUs. In \cite{10.1145/3579371.3589038, dettmers2023spqr, wei2023outlier, liu2023llm}, outliers are retained by sacrificing nearby low-importance values. In QAT settings, prior works, e.g., Q8BERT \cite{9463531} and mixed-precision schemes such as Q-BERT \cite{Shen_Dong_Ye_Ma_Yao_Gholami_Mahoney_Keutzer_2020}, show that training-aware or sensitivity-aware bit-width assignment can better preserve accuracy at low precision, and subsequent methods (e.g., TernaryBERT \cite{zhang-etal-2020-ternarybert} and BinaryBERT \cite{bai-etal-2021-binarybert}) further push quantization to ternary or even binary regimes.

\item \emph{Weight-Activation Quantization}:
This category lowers the precision of both weights and activations to reduce computation and data movement costs beyond weight-only schemes. Activation quantization is typically more challenging for LLMs because activations exhibit large dynamic ranges and can shift with prompts, sequence length, and token position.  Naively quantizing them may destabilize attention and MLP computations and amplify error during autoregressive decoding. ZeroQuant \cite{NEURIPS2022_adf7fa39} pioneers weight-activation quantization for LLMs with group-wise weight quantization and token-wise INT8 activation quantization. To handle outlier activations, LLM.int8() \cite{NEURIPS2022_c3ba4962} routes a small fraction of outlier channels through higher precision to stabilize INT8 inference without fine-tuning, while SmoothQuant \cite{pmlr-v202-xiao23c} redistributes activation magnitudes into the weights via complementary rescaling, enabling robust W8A8 and even W4A8 across diverse LLMs. Moreover, PTQ methods typically set quantization parameters via calibration methods such as min-max ranges, percentile clipping, or KL-divergence minimization \cite{Jacob_2018_CVPR, krishnamoorthi2018quantizing}. For tougher settings or domain shift, QAT integrates fake-quantization modules into the training loop so that networks explicitly learn robust scaling factors and clipping thresholds, with methods such as PACT \cite{choi2018pact} and LSQ \cite{esser2019learned} widely recognized for stabilizing low-bit training.
\end{enumerate}

\subsubsection{\textbf{Pruning}}
Pruning is another widely used LLM compression technique that removes unnecessary or redundant parameters. In general, pruning approaches can be divided into two categories, i.e., unstructured pruning and structured pruning.
\begin{enumerate}[fullwidth, itemindent=1em, label={\arabic*})]
\item \emph{Unstructured Pruning}: This form of pruning removes individual weights throughout the model without respecting architectural boundaries. Compared with classic magnitude pruning used in conventional DNNs, unstructured LLM pruning is often calibration-driven and layer-wise, because Transformer blocks are sensitive to perturbations and errors may propagate through residual connections during long-horizon generation.
For instance, SparseGPT \cite{pmlr-v202-frantar23a} is a one-shot method that removes a large fraction of weights while approximately reconstructing each layer's outputs using calibration data, achieving over 50\% sparsity in models like OPT and LLaMA without retraining. Wanda \cite{sun2023simple} combines weight magnitudes with input activation norms to drop weights that are both small and weakly activated, improving over simple magnitude pruning.
OWL \cite{10.5555/3692070.3694428} further refines pruning decisions by assigning higher pruning budgets to layers with more activation outliers through non-uniform hierarchical sparsity. Despite these advances, unstructured pruning inherently produces irregular and non-uniform weight distributions, which are poorly aligned with standard dense linear algebra kernels \cite{xia2023flash, 10120981}. Thus, realizing speedups in practice often requires specialized hardware (e.g., sparse tensor cores) or optimized software libraries (e.g., cuSPARSE\footnote{\url{https://docs.nvidia.com/cuda/cusparse/}} and oneMath\footnote{\url{https://github.com/uxlfoundation/oneMath}}).

\item \emph{Structured Pruning}: This form of pruning removes entire architectural units in a model, such as neurons/channels in MLPs, attention heads, or even whole layers, rather than individual weights. Because the remaining tensors stay dense and regularly shaped, structured pruning maps well to commodity kernels (e.g., general matrix-matrix multiplication (GEMM) and convolution) and thus delivers practical speedups on CPUs, GPUs, and NPUs.
It is especially suitable for LLMs since the Transformer architecture offers natural pruning granularity, and removing heads/hidden dimensions can further reduce attention cost and potentially shrink the KV-cache footprint, a key memory bottleneck for long-context decoding. A representative example is LLM-Pruner \cite{NEURIPS2023_44956951}, which performs task-agnostic compression by estimating component importance with first-order gradients and approximate Hessians, pruning structural groups, and then restoring accuracy with lightweight LoRA fine-tuning. Beyond attention heads and FFNs, some studies further explore pruning hidden dimensions (e.g., embedding layers, layer norm) for additional gains \cite{tao-etal-2023-structured, NEURIPS2023_ced46a50}. Sheared-LLaMA \cite{xia2023sheared} combines targeted structured pruning with dynamic batch loading to reshape pretrained models into target architectures. By adapting data sampling ratios across domains, it maintains quality while using about 3\% of the computation needed to train from scratch, making it a highly cost-effective approach.
\end{enumerate}

\subsubsection{\textbf{Knowledge Distillation}}
Knowledge distillation transfers the capabilities of a large ``teacher'' model to a smaller ``student'' model, producing a more efficient model that reduces size and deployment cost while aiming to retain performance. Distillation methods are commonly categorized as open-box or closed-box.
\begin{enumerate}[fullwidth, itemindent=1em, label={\arabic*})]
\item \emph{Open-Box Knowledge Distillation}: This approach exploits full access to the teacher model's architecture and parameters, transferring knowledge through both internal representations (e.g., hidden states, attention maps) and output logits.
    Early efforts such as MiniLM \cite{NEURIPS2020_3f5ee243} extract self-attention and value relations from the final Transformer layer to alleviate the complexity of strict layer-to-layer mapping, enabling smaller students to retain over 99\% of the teacher's accuracy with roughly half the parameters. To address teacher-student capacity mismatch, MiniLLM \cite{gu2024minillm} and TED \cite{pmlr-v202-liang23j} improve distillation robustness using reverse KL divergence and layer-wise alignment, respectively. FUSELLM \cite{wan2024knowledge} further boosts performance by fusing knowledge from multiple teachers at multiple granularities.

\item \emph{Closed-Box Knowledge Distillation}: This approach uses only the teacher's outputs and requires no access to its internal architecture or parameters, which suits proprietary or closed-source teachers (e.g., commercial LLMs). The student is trained to mimic the teacher's predictions or responses. PromptMix \cite{sahu-etal-2023-promptmix} generates labeled examples through prompting, where borderline examples improve knowledge transfer from teacher models such as GPT-3.5 to student models. For chain-of-thought reasoning, Fine-tune-CoT \cite{ho-etal-2023-large} demonstrates that large models (over 100B parameters) can distill their reasoning into students with as few as 0.3B parameters. Instruction following is another major focus, with Lion \cite{jiang-etal-2023-lion} achieving ChatGPT-level performance using a 13B-parameter model by adversarially generating complex instructions.
\end{enumerate}

\subsubsection{\textbf{Low-Rank Factorization}}
Low-rank factorization approximates a matrix $\mathbf{W} \in \mathbb{R}^{m \times n}$ with the product of two lower-rank matrices, $\mathbf{A} \in \mathbb{R}^{m \times r}$ and $\mathbf{B} \in \mathbb{R}^{r \times n}$, i.e., $\mathbf{W} \approx \mathbf{A}\mathbf{B}$, where $r \ll \min(m, n)$. This reduces parameters from $\mathcal{O}(mn)$ to $\mathcal{O}((m+n)\times r)$. ASVD \cite{yuan2023asvd} is the first work applying factorization to compress LLMs, explicitly managing activation outliers by adapting weight matrices to the activation distribution and iteratively calibrating layers according to their decomposition sensitivity. LASER \cite{sharma2023truth} further reveals that selectively removing higher-order components of weight matrices can surprisingly improve model performance while reducing complexity. DSFormer \cite{chand2023dsformer} factorizes weights into a semi-structured sparse matrix times a small dense matrix, yielding up to 40\% higher compression than standard low-rank factorization methods. Overall, low-rank factorization reduces memory footprint and computation. Since memory access often bottlenecks LLM decoding, it also lowers the number of parameters to load, speeding up generation.

\subsubsection{\textbf{Compression, accuracy, and trustworthiness trade-off}}
While quantization, pruning, distillation, and low-rank factorization facilitate edge deployment by reducing memory and computation, they can also change the generation behavior of LLMs. As compression becomes more aggressive (e.g., lower bitwidths, higher pruning ratios, or smaller distilled students), task performance may degrade and the impact is often uneven across tasks. Prompts that require strong factuality or multi-step reasoning are typically more sensitive. For generative applications, standard accuracy metrics should therefore be complemented with trustworthiness evaluations, including hallucination and factuality as well as bias and toxicity measurements \cite{lin-etal-2022-truthfulqa, nadeem-etal-2021-stereoset, gehman-etal-2020-realtoxicityprompts}. Because edge deployments commonly rely on compact or heavily quantized models, compression schemes should be designed and selected with a multi-objective view that explicitly balances inference efficiency (latency, memory, and energy) with task quality and trustworthiness. Moreover, although this survey focuses on inference rather than training, many compression techniques still require calibration or fine-tuning, such as PTQ calibration or QAT. When these updates are learned from distributed edge data, federated learning and other decentralized training frameworks \cite{9760729, piccialli2025federated} are particularly useful, as they can enable privacy-preserving adaptation while limiting data movement.

\begin{figure}[ht]
\centering
\subfigure[]{\includegraphics[width=0.3\textwidth]{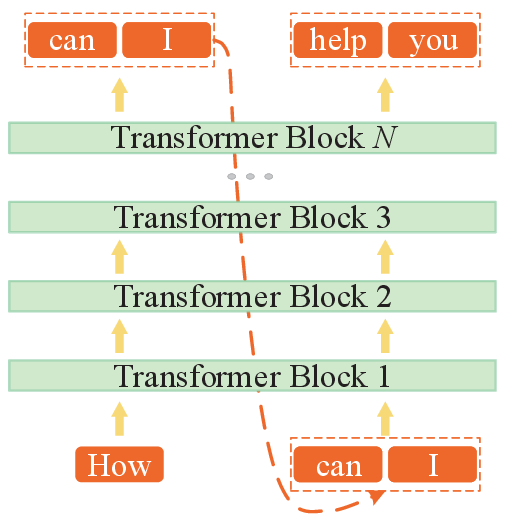}\label{fig:decoding_NonAutoregssive}}
\hspace{0.3cm}
\subfigure[]{\includegraphics[width=0.3\textwidth]{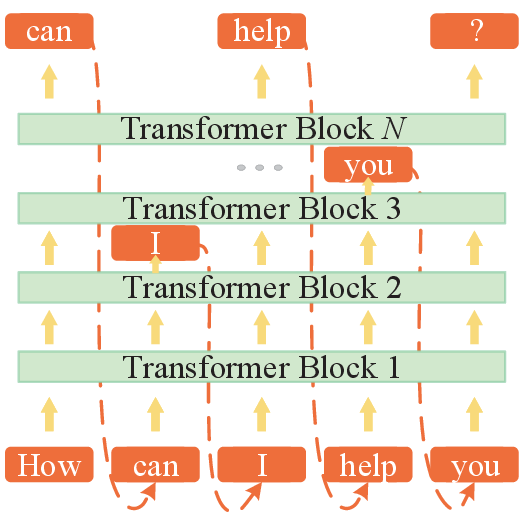}\label{fig:decoding_Early_exist}}\\
\hspace{0.05cm}
\subfigure[]{\includegraphics[width=0.45\textwidth]{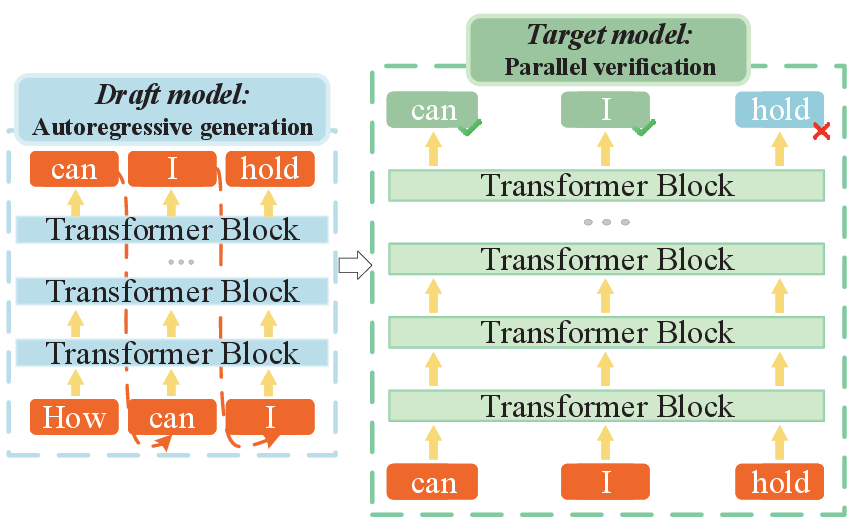}\label{fig:decoding_speculative_decoding}}
\hspace{0.3cm}
\subfigure[]{\includegraphics[width=0.45\textwidth]{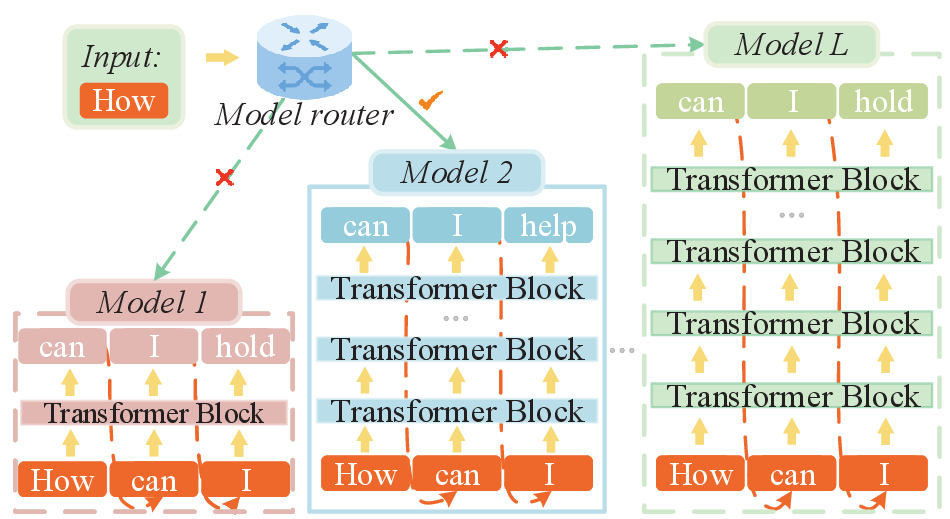}\label{fig:decoding_cascade_inference}}
\vspace{-0.1cm}
\caption{Decoding strategies:
(a) Non-autoregressive.
(b) Early exiting.
(c) Speculative decoding.
(d) Cascade inference.}
\label{fig:Decoding_algs}
\end{figure}

\subsection{Decoding Mechanisms}\label{subsec:decoding_strategy}
This subsection reviews decoding mechanisms, as illustrated in Fig. \ref{fig:Decoding_algs}, that improve the efficiency and effectiveness of LLM edge inference.
A comparison of different decoding mechanisms is provided in Table \ref{tab:decoding_compare}.
\subsubsection{\textbf{Non-autoregressive Decoding}}
Non-autoregressive decoding departs from standard autoregressive generation by generating multiple tokens in parallel, as shown in Fig. \ref{fig:decoding_NonAutoregssive}. It relaxes sequential dependencies by assuming a certain degree of conditional independence across output positions, removing the per-token feedback loop and markedly speeding up inference. However, non-autoregressive methods often lag in accuracy because they do not explicitly model token dependencies. To mitigate this limitation, knowledge distillation \cite{NEURIPS2019_f63f65b5} and source-target alignment \cite{Guo_Tan_He_Qin_Xu_Liu_2019} approaches are proposed to narrow the performance gap with autoregressive models. Moreover, semi-autoregressive decoding further extends non-autoregressive methods by modeling output dependencies \cite{ghazvininejad2019mask, gu2020fully} or iteratively refining predicted tokens \cite{lee2018deterministic}, with the goal of approaching autoregressive-level quality.
\begin{table}[ht]\small
\caption{Comparison of decoding mechanisms for LLM edge inference}
\centering
\label{tab:decoding_compare}
\begin{tabular}{|m{2.2cm}<{\centering}|m{2.5cm}<{\centering}|m{2.8cm}<{\centering}|m{3.2cm}<{\centering}|m{2.3cm}<{\centering}|}
\hline
Mechanism & Key Idea & Advantages & Limitations & Rep. works \\
\hline
Non-autoregressive decoding &  Generate tokens in parallel & Lower decoding latency, higher throughput & Possible quality drop, may need distillation/refinement & \cite{ghazvininejad2019mask, gu2020fully, lee2018deterministic} \\
\hline
Early exiting  & Stop at intermediate layers when confidence is high & Less compute per token; saves energy/latency & Needs reliable confidence/exit policy & \cite{NEURIPS2022_6fac9e31, del2023skipdecode, elhoushi2024layerskip, Zeng_Hong_Dai_Zhuang_Chen_2024}\\
\hline
Speculative decoding & Small draft model proposes, large model verifies in parallel & Speedup without retraining target model & Depends on draft accuracy, extra draft cost &  \cite{zhu2025efficient, zheng2025communication, bh_conformal, 10.1145/3662006.3662067, NEURIPS2023_6034a661}\\
\hline
Cascade inference &  Route queries across small$\rightarrow$large models; escalate if needed & Lower average cost/latency, reduces offloading/backhaul & Requires routing + multiple models; misrouting overhead & \cite{ding2024hybrid, li2020cascadebert, 11161094, NEURIPS2023_b914a8fc}\\
\hline
Reasoning \& agentic multi-LLM & Add deliberation, tool/agent loops & Improves correctness; reduces hallucinations & Higher latency/cost, coordination issues & \cite{NEURIPS2022_9d560961,wang2022self,NEURIPS2023_271db992,yao2022react,wang2024mixture,chen2024internet}\\
\hline
\end{tabular}
\end{table}

\subsubsection{\textbf{Early Exiting}}
Early exiting reduces latency and computational cost by skipping the computation of later layers instead of always executing the full model. It is commonly implemented by attaching lightweight exit heads at multiple depths and triggering an early stop based on confidence signals such as entropy or margin, as shown in Fig. \ref{fig:decoding_Early_exist}. Prior work in \cite{simoulin-crabbe-2021-many,geva2022transformer,xin2020deebert,wang-etal-2022-skipbert} observed that hidden states often saturate at intermediate layers for some tokens, so early exits can match the full model's top-1 prediction, motivating decoder-side early exiting. Building on this insight, CALM \cite{NEURIPS2022_6fac9e31} performs token-wise adaptive compute during generation using confidence- and saturation-based signals to select exit depth. However, token-level exiting can be hard to deploy with batched decoding because different sequences may require different depths, creating stragglers and KV-cache complications. SkipDecode \cite{del2023skipdecode} addresses this by designing exit behavior that remains compatible with batching and KV caching while achieving substantial speedups. More recently, LayerSkip \cite{elhoushi2024layerskip} trains LLMs for early-exit inference via layer dropout and an early-exit objective, and introduces self-speculative decoding that exits early and verifies with remaining layers without a separate draft model. Beyond heuristic thresholds, ConsistentEE \cite{Zeng_Hong_Dai_Zhuang_Chen_2024} learns exit policies via reinforcement learning to balance compute savings and accuracy. In edge deployments, exit policies should be co-optimized with compute and communication scheduling to meet end-to-end SLOs. For a broader discussion of early-exit design choices and trade-offs, we refer readers to the surveys in \cite{10.1145/3527155}.

\subsubsection{\textbf{Speculative Decoding}}
Speculative decoding mitigates the sequential bottleneck of autoregressive generation by having a small draft model propose multiple tokens per step, then letting the target LLM verify them in parallel, as shown in Fig. \ref{fig:decoding_speculative_decoding}. Only tokens that pass the target LLM's verification are accepted, preserving generation quality. This approach can significantly reduce latency without retraining or modifying the original LLM. Recent systems deploy draft and target models across heterogeneous edge nodes \cite{zhu2025efficient}, split drafting on-device with verification at the base station while transmitting only sparse logits \cite{zheng2025communication}, or use edge drafting with cloud verification plus output sparsification and quantization to cut end-to-end latency \cite{bh_conformal, 10.1145/3662006.3662067}. To increase draft-token acceptance rate, \cite{NEURIPS2023_6034a661} uses batch verification and resampling over multiple draft sequences, preserving the target distribution and delivering an additional $1.37\times$ speedup over standard speculative decoding.
However, the effectiveness of speculative decoding depends strongly on draft accuracy and the acceptance rate of predicted tokens, i.e., low-quality drafts may lead to frequent rejections and limiting the speedup of speculative decoding. Thus, balancing speed gains with reliability is essential.

\subsubsection{\textbf{Cascade Inference}}
Cascade inference handles varying user query difficulty by routing requests through a hierarchy of LLMs with increasing capacity, exiting early when a smaller model is sufficiently confident, as shown in Fig. \ref{fig:decoding_cascade_inference}. A trivial cascade runs a lightweight front-end LLM first and escalates to larger LLMs when confidence falls below a threshold. Several representative approaches demonstrate the effectiveness of this paradigm. For instance, \cite{ding2024hybrid} leverages a learning-based router to adaptively route queries between a small LLM on the end device and a large LLM in the cloud, reducing calls to the large LLM by up to 40\% without degrading response quality. Given limited memory and computation resources at the edge server, \cite{li2020cascadebert, 11161094} deploys multiple LLMs on the edge server to meet diverse user demands and uses a DRL-based method to learn the task scheduling policy for minimizing serving delay. A concurrent study \cite{NEURIPS2023_b914a8fc} jointly optimizes model multiplexing and query caching, providing theoretical analysis of optimality in minimizing inference cost. Overall, cascade inference is a promising way for improving the efficiency for LLM edge inference, where most queries can be handled locally by a compact model, while only difficult cases are offloaded to more powerful devices or the cloud. Nonetheless, designing accurate dispatching mechanisms remains a key challenge, as poor routing decisions can compromise generation quality.

\subsubsection{\textbf{Reasoning and Agentic Multi-LLM Inference}}
While model compression and faster decoding improve efficiency, generation quality and reliability remain critical, especially at the mobile edge where hallucinations and inconsistent outputs can undermine user trust and downstream actions. Recent work augments decoding with reasoning-enhanced inference that allocates extra deliberation to improve correctness, including step-by-step decomposition of problems such as chain-of-thought \cite{NEURIPS2022_9d560961}, sampling multiple reasoning traces and aggregating for consistency \cite{wang2022self}, searching over alternative reasoning trajectories \cite{NEURIPS2023_271db992}, and grounding responses through retrieval or tool use \cite{yao2022react}. These methods can reduce hallucinations through intermediate checks and evidence grounding, but often increase cost via longer contexts, multiple rollouts, or tool calls.

Building on reasoning, agentic AI extends LLM serving from single-turn generation to autonomous multi-step task execution via iterative plan-act-observe loops with memory and tools. Multi-LLM agents further improve capability and efficiency by orchestrating multiple specialized models through delegation, routing, and iterative refinement. For instance, \cite{wang2024mixture} aggregates outputs from multiple agents in layered stages to enhance generation quality, while \cite{chen2024internet} proposes a scalable framework for discovering and coordinating heterogeneous agents in a distributed, internet-like setting. These approaches naturally align with edge intelligence by enabling tiered orchestration across device, edge, and cloud resources, where lightweight on-device agents handle local and privacy-sensitive steps and more capable edge or cloud agents perform retrieval or compute-intensive reasoning, balancing quality, latency, bandwidth, and privacy. However, they also introduce new challenges, including multi-step latency accumulation, coordination over time-varying wireless links, long-lived state and memory management, and governance for autonomous actions.

\subsection{LLM Deployment at the Network Edge}\label{subsec:LLM_deployment}
In this subsection, we review LLM deployment techniques for the wireless network edge, including Transformer partitioning and placement, LLM caching and routing, and the disaggregation of prefill and decoding.

\subsubsection{\textbf{LLM Partition and Placement}}\label{subsubsec:LLM_partition}
Deploying LLMs under collaborative inference architectures (vertical, horizontal, hybrid) in Section \ref{sec:architectures} requires partitioning the LLM and placing its components across devices, edge servers, and the cloud to meet latency and throughput targets under memory, bandwidth, and energy constraints. At the edge, effective partitioning depends not only on FLOPs and compute capacity, but also on wireless channel quality, mobility, backhaul limits, and tight memory budgets. In practice, LLM partitioning strategies range from coarse Transformer block-level splits to fine-grained layer-level placements. Transformer block-wise  partitioning is the most widely used technique in LLM inference, splitting the model at the granularity of Transformer blocks, with a single Transformer block as the minimal unit. For example, \cite{pmlr-v235-jiang24f, 9996638, 10818760} proposes dynamic programming-based methods that partition an LLM into consecutive Transformer blocks and deploy the resulting partitions in vertical device-to-cloud and horizontal edge server collaborative architectures, respectively.
For horizontal settings, \cite{zhao-etal-2024-lingualinked, 10.1145/3627535.3638480} formulate LLM partitioning and placement as an integer linear program solved with Gurobi\footnote{\url{https://www.gurobi.com/solutions/gurobi-optimizer/}} to match segments to device capabilities, while \cite{10966456} casts partitioning as a matching game and finds a stable, low-latency assignment.
To further improve system performance, fine-grained techniques partition LLMs within Transformer layers, such as separating attention and MLP sublayers to fit heterogeneous memory and bandwidth budgets \cite{kafetzis2025large}, or sharding tensor dimensions of large projection matrices across devices, trading memory relief with extra inter-node communication \cite{shoeybi2019megatron}. Mixture-of-experts (MoE)-based LLMs further improve the quality-latency trade-off by activating only a few experts per token, but edge deployment is harder because expert weights are large, activations are skewed, and expert dispatch can amplify communication and load imbalance.
These issues motivate communication-aware expert placement and caching \cite{11083676, song2025mixture, 11161045}, which partition MoE models at the expert level and distribute experts across edge devices to balance memory and compute demand. In addition, routing and batching should be expert-aware to reduce switching and maximize reuse, and load balancing is crucial to mitigate stragglers under heterogeneous wireless or interconnect conditions.

\subsubsection{\textbf{LLM Caching and Selection}}\label{subsubsec:LLM_caching}
In practice, edge deployments seldom rely on a single LLM due to the varying user preferences and model capabilities. A portfolio of models with different sizes and providers can be spread across end devices, edge servers, and the cloud. Under tight memory, compute, and bandwidth budgets, LLM caching and routing jointly decide which LLM instances stay resident at each tier, when to prefetch or evict, and how to dispatch each request to meet latency-accuracy targets under wireless and capacity constraints. Unlike conventional data caching that stores immutable content objects and serves exact matches by key, LLM caching means keeping selected LLM service instances (weights, segments, and possibly adapters) resident to avoid loading and cold-start overheads and to reduce cross-tier traffic. Its objective and constraints are therefore dominated by GPU memory residency, model loading or migration cost, and accuracy-latency trade-offs, and it is naturally coupled with request routing and model selection under heterogeneous and time-varying edge resources. Towards this direction, \cite{11147210} and \cite{ding2024hybrid} propose semantic-aware and learning-based routers that orchestrate inference between a lightweight LLM on the device and a larger LLM at the edge server to balance latency and quality, while \cite{11086391, 11161094} develop DRL-based routers that dispatch user requests to appropriate edge-hosted LLMs. Moreover, \cite{10.1145/3669940.3707215} partitions an LLM into segments, replicates segments across edge devices, and formulates routing as a max-flow problem with a heuristic that improves throughput by up to 3.3$\times$ over no-replication baselines.

\subsubsection{\textbf{Disaggregated Prefill and Decoding}}\label{subsubsec:disaggregrate_pd}
Motivated by the two-phase asymmetry of LLM inference (Section \ref{subsec:LLM_edge_challenges}), disaggregated prefill and decoding separates prefill and decoding phases so each can be provisioned, parallelized, and scheduled on hardware matched to its bottlenecks. DistServe \cite{298687} routes prefill to throughput-oriented devices and decoding to memory-efficient devices, handing off prompt state to reduce cross-phase interference and improve goodput. \cite{11044734} proposes intra-sequence pipeline parallel planning for efficient prefill and outline-based parallel decoding across collaborative edge nodes to overcome single-device limits, while SplitWiser \cite{aali2025splitwiser} keeps both phases on the same device to avoid inter-device transfers and improve cache reuse, reducing network overhead and end-to-end latency. To handle fluctuations in prompt and output lengths, Arrow \cite{wu2025arrow} dynamically adjusts prefill and decoding instances using real-time cluster metrics, substantially improving resilience to traffic spikes and load variation. Overall, phase-aware disaggregation with lightweight state movement improves throughput, tail latency, and hardware efficiency. It is particularly effective at the edge, where prefill can burst to near-edge or cloud resources while decoding runs locally to deliver low token latency, save energy, and preserve privacy.

\section{Resource Management and Scheduling}\label{sec:resource_scheduling}
Beyond model optimization and deployment, efficient LLM edge inference hinges on coordinated resource management and scheduling. This section surveys cross-layer techniques spanning parallel computing, memory management, communication optimization, and their joint optimization.

\subsection{Computation Scheduling}
This subsection reviews two computation scheduling techniques, namely batching and parallel computing.
\begin{table}[ht]\small
\caption{Comparison of batching techniques for LLM edge inference}
\centering
\label{tab:batching_scheduling_compare}
\begin{tabular}{|m{1.5cm}<{\centering}|m{2.3cm}<{\centering}|m{3.3cm}<{\centering}|m{3.5cm}<{\centering}|m{2cm}<{\centering}|}
\hline
Technique & Goal & Key Idea & Overheads & Rep. works \\
\hline
Static batching  & Max throughput & Wait until batch fills, then run & High queueing delay, poor under bursty arrivals and wireless jitter & --  \\
\hline
Dynamic batching &  Balance latency and throughput & Run when batch size or timeout is met & Head-of-line delay with variable lengths, sensitive to arrival variability & \cite{10682995, 10382642, wu2023fast}\\
\hline
Continuous batching & High utilization (variable lengths) & Admit/finish requests at any decoding step & More complex scheduling and KV management & \cite{280922, 10.1145/3600006.3613165, holmes2024deepspeed} \\
\hline
Chunked prefill & Better long-prompt efficiency  & Split long prompts and overlap prefill with decoding & Needs chunk-size tuning, extra scheduler/kernel overhead  & \cite{298679, 11207452}\\
\hline
\end{tabular}
\end{table}

\subsubsection{\textbf{Batch Optimization}}
Batching groups multiple input requests for simultaneous processing, fully exploiting the parallelism of modern accelerators such as GPUs and TPUs to boost throughput and hardware utilization.
Common batching strategies are summarized below, with a comparison in Table \ref{tab:batching_scheduling_compare}.

\begin{enumerate}[fullwidth, itemindent=1em, label={\arabic*})]
\item \emph{Static Batching}: This form waits for a fixed number of requests to arrive and fills the batch before processing begins. In general, static batching works well in scenarios with controlled request rates or offline inference scenarios where all inference tasks are arrived before inference starts, but it can cause unbounded delays in the online inference scenarios where requests typically arrive irregularly.

\item \emph{Dynamic Batching}:
Dynamic batching bounds the delay of static batching by launching a batch when either a target batch size is reached or a time window expires. The batch size and time limit can be configured to balance throughput and latency according to workload demands. It is widely used in DNN inference with stochastic arrivals (e.g., \cite{10682995, 10382642}) and extended to LLM serving frameworks such as  vLLM \cite{10.1145/3600006.3613165} to balance responsiveness and throughput. However, these approaches batch requests in First-Come-First-Served (FCFS) order, which can be suboptimal because early long requests block shorter ones and inflate latency. To this end, \cite{wu2023fast} adopts a skip-join multi-level feedback queue that prioritizes high-priority requests and preempts long-running tasks to accelerate short ones, improving throughput by up to 31.4\% under latency constraints over vLLM. A key limitation is that new requests can be admitted only after the current batch finishes. Dynamic batching works well when requests have similar runtimes, as in conventional DNNs or encoder-only LLMs. For autoregressive LLMs, highly variable output lengths force all requests in a batch to wait for the longest one, causing idle GPU time and reducing efficiency.

\item \emph{Continuous Batching}:
This technique extends dynamic batching by allowing requests to join or exit a batch at any decoding step, rather than waiting for the entire batch to complete. It is well suited to autoregressive LLMs with highly variable output lengths, since it prevents shorter queries from being blocked by longer ones. By overlapping prefill and decoding and reusing cached KV states, continuous batching improves GPU utilization and reduces head-of-line blocking compared to static or dynamic batching. Orca \cite{280922} implements this approach via iteration-level scheduling and selective batching, where the former constructs batches at every decoding step to accommodate new arrivals, while the latter targets only batchable transformer operations, achieving $36.9\times$ higher throughput than conventional batching at the same latency level. Modern LLM serving frameworks such as vLLM \cite{10.1145/3600006.3613165} and DeepSpeed-FastGen \cite{holmes2024deepspeed} have adopted continuous batching as a core mechanism to achieve high throughput while maintaining interactive latencies. However, continuous batching requires sophisticated scheduling to handle challenges such as memory fragmentation in the KV cache and fairness across heterogeneous requests.

\item \emph{Chunked-Prefill}:
This technique targets the inefficiency of processing long input prompts by splitting them into smaller segments that are processed incrementally. The decoding for the first segment can start immediately while subsequent segments undergo prefill, allowing these phases to overlap and reducing peak memory usage. Building on this idea, Sarathi-Serve \cite{298679} splits prompts into near-equal sized chunks and creates stall-free schedules that add new requests in a batch without pausing ongoing decoding, achieving 3.7$\times$ higher serving capacity than vLLM. \cite{11207452} further proposes a dynamic adaptive chunk-based prefetching strategy to overcome the resource and latency limitations inherent in static chunking approaches, improving throughput by 24.4\% over Sarathi-Serve. Chunked-prefill is especially promising in edge environments because it enables serving long-context queries without overwhelming limited hardware resources. However, too many small chunks can increase scheduler and kernel overhead, making chunk-size tuning critical.
\end{enumerate}

\subsubsection{\textbf{Parallel Computing}}
\begin{figure}[ht]
\centering
\subfigure[Data Parallelism]{\includegraphics[width=0.25\textwidth]{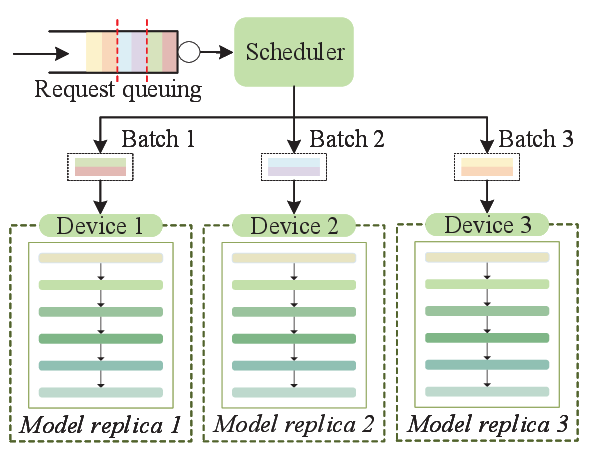}\label{fig:DP}}
\subfigure[Tensor Parallelism]{\includegraphics[width=0.36\textwidth]{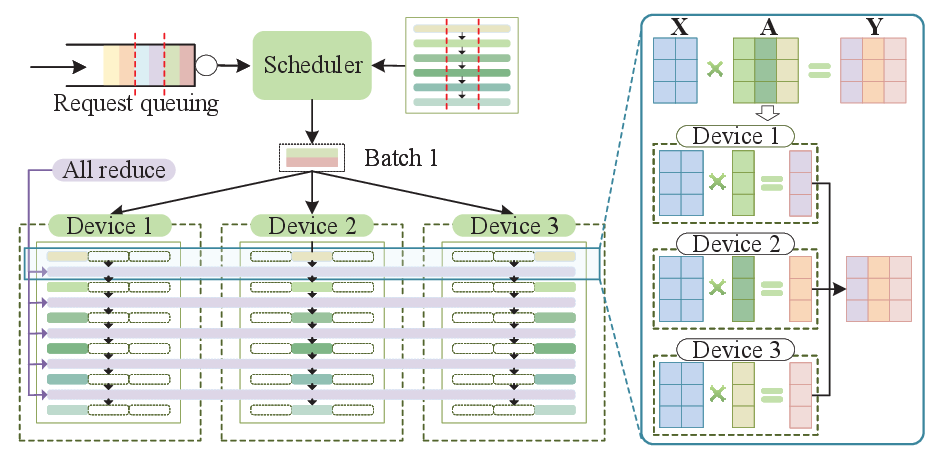}\label{fig:TP}}
\subfigure[Pipeline Parallelism]{\includegraphics[width=0.36\textwidth]{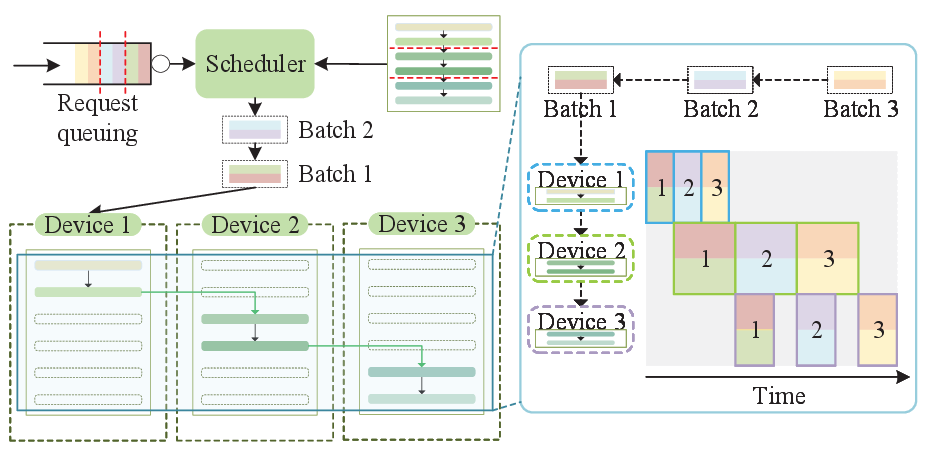}\label{fig:PP}}
\caption{Comparison of different parallelism methods.}
\label{fig:Parallel_computing}
\end{figure}

LLMs often have billions to trillions of parameters, making compute and memory key barriers to deployment at the resource-constrained edge. Parallel computing mitigates these limits by exploiting hardware and algorithmic concurrency, distributing work across cores or devices to accelerate inference. Below, we review parallelization strategies, as shown in Fig. \ref{fig:Parallel_computing}, with a comparison in Table \ref{tab:parallelism_compare}.
\begin{table}[ht]\small
\caption{Comparison of parallelism computing techniques}
\centering
\label{tab:parallelism_compare}
\begin{tabular}{|m{1.5cm}<{\centering}|m{2.6cm}<{\centering}|m{3.3cm}<{\centering}|m{3.6cm}<{\centering}|m{2cm}<{\centering}|}
\hline
Technique & Goal & Key Idea & Overheads & Rep. works \\
\hline
DP  & Scale throughput & Replicate model, split requests across replicas & Needs load balancing, routing overhead across nodes & \cite{9355301,nvidia_fast_transformer}  \\
\hline
TP &  Fit/accelerate large layers & Split matrix operations across devices & Communication-heavy, performance depends on interconnect quality & \cite{NEURIPS2024_0f4d1fc0, 10.1145/3620666.3651357,shoeybi2019megatron}\\
\hline
PP & Fit model across devices, overlap stages & Split layers into stages, micro-batch pipeline & Bubble overhead, activation transfers between stages & \cite{pmlr-v235-jiang24f, 10818760, 10966456} \\
\hline
Hybrid parallelism & Combine scalability + flexibility  & Combine DP/TP/PP to match heterogeneity & More complex orchestration, higher coordination overhead & \cite{280874,fan2021dapple}\\
\hline
\end{tabular}
\end{table}
\begin{enumerate}[fullwidth, itemindent=1em, label={\arabic*})]
\item \emph{Data Parallelism (DP)}:
DP replicates the full model across multiple devices and distributes inference requests such that each device processes its assigned inputs independently \cite{9355301,nvidia_fast_transformer}, as illustrated in Fig. \ref{fig:DP}. It is simple to implement with minimal framework changes, but edge deployments require careful workload scheduling and batching to meet performance targets under the constraints in Section \ref{subsec:LLM_edge_challenges}.

\item \emph{Tensor Parallelism (TP)}:
TP shards compute-intensive LLM operations (e.g., matrix multiplications) across devices, then merges the partial results using collectives such as all-reduce or all-gather \cite{NEURIPS2024_0f4d1fc0, 10.1145/3620666.3651357, shoeybi2019megatron}. For instance, Fig. \ref{fig:TP} shows a case where three devices collaboratively computing $\mathbf{X} \times \mathbf{A} = \mathbf{Y}$ by sharding $\mathbf{A}$ across the devices in a column-wise manner (row-wise is also possible) and then aggregating the partial results. By distributing large computations, TP accelerates inference and reduces per-device memory footprint. Nevertheless, frequent inter-device communication can impose significant overhead, and poor partitioning strategies may lead to load imbalance and reduced efficiency.

\item \emph{Pipeline Parallelism (PP)}:
PP partitions an LLM into multiple stages, with each stage comprising one or more consecutive model layers. Each device is assigned a pipeline stage and is responsible for executing its corresponding layers \cite{hu2021pipeline, 10.1145/3636534.3649363, 10.1145/3627535.3638480}, as shown in Fig. \ref{fig:PP}. During inference, input requests are partitioned into micro-batches that sequentially traverse the pipeline stages, with each device forwarding its computed activations to the device responsible for the next stage. This arrangement reduces per-device memory usage by distributing model layers across devices and can improve throughput by overlapping computation across micro-batches. Motivated by these advantages, PP has been widely investigated for horizontal edge collaborative inference scenarios, such as \cite{pmlr-v235-jiang24f, 10818760, 10966456}.

\item \emph{Hybrid Parallelism}:
Hybrid parallelism combines DP, TP, and PP to leverage their complementary strengths while alleviating individual limitations \cite{280874,fan2021dapple}. For instance, within a device group, TP can be applied to split large matrix multiplications across GPUs, while PP distributes layers across groups, and DP replicates these groups to process independent requests in parallel. This multi-dimensional design enables scaling to extremely large LLMs that exceed the memory capacity or compute limits of any single parallelism method alone. At the wireless edge, hybrid schemes help handle device heterogeneity and network variability by balancing work within high-bandwidth local clusters with TP or PP while using DP across geographically distributed nodes to raise throughput.
\end{enumerate}

\subsection{Memory Management}
The runtime memory footprint in LLM inference mainly stems from model weights, activations, and the KV cache. Weights are static once deployed and activations are relatively small and predictable, whereas the KV cache grows with context length during autoregressive decoding and often becomes the dominant term. Effective memory management is therefore essential, otherwise context length, batch size, and throughput are capped. Below we review two complementary mechanisms, i.e., memory allocation and memory offloading, with a comparison in Table \ref{tab:memory_management_compare}.

\begin{table}[ht]\small
\caption{Comparison of memory management techniques}
\centering
\label{tab:memory_management_compare}
\begin{tabular}{|m{2cm}<{\centering}|m{2.2cm}<{\centering}|m{3.5cm}<{\centering}|m{3.8cm}<{\centering}|m{1.6cm}<{\centering}|}
\hline
Technique & Goal & Key Idea & Overheads & Rep. works \\
\hline
Fixed KV pre-allocation  & Avoid OOM, simplify runtime & Reserve contiguous KV for max length & Wastes memory, lowers concurrency with variable lengths & \cite{pmlr-v235-jiang24f, 9996638, 10818760, shoeybi2019megatron, 11083676, 11161045}  \\
\hline
Paged KV cache &  Reduce fragmentation & Store KV in noncontiguous ``pages'' & Allocator/bookkeeping overhead, irregular access & \cite{10.1145/3600006.3613165}\\
\hline
Paged KV kernel and layout co-design & Improve paged efficiency &Co-design KV layout and attention kernels & Implementation complexity, hardware-specific tuning & \cite{ye2025flashinfer} \\
\hline
Prefix reuse / tree attention & Reuse shared KV  & Share prefixes across candidates/branches & Extra control logic, workload dependent & \cite{10.1145/3620666.3651335}\\
\hline
Token-level KV management & Fine-grained KV utilization & Manage KV at token granularity to reduce waste & Higher bookkeeping overhead, gains depend on workload & \cite{chen2025pre, zhou2024dynamickv, wu-etal-2025-tokenselect} \\
\hline
Memory offloading & Fit larger models/ context & Place weights/activations/KV across devices & Data transfer overhead, sensitive to bandwidth & \cite{pmlr-v202-sheng23a, 10949701}\\
\hline
\end{tabular}
\end{table}

\subsubsection{\textbf{Memory Allocation}}
Considering the KV cache grows and shrinks dynamically during decoding, many LLM edge inference studies, e.g., \cite{pmlr-v235-jiang24f, 9996638, 10818760, kafetzis2025large, shoeybi2019megatron, zhang2025communication, 11083676, song2025mixture, 11161045},  pre-allocate a contiguous piece of memory with a fixed maximum sequence length to avoid OOM errors. However, this wastes significant memory by holding reserved memory until each request completes, blocking reuse and reducing effective system capacity. To mitigate fragmentation and overprovisioning, vLLM \cite{10.1145/3600006.3613165} introduces PagedAttention, which partitions the KV cache into noncontiguous pages and stores it in a paged format similar to an operating system, improving throughput by $2$-$4\times$ over Orca. To counter the irregular memory accesses introduced by paged layouts, FlashInfer \cite{ye2025flashinfer} co-designs KV-cache data layouts and attention-kernel access patterns for paged caches, efficiently mapping virtual to physical pages to restore contiguous, vectorized loads. SpecInfer \cite{10.1145/3620666.3651335} uses tree attention with depth-first traversal to reuse shared prefixes across multiple output sequences and eliminate redundant KV allocations. \cite{chen2025pre, zhou2024dynamickv, wu-etal-2025-tokenselect} push granularity further by introducing token-level KV cache management to reduce waste. However, these fragmented memory management introduces allocator and bookkeeping overheads. When other optimizations already increase batch size, the additional throughput gains can be limited while latency overhead becomes significant.

\subsubsection{\textbf{Memory Offloading}}
While memory allocation reduces the footprint on a single device, the memory demands of LLM inference often exceed the capacity of one device. Consequently, memory offloading that distribute weights, activations, and the KV cache across tiers such as CPU memory, local storage, and even remote devices are essential to meet capacity goals \cite{jiang2024neo}. On a single machine, FlexGen \cite{pmlr-v202-sheng23a} formulates a linear program to place weights, activations, and KV across GPU, CPU, and disk, enabling single-GPU inference of OPT-175B under tight budgets and achieving a generation throughput of 1 token/s with a batch size of 144 for the first time.
TightLLM \cite{10949701} uses an adaptive offloading policy that balances data transfers with compute, recomputes parts of the KV cache to overlap data movement, and slices weight loading across batches to amortize loading overhead, significantly improving throughput. Beyond a single device, weights, activations, and KV caches can be offloaded to peer edge devices or even to the cloud.

\subsection{Communication Efficiency Enhancement}
For LLM edge inference, communication is often the main bottleneck for latency and energy targets because prompts, retrieved context, and intermediate activations are transmitted among devices, edge servers, and the cloud.
The bandwidth-limited uplinks and time-varying channels affected by fading, interference, mobility, and blockage can increase delay and retransmissions. These constraints motivate communication-aware designs beyond generic compression. Below, we review three promising techniques to address this bottleneck.

\subsubsection{\textbf{Semantic Communication}}
In LLM edge inference, end devices may need to transmit high-volume data, such as prompts and activations, to peer devices or edge servers, leading to substantial communication overhead, especially with long prompts and large prefill activations. Semantic communication reduces this cost by transmitting compact, task-aware representations of raw inputs, intermediate features, or textual context instead of source-agnostic bitstreams, while largely preserving downstream utility and generation quality \cite{10832517, 10949629}. In extended reality (XR) applications and multimodal mobile assistants, this helps because compact semantics shrink uplink payloads and reduce end-to-end latency. In vehicle-to-everything (V2X) perception and cooperative sensing, this helps because exchanging only task-relevant information scales better under multi-user contention. However, designing semantic encoders and decoders that handle data and device heterogeneity, time-varying wireless conditions, and privacy constraints while meeting strict latency, accuracy, and energy targets remains challenging and warrants further investigation.

\subsubsection{\textbf{Over-the-Air Computation}}
Over-the-air computation (AirComp) exploits the analog superposition property of wireless multiple access channels, allowing many devices to transmit simultaneously so that the receiver directly obtains a desired function of their signals, typically a weighted sum. Recent work in \cite{zhang2025communication} applies AirComp to distributed on-device LLM inference with tensor parallelism and shows that analog all-reduce can significantly reduce end-to-end inference delay compared with digital baselines. AirComp is particularly useful for industrial IoT and smart factories, where many devices can aggregate updates or intermediate results in one shot, reducing scheduling overhead and latency. For UAV swarms and V2X collaboration, the simultaneous transmissions can accelerate collective inference as many participants must synchronize frequently. However, AirComp is sensitive to synchronization errors, imperfect channel estimates, and hardware nonidealities. It therefore requires careful calibration, power control, and device grouping.

\subsubsection{\textbf{Spatial-domain Technologies}}
In addition to semantic communication and AirComp, spatial-domain technologies such as multiple-input multiple-output (MIMO) \cite{6736761, 6798744}, reconfigurable intelligent surfaces (RIS) \cite{8741198}, and programmable antennas \cite{9264694} can shape wavefronts to increase spectral efficiency, enhance reliability, and shorten end-to-end transmission time between edge devices. These gains translate directly to faster transmission of prompts and activations between edge devices, and lower energy consumption. They are particularly valuable in dense, multi-user deployments, where massive MIMO enables spatial multiplexing of numerous uplink transmissions, and in high-frequency or blockage-prone environments, where RIS and antenna reconfiguration can mitigate blockages for mobile XR or V2X links. However, spatial-domain techniques increase system complexity because they require accurate and timely channel state acquisition, pilot and control overhead, RIS reconfiguration latency, and hardware calibration, all of which must be carefully managed to preserve end-to-end latency benefits.

\subsection{Joint Optimization of Communication, Computation, and Memory}
LLM edge inference is constrained by tightly coupled budgets of communication, computation, and memory. Communication cost depends on what is transmitted (prompts, activations, KV state) and how often, computation depends on batching, parallelism, and per-token workload, and memory limits model size, context length, and concurrency. Optimizing any single resource in isolation often harms another. For example, larger batches improve accelerator utilization but increase KV memory pressure and queueing delay, while offloading reduces on-device compute but can raise communication latency. These couplings motivate coordinated optimization across all three resources to meet latency, throughput, and energy targets. Prior work in \cite{zhu2025efficient} combines speculative decoding with pipeline parallelism to overlap drafting and verification across concurrent tasks, and then uses dynamic programming to jointly optimize speculation length, batching, and communication and memory allocation. In \cite{10759588}, batching, model quantization, and communication and computation provisioning are co-optimized to improve throughput on resource-constrained edge servers. \cite{gond2025tokenweave} further overlaps communication and computation by splitting each batch into two wave-aware token subsets. Recent works also show that phase disaggregation enables more flexible joint optimization. DistServe \cite{298687} separates prefill and decode onto different GPU pools and co-optimizes phase-specific resource allocation and parallelism under latency constraints while accounting for disaggregation-induced communication. Follow-up work further explores fine-grained scheduling and resource partitioning for phase-disaggregated serving \cite{10.1145/3695053.3730999, hong2025semi}. Despite the progress, joint optimization of communication, computation, and memory for LLM edge inference remains at an early stage. Further work is needed on online and robust optimization under time-varying wireless channels, cost models that capture prefill and decode asymmetry and KV growth, and cross-layer mechanisms that enforce strict latency and energy targets without sacrificing service quality.

\section{Evaluation Methodology}\label{sec:evaluation_method}
This section outlines the evaluation methodology for LLM edge inference, covering metrics, models, datasets, and platforms to support rigorous evaluation and deployment.

\subsection{Metrics}

\subsubsection{\textbf{Latency}}
Latency measures how quickly an LLM inference system responds and delivers results to a user request.
The key metrics to measure latency are:
\begin{enumerate}[fullwidth, itemindent=1em, label={\arabic*})]
\item \emph{Time to First Token (TTFT)}: TTFT \cite{298687,10609649,aali2025splitwiser,10946802} is the elapsed time from when an LLM request is accepted by the serving system to when the first output token is generated (or begins streaming to the user). It captures how quickly the system starts responding. TTFT typically includes (i) request queueing, (ii) prefill/encoding (e.g., tokenization and attention over the full prompt to build the KV cache), and (iii) network/coordination delays in distributed edge execution. TTFT generally increases with prompt length since the full input must be processed before decoding can begin.

\item \emph{End-to-End Latency (E2EL)}: E2EL \cite{280874,10609649,wu2023fast,280922,aali2025splitwiser,10.1145/3627535.3638480} is the elapsed time from request submission to receiving the complete response, and thus directly determines perceived responsiveness. It spans the full serving pipeline, including queueing, batching/scheduling, prefill and decoding, post-processing, and any network latency in split or multi-edge deployments. Notably, a low TTFT does not imply a low E2EL, since slow token generation can still dominate the time to completion.

\item \emph{Token Generation Time (TGT)}: TGT \cite{ye2025flashinfer} is the elapsed time to stream all output tokens after the first token is produced. It excludes TTFT and measures only the steady generation phase. Formally, $\text{TGT}=\text{E2EL}- \text{TTFT}$.

\item \emph{Time per Output Token (TPOT)}: TPOT \cite{298687, 298679, wu2023fast,aali2025splitwiser} is the average inter-token latency during the steady generation phase, i.e., after the first token is produced.
It captures how quickly the serving system emits subsequent tokens. Lower TPOT corresponds to faster token production and higher throughput.
\end{enumerate}

\subsubsection{\textbf{Throughput}}
Throughput characterizes how much useful work an LLM inference system delivers per unit time \cite{298679} and is typically reported in three forms:
\begin{enumerate}[fullwidth, itemindent=1em, label={\arabic*})]
\item \emph{Tokens per Second (TPS)}: TPS \cite{280922,10818760,pmlr-v202-sheng23a,wu2023fast,10.1145/3627535.3638480} quantifies system throughput, i.e., the number of tokens processed per second across all concurrently served requests over a measurement window. In practice, TPS often reported separately for prefill (prompt processing) and decode (iterative generation), as the two phases exhibit different bottlenecks. TPS generally increases with larger batch sizes and higher parallelism, but may come at the cost of higher queueing delay and worse tail latency.
\item \emph{Requests per Second (RPS)}: RPS \cite{10759588,10591707,mudvari2024splitllm,9996638} measures the number of requests an LLM serving system completes per second over a measurement interval, reflecting its capacity to sustain concurrent workloads. Unlike TPS, which scales with response length, RPS is request-centric and thus depends on the mix of prompt/output lengths, decoding settings (e.g., max tokens), and system policies such as batching, scheduling, and admission control.
\item \emph{End-to-End Throughput (E2ET)}: E2ET \cite{10759588,280922} quantifies the number of tasks completed per unit time. It reflects the effective throughput of the entire serving pipeline, from request arrival to response completion, including admission and queueing, batching and scheduling, prefill and decoding, post-processing, and any communication overhead in split or multi-edge deployments.
\end{enumerate}

\subsubsection{\textbf{Goodput}}
Goodput extends the classic networking notion of ``useful throughput'' to LLM serving by measuring how many requests per second the LLM successfully completes while meeting predefined SLOs, such as TTFT, end-to-end latency, and timeout constraints \cite{wang2024revisiting, 298687}. By construction, goodput $\le$ throughput, since it excludes errors, timeouts, policy blocks, and responses that violate the SLO. Goodput is particularly important at the edge because limited compute and memory, constrained bandwidth, energy budgets, and network variability can cause queueing and tail-latency spikes, reducing the share of SLO-compliant requests even when raw TPS or RPS is high. Improving goodput therefore requires SLO-aware control rather than simply maximizing utilization. Common approaches include admission control and load shedding to avoid queue buildup, latency-aware batching and scheduling, KV-cache-aware concurrency limits and cache management, adaptive model or precision selection and speculative decoding to speed generation under tight budgets, and offloading or split execution that trades compute savings for communication delay.

\subsubsection{\textbf{Energy Efficiency}}
Energy efficiency quantifies how much electrical energy is consumed to produce a response, which can be measured as following metrics:
\begin{enumerate}[fullwidth, itemindent=1em, label={\arabic*})]
  \item \emph{Tokens per Joule (TPJ)}: It characterizes the number of output tokens delivered per unit energy \cite{10.1145/3721146.3721953}.
  \item \emph{Requests per Joule (RPJ)}: It measures the number of fully completed requests per unit energy \cite{li-etal-2024-sprout}.
  \item \emph{Total Energy Consumption}: It measures the absolute electrical energy consumed by the LLM inference system to process a given workload \cite{10946802, 9352968, 10978053}.
\end{enumerate}
Conceptually, TPJ and RPJ are the energy-normalized counterparts of TPS and RPS, i.e., TPS/RPS express work per unit time, whereas TPJ/RPJ express work per unit energy. For completeness, energy analogs of latency metrics can also be defined, such as energy to first token and energy per request, as well as composite energy-delay measures. These metrics are crucial because edge deployments are constrained by battery capacity and local power budgets. ETSI MEC standards \cite{etsi_mec_ieg_006_2017} treat energy and delay as coequal KPIs for evaluating edge systems, underscoring the importance of energy-efficient LLM serving.

\subsubsection{\textbf{Memory}}
Unlike compute limits, memory pressure affects feasibility (whether the model and KV cache can reside on-device), speed (decoding stages that are bandwidth- and cache-bound), and stability (OOM failures). Below, we summarize the key memory metrics that govern edge deployability and performance.
\begin{enumerate}[fullwidth, itemindent=1em, label={\arabic*})]
\item \emph{Model Footprint}: It is the static memory required to store model parameters on-device \cite{pmlr-v202-sheng23a,10.1145/3600006.3613165,aali2025splitwiser}. It determines basic deployability, affects cold-start time, and limits the memory available for runtime state. The footprint is governed by model architecture (e.g., hidden size and number of layers), parameter precision (e.g., FP16, INT8), and compression methods such as quantization, pruning, and low-rank factorization.

\item \emph{KV Cache Footprint}: It is the memory required to store key and value tensors for all processed tokens during inference \cite{10.1145/3600006.3613165,aali2025splitwiser,pmlr-v202-sheng23a}. It scales roughly linearly with context length (prompt plus generated tokens), the number of concurrent requests, model architectural factors, and it is proportional to the storage precision. The footprint can be moderated by KV compression, offloading to slower memory, and eviction or sliding-window policies that cap growth at the cost of long-range context.

\item \emph{Peak Inference Memory}: It characterizes the maximum resident memory observed during the two main phases of LLM serving, i.e., prefill and decoding \cite{pmlr-v202-sheng23a,10.1145/3600006.3613165,aali2025splitwiser}. Prefill peaks are largely driven by activations and temporary kernel workspaces (e.g., attention and GEMM buffers), while decoding peaks are typically dominated by the KV cache and attention buffers. Peak inference memory therefore determines the minimum device capacity needed to avoid OOM and constrains the achievable concurrency for a given context length.
\end{enumerate}

\subsubsection{\textbf{Inference Quality}}
Beyond efficiency (latency, memory, energy),  LLM edge inference systems must also be evaluated for output quality. Because deployment optimizations and compression can change model behavior, designs should explicitly balance efficiency and quality. We summarize key quality metrics below.
\begin{enumerate}[fullwidth, itemindent=1em, label={\arabic*})]
\item \emph{Task quality metrics}:
These metrics quantify how often model outputs match the expected answer under a benchmark protocol. For classification-style benchmarks, accuracy \cite{hendrycks2020measuring, zellers-etal-2019-hellaswag, clark2018think} is the fraction of examples predicted correctly. For question answering, Exact Match (EM) \cite{cobbe2021training, amini-etal-2019-mathqa} is the fraction of predictions that exactly match the reference answer after normalization, and token-level F1 \cite{bai-etal-2024-longbench} is the harmonic mean of token precision and recall between prediction and reference. For generation tasks, ROUGE and BLEU \cite{narayan-etal-2018-dont, NIPS2015_afdec700, bai-etal-2024-longbench} measure $n$-gram overlap between generated and reference texts. For code generation, pass@k \cite{chen2021evaluating, austin2021program} is the probability that at least one of $k$ sampled programs passes unit tests.

\item \emph{Hallucination}:
Hallucinations are fluent outputs that are incorrect or unsupported by the given context \cite{lin-etal-2022-truthfulqa}. They are often measured by hallucination rate (the fraction of outputs with false or unsupported claims) or dataset-specific factuality scores (the fraction judged factually correct), and are especially important for compact or aggressively compressed edge models.

\item \emph{Bias and toxicity}:
Bias evaluations quantify stereotypical associations or performance disparities across demographic groups, often reported as a stereotype score or group disparity (performance gap between groups) \cite{nadeem-etal-2021-stereoset}. Toxicity is typically reported as a toxicity rate, defined as the fraction of outputs flagged as toxic by a classifier or human annotation \cite{gehman-etal-2020-realtoxicityprompts}.
\end{enumerate}

Optimizing a single metric in isolation can be misleading for edge LLM inference because the metrics are tightly coupled. Batching and parallelism may increase TPS or RPS, but often increase queueing delay and worsen tail TTFT or E2EL. Higher concurrency or longer context increases memory demand, raising TPOT and potentially causing latency spikes or OOM failures. Higher utilization can boost raw throughput while reducing goodput when more requests violate SLOs. Compression and acceleration can improve latency, throughput, or energy efficiency, but may degrade output quality or robustness depending on the configuration. Offloading or split execution can reduce on-device compute and energy, yet introduces communication delay and variability that can harm tail latency. These interactions motivate SLO- and resource-constrained optimization to identify Pareto-efficient operating points that balance latency, goodput, energy, and memory under edge hardware and network constraints.
\begin{table}[ht]\small
\caption{Representative Open LLMs for Edge Inference Evaluation}
\centering
\label{tab:evaluation_LLMS}
\begin{tabular}{|m{3.6cm}<{\centering}|m{1.5cm}<{\centering}|m{1.3cm}<{\centering}|m{6.6cm}<{\centering}|}
\hline
Model Family & Scales (B) & Provider & Key Features \\
\hline
LLaMA / Variants \cite{touvron2023llama}  & 1-70  & Meta &  General-purpose baselines with broad tooling support, long-context variants available \\
\hline
Mistral / Mixtral \cite{jiang2023mistral7b, jiang2024mixtral} & 7-8 (dense), 8$\times$7 (MoE) & Mistral AI & Optimized attention mechanisms, MoE architecture \\
\hline
Gemma / Gemma 2 \cite{team2024gemma} & 2-27 & Google & Compact architectures with optimized kernels, compatible with speculative decoding \\
\hline
Qwen / Qwen2 \cite{team2024qwen2, hui2024qwen2, yang2025qwen2} & 0.5-72 & Alibaba & Multilingual and tool-use capabilities, long-context variants available \\
\hline
Phi-2 / Phi-3 \cite{abdin2024phi} & 2-7 & Microsoft & Distilled-style models, friendly for edge devices \\
\hline
OPT \cite{zhang2022opt} & 1-66 & Meta & Transparent architectures and training details \\
\hline
TinyLLaMA / Variants \cite{zhang2024tinyllama} & 1-3 & Community & Mobile/embedded focus, support aggressive quantization\\
\hline
DeepSeek-Coder \cite{guo2024deepseek} & 1.3, 6.7, 33 & DeepSeek & Code-specialized models with domain-adapted tokenizers, long-context variants available \\
\hline
\end{tabular}
\end{table}

\subsection{Models and Datasets}
The choice of models and datasets is central to evaluating LLM inference on edge platforms, since both directly determine the representativeness and comparability of results.

On the model side, most evaluation studies focus on widely adopted open-source families such as LLaMA \cite{touvron2023llama}, Mistral (including Mixtral) \cite{jiang2023mistral7b, jiang2024mixtral}, Gemma \cite{team2024gemma}, and Qwen \cite{team2024qwen2, hui2024qwen2, yang2025qwen2}. Table \ref{tab:evaluation_LLMS} lists representative models commonly used for inference evaluation, all available on Hugging Face. Their architectural features make them particularly relevant for different aspects of inference evaluation on edge platforms. For example, the availability of long-context variants supports systematic testing of retrieval, summarization, and memory scaling.
Lightweight distilled or quantized models are well suited for evaluating trade-offs between accuracy, latency, and energy efficiency under mobile or embedded constraints. MoE models (e.g., Mixtral \cite{jiang2024mixtral}) highlight tokens-per-watt efficiency and reveal how expert activation affects KV-cache growth and scheduling, making them valuable for throughput and concurrency evaluation.

In addition to models, evaluation of LLM inference at the edge relies on datasets that capture both task semantics and workload characteristics. General reasoning and knowledge are typically measured using MMLU \cite{hendrycks2020measuring}, HellaSwag \cite{zellers-etal-2019-hellaswag}, and ARC \cite{clark2018think}, which feature short prompts and short outputs suited to classification-style evaluation. Mathematical reasoning datasets such as GSM8K \cite{cobbe2021training} and MathQA \cite{amini-etal-2019-mathqa} often require CoT prompting, producing longer outputs that stress both accuracy and inter-token latency. Code generation benchmarks like HumanEval \cite{chen2021evaluating} and MBPP \cite{austin2021program} follow a short-input/long-output pattern, making them appropriate for assessing throughput. For long-context evaluation, suites such as LongBench \cite{bai-etal-2024-longbench} and RULER \cite{hsieh2024ruler} substantially extend prompt length, enabling systematic study of memory footprint and KV-cache scaling. Summarization tasks (e.g., CNN/DailyMail \cite{NIPS2015_afdec700}, XSum \cite{narayan-etal-2018-dont}) represent long-input/long-output scenarios that reflect document-level generation workloads. Table \ref{tab:evaluation_datsets} summarizes representative datasets frequently used in evaluation. While not exhaustive, this collection highlights the most widely adopted benchmarks and illustrates how different dataset types align with specific evaluation dimensions.

\begin{table}[ht]\small
\caption{Representative Datasets for LLM Inference Evaluation}
\begingroup\footnotesize\centering
Notes: Sequence classes are defined based on input and output length. Short input: prompt length $\le$ 1000 tokens; long input: $>$ 1000.
Short output: output length $\le$ 128 tokens, long output: $>$ 128.
Accordingly, $S \leftrightarrow S$ denotes short input \& short output, $S \leftrightarrow L$ short input \& long output,
$L \leftrightarrow S$ long input \& short output, and $L \leftrightarrow L$ long input \& long output.
\endgroup
\centering
\label{tab:evaluation_datsets}
\begin{tabular}{|m{4.2cm}<{\centering}|m{3.2cm}<{\centering}|m{3.8cm}<{\centering}|m{1.7cm}<{\centering}|}
\hline
Dataset & Task Focus & Prompt length / Output length & Sequence Class\\
\hline
MMLU \cite{hendrycks2020measuring} & Knowledge/reasoning  &  50-400 / 1-10 & $S \leftrightarrow S$ \\
\hline
HellaSwag / PIQA \cite{zellers-etal-2019-hellaswag} &  Commonsense & 50-200 / 1-5 & $S \leftrightarrow S$ \\
\hline
ARC-C / ARC-E \cite{clark2018think} & Grade-school reasoning  & 50-300 / 1-10 & $S \leftrightarrow S$ \\
\hline
TruthfulQA \cite{lin-etal-2022-truthfulqa} & Factuality & 50-200 / 5-50 & $S \leftrightarrow S$ \\
\hline
XSum \cite{narayan-etal-2018-dont} & Summarization & 400-1k / 20-80 & $S \leftrightarrow S$ \\
\hline
GSM8K \cite{cobbe2021training} & Math word problems  & 30-200 / 50-300 (CoT) & $S \leftrightarrow L$ \\
\hline
MathQA \cite{amini-etal-2019-mathqa} & Math QA & 50-300 / 30-200 & $S \leftrightarrow L$ \\
\hline
HumanEval \cite{chen2021evaluating} & Code generation  & 30-200 / 50-300 & $S \leftrightarrow L$ \\
\hline
MBPP \cite{austin2021program} & Code (beginner) & 30-150 / 50-250 & $S \leftrightarrow L$ \\
\hline
LongBench (Retrieval/QA) \cite{bai-etal-2024-longbench} & Long-context retrieval & 2k-16k / 20-150 & $L \leftrightarrow S$ \\
\hline
LongBench (Summ.) \cite{bai-etal-2024-longbench} & Summarization  & 4k-16k / 150-400 & $L \leftrightarrow L$\\
\hline
RULER \cite{hsieh2024ruler} & Context-length stress  & 4k-32k+ / 5-200 & $L \leftrightarrow S$ / $L \leftrightarrow L$ \\
\hline
CNN/DailyMail \cite{NIPS2015_afdec700} & Summarization (news)  & 600-1.5k / 120-300 & $L \leftrightarrow L$ \\
\hline
\end{tabular}
\end{table}

\subsection{Evaluation Platforms}
Evaluating LLM edge inference requires both simulation frameworks and real-device testbeds, as each offers complementary insights. Simulations enable rapid, low-cost exploration of design alternatives under controlled conditions, whereas physical deployments validate performance under realistic constraints such as wireless resources, energy, memory. Table \ref{tab:evaluation_platform} summarizes the main evaluation platforms.
\begin{table}[ht]\small
\caption{Representative Evaluation platforms and software stacks for edge LLM inference}
\centering
\label{tab:evaluation_platform}
\begin{tabular}{|m{3.8cm}<{\centering}|m{3.8cm}<{\centering}|m{5.8cm}<{\centering}|}
\hline
Category & Rep. tools/platforms & Evaluation focus \\
\hline
Hardware simulation & gem5~\cite{10.1145/2024716.2024718}, Timeloop~\cite{8695666}, NVSim~\cite{6218223} & Processor/accelerator pipelines and memory hierarchy behavior\\
\hline
System simulation or profiling & TVM~\cite{222575}, PyTorch profilers, MLPerf~\cite{9001257} & Kernel scheduling, operator fusion, quantization effects, backend behavior \\
\hline
Network emulation & ns-3~\cite{henderson2008network}, Mininet~\cite{10.1145/1868447.1868466} & Split/distributed inference under bandwidth, RTT, and packet loss constraints \\
\hline
Real-device testbeds (hardware) & Smartphones; Raspberry Pi; NVIDIA Jetson Orin Nano/AGX;  GPUs/accelerators & End-to-end performance under realistic compute, memory, energy, and thermal constraints \\
\hline
Real-device software stack & llama.cpp~\cite{Georgi_llamacpp}, TensorRT-LLM~\cite{nvidia_tensorrt_llm}, vLLM~\cite{10.1145/3600006.3613165} &
Execution efficiency and system-level behavior across backends and accelerators \\
\hline
\end{tabular}
\end{table}

\textbf{Simulation-Based Evaluation}:
Simulation is widely used to explore hardware and system design choices prior to deployment. At the hardware level, simulators such as gem5 \cite{10.1145/2024716.2024718}, Timeloop \cite{8695666}, and NVSim \cite{6218223} enable detailed modeling of processor pipelines, accelerator dataflows, and memory hierarchies, providing cycle-accurate estimates of latency and energy consumption. At the system level, frameworks including TVM \cite{222575}, MLPerf \cite{9001257}, and profiling tools within PyTorch support evaluation of kernel scheduling, quantization effects, and operator fusion strategies across heterogeneous backends. For distributed or split inference scenarios, network emulators such as ns-3 \cite{henderson2008network} and Mininet \cite{10.1145/1868447.1868466} are widely adopted to study sensitivity to bandwidth limitations, round-trip time (RTT), and packet loss.

\textbf{Real-Device Evaluation}:
It is indispensable for capturing end-to-end performance, energy consumption, and user experience under realistic operating conditions. Representative hardware platforms span a wide spectrum of form factors: smartphones, edge AI boards (e.g., Raspberry Pi and NVIDIA Jetson Orin Nano), and edge servers equipped with GPUs or accelerators.
On the software side, a variety of inference engines are employed, including llama.cpp \cite{Georgi_llamacpp}, TensorRT-LLM \cite{nvidia_tensorrt_llm}, vLLM \cite{10.1145/3600006.3613165}, and ONNX Runtime \cite{onnx_runtime_mobile}. These are often complemented by vendor-specific SDKs such as Apple Core ML \cite{Apple_coreml} and Qualcomm AI Engine Direct\cite{Qualcomm_ai_engine}, which provide optimized execution paths for on-device accelerators. Benchmarking harnesses such as MLPerf \cite{9001257}, Hugging Face Optimum \cite{Huggingface_optim}, and custom token-latency profilers are widely used to quantify key metrics, including TTFT, TPS, and memory footprint across standardized datasets.

\section{Furture Research Directions}\label{sec:future_research}
As LLM inference at the network edge rapidly evolves, it confronts a set of open challenges that define the next wave of research. Below we outline three future promising directions.

\textbf{Scalable Inference for Multi-Model and Multi-modal LLMs}:
Most existing LLM serving frameworks are optimized for a single, text-only model, which limits perception and application scope. In practice, edge deployments increasingly require multimodal LLMs, e.g., Qwen2-VL \cite{team2024qwen2}, LLaVA-1.5 \cite{Liu_2024_CVPR}, that integrate text with images, audio, video, and sensor streams, as well as multi-model pipelines that combine a general-purpose LLM (e.g., LLaMA \cite{touvron2023llama}) with compact specialists for retrieval or tool use. Supporting these settings calls for end-to-end co-design across model selection, data pipelines, orchestration, scheduling, and runtime placement to meet tight edge budgets on memory, energy, and compute while maintaining service quality. From a wireless communications perspective, scalable multi-model/multimodal inference can enable emerging services such as XR and interactive agents, V2X and intent understanding, and UAV/robot coordination. It can also enhance network intelligence by jointly reasoning over heterogeneous telemetry (e.g., logs, traffic traces, and spectrum measurements) for anomaly detection and policy-driven configuration.

\textbf{Secure LLM Edge Inference}:
Security threats widely observed in cloud LLM serving, such as prompt injection, jailbreaks, and data leakage, are often amplified at the wireless edge, where devices are user-owned, physically accessible, intermittently connected, and frequently shared in multi-tenant settings \cite{9364272}. From an edge threat model perspective, attacks can be classified into: 1) Integrity attacks that manipulate behavior (e.g., prompt injection, jailbreaks, and tool abuse). 2) Privacy leakage of prompts, retrieved context, outputs, and runtime states (e.g., KV caches), including leakage via logs or improper state retention. 3) Model extraction and theft via repeated queries or compromised storage of weights/adapters. and 4) Side-channel leakage (e.g., timing and cache/memory contention) that is exacerbated by co-resident multi-tenancy. Conventional data-center defenses such as role-based access control \cite{ganie2025securing} and multi-factor authentication \cite{rehman2025claf} provide a baseline, but edge deployments additionally require edge-tailored mitigations, including prompt/retrieval sanitization with strict tool permissioning, encrypted communication and secure storage, zeroization with explicit retention policies, and side-channel-aware resource management. These protections are particularly critical in regulated domains such as healthcare and finance, and are a prerequisite for deploying LLMs in wireless workflows such as enterprise edge copilots and network automation.

\textbf{Green LLM Edge Inference}:
With the ubiquitous deployment of LLMs, the inference demand is ever-expanding. Since these models are both compute- and memory-intensive, energy availability is now a primary bottleneck for capacity growth in data centers.
This problem is further amplified at edge environments as energy is the binding limit for edge devices such as phones and embedded boards. A public estimation \cite{Suchi_AIgold} report that an average ChatGPT query consumes about 0.34 Wh, and that daily usage is comparable to the electricity consumed by roughly 180000 U.S. households, underscoring a looming ``energy wall" for LLM serving. This makes green, sustainable LLM edge inference a necessity, otherwise models that meet latency targets in controlled tests will throttle in the field, drain batteries, or become too expensive and carbon-intensive to scale. From a wireless communications perspective, green LLM edge inference is essential for enabling always-on, sustainable on-device applications such as radio access network automation, intelligent network controllers, and wearables and body area networks.

\section{Conclusion}\label{sec:conclusion}
This survey presents a comprehensive overview of recent advances in enabling LLM edge inference, systematically exploring system architectures, including single-edge-node, vertical, horizontal, and hybrid collaboration. Meanwhile, we review emerging techniques across architecture, model optimization, deployment, and resource management that improve performance under tight memory, bandwidth, and energy budgets.
We also curate methodologies for rigorous, reproducible evaluation and highlight future directions toward greater scalability, security, and sustainability.

\bibliographystyle{ACM-Reference-Format}
\bibliography{references}


\begin{thebibliography}{211}


\ifx \showCODEN    \undefined \def \showCODEN     #1{\unskip}     \fi
\ifx \showISBNx    \undefined \def \showISBNx     #1{\unskip}     \fi
\ifx \showISBNxiii \undefined \def \showISBNxiii  #1{\unskip}     \fi
\ifx \showISSN     \undefined \def \showISSN      #1{\unskip}     \fi
\ifx \showLCCN     \undefined \def \showLCCN      #1{\unskip}     \fi
\ifx \shownote     \undefined \def \shownote      #1{#1}          \fi
\ifx \showarticletitle \undefined \def \showarticletitle #1{#1}   \fi
\ifx \showURL      \undefined \def \showURL       {\relax}        \fi
\providecommand\bibfield[2]{#2}
\providecommand\bibinfo[2]{#2}
\providecommand\natexlab[1]{#1}
\providecommand\showeprint[2][]{arXiv:#2}

\bibitem[Aali et~al\mbox{.}(2025)]%
        {aali2025splitwiser}
\bibfield{author}{\bibinfo{person}{Asad Aali}, \bibinfo{person}{Adney Cardoza},
  {and} \bibinfo{person}{Melissa Capo}.} \bibinfo{year}{2025}\natexlab{}.
\newblock \showarticletitle{Splitwiser: Efficient LM inference with constrained
  resources}.
\newblock \bibinfo{journal}{\emph{arXiv preprint arXiv:2505.03763}}
  (\bibinfo{year}{2025}).
\newblock


\bibitem[Abdin et~al\mbox{.}(2024)]%
        {abdin2024phi}
\bibfield{author}{\bibinfo{person}{Marah Abdin}, \bibinfo{person}{Jyoti Aneja},
  \bibinfo{person}{Harkirat Behl}, \bibinfo{person}{S{\'e}bastien Bubeck},
  \bibinfo{person}{Ronen Eldan}, \bibinfo{person}{Suriya Gunasekar},
  \bibinfo{person}{Michael Harrison}, \bibinfo{person}{Russell~J Hewett},
  \bibinfo{person}{Mojan Javaheripi}, {et~al\mbox{.}}}
  \bibinfo{year}{2024}\natexlab{}.
\newblock \showarticletitle{Phi-4 technical report}.
\newblock \bibinfo{journal}{\emph{arXiv preprint arXiv:2412.08905}}
  (\bibinfo{year}{2024}).
\newblock


\bibitem[Achiam et~al\mbox{.}(2023)]%
        {achiam2023gpt}
\bibfield{author}{\bibinfo{person}{Josh Achiam}, \bibinfo{person}{Steven
  Adler}, \bibinfo{person}{Sandhini Agarwal}, \bibinfo{person}{Lama Ahmad},
  \bibinfo{person}{Ilge Akkaya}, \bibinfo{person}{Florencia~Leoni Aleman},
  \bibinfo{person}{Diogo Almeida}, \bibinfo{person}{Janko Altenschmidt},
  {et~al\mbox{.}}} \bibinfo{year}{2023}\natexlab{}.
\newblock \showarticletitle{Gpt-4 technical report}.
\newblock \bibinfo{journal}{\emph{arXiv preprint arXiv:2303.08774}}
  (\bibinfo{year}{2023}).
\newblock


\bibitem[Agrawal et~al\mbox{.}(2024)]%
        {298679}
\bibfield{author}{\bibinfo{person}{Amey Agrawal}, \bibinfo{person}{Nitin
  Kedia}, \bibinfo{person}{Ashish Panwar}, \bibinfo{person}{Jayashree Mohan},
  \bibinfo{person}{Nipun Kwatra}, \bibinfo{person}{Bhargav Gulavani},
  {et~al\mbox{.}}} \bibinfo{year}{2024}\natexlab{}.
\newblock \showarticletitle{Taming {Throughput-Latency} Tradeoff in {LLM}
  Inference with {Sarathi-Serve}}. In \bibinfo{booktitle}{\emph{Proc. USENIX
  OSDI 24}}. \bibinfo{pages}{117--134}.
\newblock
\showISBNx{978-1-939133-40-3}


\bibitem[Amini et~al\mbox{.}(2019)]%
        {amini-etal-2019-mathqa}
\bibfield{author}{\bibinfo{person}{Aida Amini}, \bibinfo{person}{Saadia
  Gabriel}, {et~al\mbox{.}}} \bibinfo{year}{2019}\natexlab{}.
\newblock \showarticletitle{{M}ath{QA}: Towards Interpretable Math Word Problem
  Solving with Operation-Based Formalisms}. In \bibinfo{booktitle}{\emph{Proc.
  Conf. North {A}merican Chapter of ACL}}. \bibinfo{pages}{2357--2367}.
\newblock


\bibitem[{Apple}(2017)]%
        {Apple_coreml}
\bibfield{author}{\bibinfo{person}{{Apple}}.} \bibinfo{year}{2017}\natexlab{}.
\newblock \bibinfo{booktitle}{\emph{Core ML Tools}}.
\newblock
\urldef\tempurl%
\url{https://github.com/apple/coremltools}
\showURL{%
\tempurl}


\bibitem[Apple(2024)]%
        {apple_pcc_2024}
\bibfield{author}{\bibinfo{person}{Apple}.} \bibinfo{year}{2024}\natexlab{}.
\newblock \bibinfo{title}{Private Cloud Compute: A new frontier for AI privacy
  in the cloud}.
\newblock
  \bibinfo{howpublished}{\url{https://security.apple.com/blog/private-cloud-compute/}}.
\newblock


\bibitem[Austin et~al\mbox{.}(2021)]%
        {austin2021program}
\bibfield{author}{\bibinfo{person}{Jacob Austin}, \bibinfo{person}{Augustus
  Odena}, \bibinfo{person}{Maxwell Nye}, \bibinfo{person}{Maarten Bosma},
  \bibinfo{person}{Henryk Michalewski}, \bibinfo{person}{David Dohan},
  \bibinfo{person}{Ellen Jiang}, \bibinfo{person}{Carrie Cai}, {et~al\mbox{.}}}
  \bibinfo{year}{2021}\natexlab{}.
\newblock \showarticletitle{Program synthesis with large language models}.
\newblock \bibinfo{journal}{\emph{arXiv preprint arXiv:2108.07732}}
  (\bibinfo{year}{2021}).
\newblock


\bibitem[Bai et~al\mbox{.}(2024a)]%
        {bai2024beyond}
\bibfield{author}{\bibinfo{person}{Guangji Bai}, \bibinfo{person}{Zheng Chai},
  \bibinfo{person}{Chen Ling}, \bibinfo{person}{Shiyu Wang},
  \bibinfo{person}{Jiaying Lu}, \bibinfo{person}{Nan Zhang},
  \bibinfo{person}{Tingwei Shi}, \bibinfo{person}{Ziyang Yu}, {et~al\mbox{.}}}
  \bibinfo{year}{2024}\natexlab{a}.
\newblock \showarticletitle{Beyond efficiency: A systematic survey of
  resource-efficient large language models}.
\newblock \bibinfo{journal}{\emph{arXiv preprint arXiv:2401.00625}}
  (\bibinfo{year}{2024}).
\newblock


\bibitem[Bai et~al\mbox{.}(2021)]%
        {bai-etal-2021-binarybert}
\bibfield{author}{\bibinfo{person}{Haoli Bai}, \bibinfo{person}{Wei Zhang},
  \bibinfo{person}{Lu Hou}, \bibinfo{person}{Lifeng Shang},
  \bibinfo{person}{Jin Jin}, \bibinfo{person}{Xin Jiang}, \bibinfo{person}{Qun
  Liu}, \bibinfo{person}{Michael Lyu}, {and} \bibinfo{person}{Irwin King}.}
  \bibinfo{year}{2021}\natexlab{}.
\newblock \showarticletitle{{B}inary{BERT}: Pushing the Limit of {BERT}
  Quantization}. In \bibinfo{booktitle}{\emph{Proc. ACL}}.
  \bibinfo{pages}{4334--4348}.
\newblock


\bibitem[Bai et~al\mbox{.}(2024b)]%
        {bai-etal-2024-longbench}
\bibfield{author}{\bibinfo{person}{Yushi Bai}, \bibinfo{person}{Xin Lv},
  \bibinfo{person}{Jiajie Zhang}, \bibinfo{person}{Hongchang Lyu},
  \bibinfo{person}{Jiankai Tang}, \bibinfo{person}{Zhidian Huang},
  \bibinfo{person}{Zhengxiao Du}, {et~al\mbox{.}}}
  \bibinfo{year}{2024}\natexlab{b}.
\newblock \showarticletitle{{L}ong{B}ench: A Bilingual, Multitask Benchmark for
  Long Context Understanding}. In \bibinfo{booktitle}{\emph{Annual Meeting of
  the ACL}}. \bibinfo{pages}{3119--3137}.
\newblock


\bibitem[Bao et~al\mbox{.}(2025)]%
        {11147210}
\bibfield{author}{\bibinfo{person}{Rui Bao}, \bibinfo{person}{Nan Xue},
  \bibinfo{person}{Yaping Sun}, {and} \bibinfo{person}{Zhiyong Chen}.}
  \bibinfo{year}{2025}\natexlab{}.
\newblock \showarticletitle{Dynamic Quality-Latency Aware Routing for LLM
  Inference in Wireless Edge-Device Networks}. In
  \bibinfo{booktitle}{\emph{IEEE Int. Conf. Commun. China}}.
  \bibinfo{pages}{1--6}.
\newblock


\bibitem[Bhattacharjee et~al\mbox{.}(2025)]%
        {bh_conformal}
\bibfield{author}{\bibinfo{person}{Payel Bhattacharjee},
  \bibinfo{person}{Fengwei Tian}, {et~al\mbox{.}}}
  \bibinfo{year}{2025}\natexlab{}.
\newblock \showarticletitle{Conformal Sparsification for Bandwidth-Efficient
  Edge-Cloud Speculative Decoding}. In \bibinfo{booktitle}{\emph{Proc. NeurIPS
  Workshop: AI and ML for Next-Generation Wireless Commun. and Netw.}}
\newblock


\bibitem[Binkert et~al\mbox{.}(2011)]%
        {10.1145/2024716.2024718}
\bibfield{author}{\bibinfo{person}{Nathan Binkert}, \bibinfo{person}{Bradford
  Beckmann}, \bibinfo{person}{Gabriel Black}, \bibinfo{person}{Steven~K.
  Reinhardt}, \bibinfo{person}{Ali Saidi}, \bibinfo{person}{Arkaprava Basu},
  \bibinfo{person}{Joel Hestness}, \bibinfo{person}{Derek~R. Hower},
  {et~al\mbox{.}}} \bibinfo{year}{2011}\natexlab{}.
\newblock \showarticletitle{The gem5 simulator}.
\newblock \bibinfo{journal}{\emph{SIGARCH Comput. Archit. News}}
  \bibinfo{volume}{39} (\bibinfo{year}{2011}), \bibinfo{pages}{1–7}.
\newblock


\bibitem[Borzunov et~al\mbox{.}(2023)]%
        {NEURIPS2023_28bf1419}
\bibfield{author}{\bibinfo{person}{Alexander Borzunov}, \bibinfo{person}{Max
  Ryabinin}, \bibinfo{person}{Artem Chumachenko}, \bibinfo{person}{Dmitry
  Baranchuk}, {et~al\mbox{.}}} \bibinfo{year}{2023}\natexlab{}.
\newblock \showarticletitle{Distributed Inference and Fine-tuning of Large
  Language Models Over The Internet}. In \bibinfo{booktitle}{\emph{Adv. Neural
  Infor. Process. Syst.}}, Vol.~\bibinfo{volume}{36}.
  \bibinfo{pages}{12312--12331}.
\newblock


\bibitem[Brown et~al\mbox{.}(2020)]%
        {NEURIPS2020_1457c0d6}
\bibfield{author}{\bibinfo{person}{Tom Brown}, \bibinfo{person}{Benjamin Mann},
  \bibinfo{person}{Nick Ryder}, \bibinfo{person}{Melanie Subbiah},
  \bibinfo{person}{Jared~D Kaplan}, \bibinfo{person}{Prafulla Dhariwal},
  \bibinfo{person}{Arvind Neelakantan}, {et~al\mbox{.}}}
  \bibinfo{year}{2020}\natexlab{}.
\newblock \showarticletitle{Language Models are Few-Shot Learners}. In
  \bibinfo{booktitle}{\emph{Adv. Neural Infor. Process. Syst.}},
  Vol.~\bibinfo{volume}{33}. \bibinfo{pages}{1877--1901}.
\newblock


\bibitem[Burke(2023)]%
        {google_gemini_nano_2025}
\bibfield{author}{\bibinfo{person}{Dave Burke}.}
  \bibinfo{year}{2023}\natexlab{}.
\newblock \bibinfo{title}{A New Foundation for AI on Android}.
\newblock
  \bibinfo{howpublished}{\url{https://android-developers.googleblog.com/2023/12/a-new-foundation-for-ai-on-android.html}}.
\newblock


\bibitem[Chand et~al\mbox{.}(2023)]%
        {chand2023dsformer}
\bibfield{author}{\bibinfo{person}{Rahul Chand}, \bibinfo{person}{Yashoteja
  Prabhu}, {and} \bibinfo{person}{Pratyush Kumar}.}
  \bibinfo{year}{2023}\natexlab{}.
\newblock \showarticletitle{Dsformer: Effective compression of
  text-transformers by dense-sparse weight factorization}.
\newblock \bibinfo{journal}{\emph{arXiv preprint arXiv:2312.13211}}
  (\bibinfo{year}{2023}).
\newblock


\bibitem[Chen et~al\mbox{.}(2025a)]%
        {chen2025pre}
\bibfield{author}{\bibinfo{person}{Junyi Chen}, \bibinfo{person}{Shihao Bai},
  \bibinfo{person}{Zaijun Wang}, \bibinfo{person}{Siyu Wu},
  \bibinfo{person}{Chuheng Du}, \bibinfo{person}{Hailong Yang},
  \bibinfo{person}{Ruihao Gong}, \bibinfo{person}{Shengzhong Liu},
  {et~al\mbox{.}}} \bibinfo{year}{2025}\natexlab{a}.
\newblock \showarticletitle{{Pre$^{3}$}: Enabling Deterministic Pushdown
  Automata for Faster Structured LLM Generation}. In
  \bibinfo{booktitle}{\emph{Proc. ACL}}.
\newblock


\bibitem[Chen et~al\mbox{.}(2021)]%
        {chen2021evaluating}
\bibfield{author}{\bibinfo{person}{Mark Chen}, \bibinfo{person}{Jerry Tworek},
  \bibinfo{person}{Heewoo Jun}, \bibinfo{person}{Qiming Yuan},
  \bibinfo{person}{Henrique Ponde De~Oliveira Pinto}, \bibinfo{person}{Jared
  Kaplan}, \bibinfo{person}{Harri Edwards}, \bibinfo{person}{Yuri Burda},
  {et~al\mbox{.}}} \bibinfo{year}{2021}\natexlab{}.
\newblock \showarticletitle{Evaluating large language models trained on code}.
\newblock \bibinfo{journal}{\emph{arXiv preprint arXiv:2107.03374}}
  (\bibinfo{year}{2021}).
\newblock


\bibitem[Chen et~al\mbox{.}(2018)]%
        {222575}
\bibfield{author}{\bibinfo{person}{Tianqi Chen}, \bibinfo{person}{Thierry
  Moreau}, \bibinfo{person}{Ziheng Jiang}, \bibinfo{person}{Lianmin Zheng},
  \bibinfo{person}{Eddie Yan}, \bibinfo{person}{Haichen Shen},
  \bibinfo{person}{Meghan Cowan}, {et~al\mbox{.}}}
  \bibinfo{year}{2018}\natexlab{}.
\newblock \showarticletitle{{TVM}: An Automated {End-to-End} Optimizing
  Compiler for Deep Learning}. In \bibinfo{booktitle}{\emph{Proc. USENIX
  OSDI}}. \bibinfo{pages}{578--594}.
\newblock
\showISBNx{978-1-939133-08-3}


\bibitem[Chen et~al\mbox{.}(2024)]%
        {chen2024internet}
\bibfield{author}{\bibinfo{person}{Weize Chen}, \bibinfo{person}{Ziming You},
  \bibinfo{person}{Ran Li}, \bibinfo{person}{Yitong Guan},
  \bibinfo{person}{Chen Qian}, \bibinfo{person}{Chenyang Zhao},
  \bibinfo{person}{Cheng Yang}, {et~al\mbox{.}}}
  \bibinfo{year}{2024}\natexlab{}.
\newblock \showarticletitle{Internet of agents: Weaving a web of heterogeneous
  agents for collaborative intelligence}.
\newblock \bibinfo{journal}{\emph{arXiv preprint arXiv:2407.07061}}
  (\bibinfo{year}{2024}).
\newblock


\bibitem[Chen et~al\mbox{.}(2025b)]%
        {chen2025adaptive}
\bibfield{author}{\bibinfo{person}{Yuxuan Chen} {et~al\mbox{.}}}
  \bibinfo{year}{2025}\natexlab{b}.
\newblock \showarticletitle{Adaptive layer splitting for wireless large
  language model inference in edge computing: a model-based reinforcement
  learning approach}.
\newblock \bibinfo{journal}{\emph{Frontiers of Infor. Technol. \& Electronic
  Eng.}} \bibinfo{volume}{26}, \bibinfo{number}{2} (\bibinfo{year}{2025}),
  \bibinfo{pages}{278--292}.
\newblock


\bibitem[Chen et~al\mbox{.}(2023)]%
        {10089235}
\bibfield{author}{\bibinfo{person}{Zhixiong Chen}, \bibinfo{person}{Wenqiang
  Yi}, \bibinfo{person}{Yuanwei Liu}, {and} \bibinfo{person}{Arumugam
  Nallanathan}.} \bibinfo{year}{2023}\natexlab{}.
\newblock \showarticletitle{Knowledge-Aided Federated Learning for
  Energy-Limited Wireless Networks}.
\newblock \bibinfo{journal}{\emph{IEEE Trans. Commun.}} \bibinfo{volume}{71},
  \bibinfo{number}{6} (\bibinfo{year}{2023}), \bibinfo{pages}{3368--3386}.
\newblock


\bibitem[Choi et~al\mbox{.}(2018)]%
        {choi2018pact}
\bibfield{author}{\bibinfo{person}{Jungwook Choi}, \bibinfo{person}{Zhuo Wang},
  \bibinfo{person}{Swagath Venkataramani}, \bibinfo{person}{Pierce I-Jen
  Chuang}, {et~al\mbox{.}}} \bibinfo{year}{2018}\natexlab{}.
\newblock \showarticletitle{Pact: Parameterized clipping activation for
  quantized neural networks}.
\newblock \bibinfo{journal}{\emph{arXiv preprint arXiv:1805.06085}}
  (\bibinfo{year}{2018}).
\newblock


\bibitem[Clark et~al\mbox{.}(2018)]%
        {clark2018think}
\bibfield{author}{\bibinfo{person}{Peter Clark}, \bibinfo{person}{Isaac
  Cowhey}, \bibinfo{person}{Oren Etzioni}, \bibinfo{person}{Tushar Khot},
  \bibinfo{person}{Ashish Sabharwal}, {et~al\mbox{.}}}
  \bibinfo{year}{2018}\natexlab{}.
\newblock \showarticletitle{Think you have solved question answering? try arc,
  the ai2 reasoning challenge}.
\newblock \bibinfo{journal}{\emph{arXiv preprint arXiv:1803.05457}}
  (\bibinfo{year}{2018}).
\newblock


\bibitem[Cobbe et~al\mbox{.}(2021)]%
        {cobbe2021training}
\bibfield{author}{\bibinfo{person}{Karl Cobbe}, \bibinfo{person}{Vineet
  Kosaraju}, \bibinfo{person}{Mohammad Bavarian}, \bibinfo{person}{Mark Chen},
  \bibinfo{person}{Heewoo Jun}, \bibinfo{person}{Lukasz Kaiser},
  \bibinfo{person}{Matthias Plappert}, \bibinfo{person}{Jerry Tworek},
  {et~al\mbox{.}}} \bibinfo{year}{2021}\natexlab{}.
\newblock \showarticletitle{Training verifiers to solve math word problems}.
\newblock \bibinfo{journal}{\emph{arXiv preprint arXiv:2110.14168}}
  (\bibinfo{year}{2021}).
\newblock


\bibitem[Del~Corro et~al\mbox{.}(2023)]%
        {del2023skipdecode}
\bibfield{author}{\bibinfo{person}{Luciano Del~Corro}, \bibinfo{person}{Allie
  Del~Giorno}, \bibinfo{person}{Sahaj Agarwal}, \bibinfo{person}{Bin Yu},
  {et~al\mbox{.}}} \bibinfo{year}{2023}\natexlab{}.
\newblock \showarticletitle{Skipdecode: Autoregressive skip decoding with
  batching and caching for efficient llm inference}.
\newblock \bibinfo{journal}{\emph{arXiv preprint arXiv:2307.02628}}
  (\bibinfo{year}{2023}).
\newblock


\bibitem[Dettmers et~al\mbox{.}(2022)]%
        {NEURIPS2022_c3ba4962}
\bibfield{author}{\bibinfo{person}{Tim Dettmers}, \bibinfo{person}{Mike Lewis},
  \bibinfo{person}{Younes Belkada}, {and} \bibinfo{person}{Luke Zettlemoyer}.}
  \bibinfo{year}{2022}\natexlab{}.
\newblock \showarticletitle{GPT3.int8(): 8-bit Matrix Multiplication for
  Transformers at Scale}. In \bibinfo{booktitle}{\emph{Proc. Adv. Neural Infor.
  Process. Syst. (NeurIPS)}}, Vol.~\bibinfo{volume}{35}.
  \bibinfo{pages}{30318--30332}.
\newblock


\bibitem[Dettmers et~al\mbox{.}(2023)]%
        {dettmers2023spqr}
\bibfield{author}{\bibinfo{person}{Tim Dettmers}, \bibinfo{person}{Ruslan
  Svirschevski}, \bibinfo{person}{Vage Egiazarian}, \bibinfo{person}{Denis
  Kuznedelev}, {et~al\mbox{.}}} \bibinfo{year}{2023}\natexlab{}.
\newblock \showarticletitle{Spqr: A sparse-quantized representation for
  near-lossless llm weight compression}.
\newblock \bibinfo{journal}{\emph{arXiv preprint arXiv:2306.03078}}
  (\bibinfo{year}{2023}).
\newblock


\bibitem[Devlin et~al\mbox{.}(2019)]%
        {devlin-etal-2019-bert}
\bibfield{author}{\bibinfo{person}{Jacob Devlin}, \bibinfo{person}{Ming-Wei
  Chang}, \bibinfo{person}{Kenton Lee}, {and} \bibinfo{person}{Kristina
  Toutanova}.} \bibinfo{year}{2019}\natexlab{}.
\newblock \showarticletitle{{BERT}: Pre-training of Deep Bidirectional
  Transformers for Language Understanding}. In \bibinfo{booktitle}{\emph{Proc.
  ACL}}. \bibinfo{pages}{4171--4186}.
\newblock


\bibitem[Ding et~al\mbox{.}(2024)]%
        {ding2024hybrid}
\bibfield{author}{\bibinfo{person}{Dujian Ding}, \bibinfo{person}{Ankur
  Mallick}, \bibinfo{person}{Chi Wang}, \bibinfo{person}{Robert Sim},
  \bibinfo{person}{Subhabrata Mukherjee}, \bibinfo{person}{Victor Ruhle},
  {et~al\mbox{.}}} \bibinfo{year}{2024}\natexlab{}.
\newblock \showarticletitle{Hybrid llm: Cost-efficient and quality-aware query
  routing}. In \bibinfo{booktitle}{\emph{Proc. Int. Conf. Learn. Represent.
  (ICLR)}}. \bibinfo{pages}{1–23}.
\newblock


\bibitem[Ding et~al\mbox{.}(2025a)]%
        {10949629}
\bibfield{author}{\bibinfo{person}{Guangyao Ding}, \bibinfo{person}{Huiguo
  Gao}, \bibinfo{person}{Shengli Liu}, {and} \bibinfo{person}{Guanding Yu}.}
  \bibinfo{year}{2025}\natexlab{a}.
\newblock \showarticletitle{Multi-Stage Semantic Communication for Low-Latency
  Edge Inference}.
\newblock \bibinfo{journal}{\emph{IEEE Trans. Cogn. Commun. and Netw.}}
  (\bibinfo{year}{2025}), \bibinfo{pages}{1--1}.
\newblock


\bibitem[Ding et~al\mbox{.}(2025b)]%
        {11207452}
\bibfield{author}{\bibinfo{person}{Yu Ding}, \bibinfo{person}{Jingxuan Zhao},
  \bibinfo{person}{Zhengong Cai}, {et~al\mbox{.}}}
  \bibinfo{year}{2025}\natexlab{b}.
\newblock \showarticletitle{Adaptoserve: An Efficient System for Supporting
  Adaptive Chunked-Prefills in LLM Inference}. In
  \bibinfo{booktitle}{\emph{Proc. IEEE Int. Conf. High Perf. Comput. and
  Commun. (HPCC)}}. \bibinfo{pages}{1--9}.
\newblock


\bibitem[Dong et~al\mbox{.}(2022)]%
        {dong2022survey}
\bibfield{author}{\bibinfo{person}{Qingxiu Dong}, \bibinfo{person}{Lei Li},
  \bibinfo{person}{Damai Dai}, \bibinfo{person}{Ce Zheng},
  \bibinfo{person}{Jingyuan Ma}, \bibinfo{person}{Rui Li},
  \bibinfo{person}{Heming Xia}, \bibinfo{person}{Jingjing Xu},
  \bibinfo{person}{Zhiyong Wu}, \bibinfo{person}{Tianyu Liu}, {et~al\mbox{.}}}
  \bibinfo{year}{2022}\natexlab{}.
\newblock \showarticletitle{A survey on in-context learning}.
\newblock \bibinfo{journal}{\emph{arXiv preprint arXiv:2301.00234}}
  (\bibinfo{year}{2022}).
\newblock


\bibitem[Dong et~al\mbox{.}(2012)]%
        {6218223}
\bibfield{author}{\bibinfo{person}{Xiangyu Dong}, \bibinfo{person}{Cong Xu},
  {et~al\mbox{.}}} \bibinfo{year}{2012}\natexlab{}.
\newblock \showarticletitle{NVSim: A Circuit-Level Performance, Energy, and
  Area Model for Emerging Nonvolatile Memory}.
\newblock \bibinfo{journal}{\emph{IEEE Trans. Computer-Aided Design Integr.
  Circuits Syst.}} \bibinfo{volume}{31}, \bibinfo{number}{7}
  (\bibinfo{year}{2012}), \bibinfo{pages}{994--1007}.
\newblock


\bibitem[Elhoushi et~al\mbox{.}(2024)]%
        {elhoushi2024layerskip}
\bibfield{author}{\bibinfo{person}{Mostafa Elhoushi}, \bibinfo{person}{Akshat
  Shrivastava}, \bibinfo{person}{Diana Liskovich}, \bibinfo{person}{Basil
  Hosmer}, \bibinfo{person}{Bram Wasti}, \bibinfo{person}{Liangzhen Lai},
  \bibinfo{person}{Anas Mahmoud}, {et~al\mbox{.}}}
  \bibinfo{year}{2024}\natexlab{}.
\newblock \showarticletitle{Layerskip: Enabling early exit inference and
  self-speculative decoding}. In \bibinfo{booktitle}{\emph{Proc. ACL}}.
  \bibinfo{pages}{12622--12642}.
\newblock


\bibitem[Esser et~al\mbox{.}(2019)]%
        {esser2019learned}
\bibfield{author}{\bibinfo{person}{Steven~K Esser}, \bibinfo{person}{Jeffrey~L
  McKinstry}, \bibinfo{person}{Deepika Bablani}, \bibinfo{person}{Rathinakumar
  Appuswamy}, {and} \bibinfo{person}{Dharmendra~S Modha}.}
  \bibinfo{year}{2019}\natexlab{}.
\newblock \showarticletitle{Learned step size quantization}.
\newblock \bibinfo{journal}{\emph{arXiv preprint arXiv:1902.08153}}
  (\bibinfo{year}{2019}).
\newblock


\bibitem[{ETSI ISG}(2017)]%
        {etsi_mec_ieg_006_2017}
\bibfield{author}{\bibinfo{person}{{ETSI ISG}}.}
  \bibinfo{year}{2017}\natexlab{}.
\newblock \bibinfo{booktitle}{\emph{{Mobile Edge Computing; Market
  Acceleration; MEC Metrics Best Practice and Guidelines}}}.
\newblock \bibinfo{type}{ETSI GS MEC-IEG 006 V1.1.1}.
  \bibinfo{institution}{ETSI}.
\newblock
\urldef\tempurl%
\url{https://www.etsi.org/deliver/etsi_gs/mec-ieg/001_099/006/01.01.01_60/gs_mec-ieg006v010101p.pdf}
\showURL{%
\tempurl}


\bibitem[Fan et~al\mbox{.}(2021)]%
        {fan2021dapple}
\bibfield{author}{\bibinfo{person}{Shiqing Fan}, \bibinfo{person}{Yi Rong},
  {et~al\mbox{.}}} \bibinfo{year}{2021}\natexlab{}.
\newblock \showarticletitle{DAPPLE: A pipelined data parallel approach for
  training large models}. In \bibinfo{booktitle}{\emph{Proceedings of the 26th
  ACM SIGPLAN Symposium on Principles and Practice of Parallel Programming}}.
  \bibinfo{pages}{431--445}.
\newblock


\bibitem[Feng et~al\mbox{.}(2025a)]%
        {10.1145/3695053.3730999}
\bibfield{author}{\bibinfo{person}{Jingqi Feng}, \bibinfo{person}{Yukai Huang},
  \bibinfo{person}{Rui Zhang}, {et~al\mbox{.}}}
  \bibinfo{year}{2025}\natexlab{a}.
\newblock \showarticletitle{WindServe: Efficient Phase-Disaggregated LLM
  Serving with Stream-based Dynamic Scheduling}. In
  \bibinfo{booktitle}{\emph{Proc. Annual Int. Symp. Comput. Arch.}}
  \emph{(\bibinfo{series}{ISCA})}. \bibinfo{pages}{1283--1295}.
\newblock
\showISBNx{9798400712616}


\bibitem[Feng et~al\mbox{.}(2025b)]%
        {11160773}
\bibfield{author}{\bibinfo{person}{Zideng Feng}, \bibinfo{person}{Lu Lu},
  \bibinfo{person}{Qin Li}, \bibinfo{person}{Yuhao Chai},
  \bibinfo{person}{Zhenyu Zhang}, {et~al\mbox{.}}}
  \bibinfo{year}{2025}\natexlab{b}.
\newblock \showarticletitle{Distributed Inference Optimization for Large
  Language Model in Edge-Cloud Collaborative Networks}. In
  \bibinfo{booktitle}{\emph{Proc. IEEE Int. Conf. Commun.}}
  \bibinfo{pages}{6161--6166}.
\newblock


\bibitem[Frantar and Alistarh(2023)]%
        {pmlr-v202-frantar23a}
\bibfield{author}{\bibinfo{person}{Elias Frantar} {and} \bibinfo{person}{Dan
  Alistarh}.} \bibinfo{year}{2023}\natexlab{}.
\newblock \showarticletitle{{S}parse{GPT}: Massive Language Models Can be
  Accurately Pruned in One-Shot}. In \bibinfo{booktitle}{\emph{Proc. Int. Conf.
  Mach. Learn. (ICML)}}, Vol.~\bibinfo{volume}{202}.
  \bibinfo{pages}{10323--10337}.
\newblock


\bibitem[Ganie(2025)]%
        {ganie2025securing}
\bibfield{author}{\bibinfo{person}{Aadil~Gani Ganie}.}
  \bibinfo{year}{2025}\natexlab{}.
\newblock \showarticletitle{Securing AI Agents: Implementing Role-Based Access
  Control for Industrial Applications}.
\newblock \bibinfo{journal}{\emph{arXiv:2509.11431}} (\bibinfo{year}{2025}).
\newblock


\bibitem[Gehman et~al\mbox{.}(2020)]%
        {gehman-etal-2020-realtoxicityprompts}
\bibfield{author}{\bibinfo{person}{Samuel Gehman}, \bibinfo{person}{Suchin
  Gururangan}, \bibinfo{person}{Maarten Sap}, {et~al\mbox{.}}}
  \bibinfo{year}{2020}\natexlab{}.
\newblock \showarticletitle{{R}eal{T}oxicity{P}rompts: Evaluating Neural Toxic
  Degeneration in Language Models}. In \bibinfo{booktitle}{\emph{Proc. EMNLP}}.
  \bibinfo{pages}{3356--3369}.
\newblock


\bibitem[Gerganov(2023)]%
        {Georgi_llamacpp}
\bibfield{author}{\bibinfo{person}{Georgi Gerganov}.}
  \bibinfo{year}{2023}\natexlab{}.
\newblock \bibinfo{booktitle}{\emph{llama.cpp}}.
\newblock
\urldef\tempurl%
\url{https://github.com/ggml-org/llama.cpp}
\showURL{%
\tempurl}


\bibitem[Geva et~al\mbox{.}(2022)]%
        {geva2022transformer}
\bibfield{author}{\bibinfo{person}{Mor Geva}, \bibinfo{person}{Avi Caciularu},
  \bibinfo{person}{Kevin~Ro Wang}, {and} \bibinfo{person}{Yoav Goldberg}.}
  \bibinfo{year}{2022}\natexlab{}.
\newblock \showarticletitle{Transformer feed-forward layers build predictions
  by promoting concepts in the vocabulary space}.
\newblock \bibinfo{journal}{\emph{arXiv preprint arXiv:2203.14680}}
  (\bibinfo{year}{2022}).
\newblock


\bibitem[Ghazvininejad et~al\mbox{.}(2019)]%
        {ghazvininejad2019mask}
\bibfield{author}{\bibinfo{person}{Marjan Ghazvininejad}, \bibinfo{person}{Omer
  Levy}, \bibinfo{person}{Yinhan Liu}, {and} \bibinfo{person}{Luke
  Zettlemoyer}.} \bibinfo{year}{2019}\natexlab{}.
\newblock \showarticletitle{Mask-predict: Parallel decoding of conditional
  masked language models}.
\newblock \bibinfo{journal}{\emph{arXiv preprint arXiv:1904.09324}}
  (\bibinfo{year}{2019}).
\newblock


\bibitem[Goldsmith(2005)]%
        {goldsmith2005wireless}
\bibfield{author}{\bibinfo{person}{Andrea Goldsmith}.}
  \bibinfo{year}{2005}\natexlab{}.
\newblock \bibinfo{booktitle}{\emph{Wireless communications}}.
\newblock \bibinfo{publisher}{Cambridge university press}.
\newblock


\bibitem[Gond et~al\mbox{.}(2025)]%
        {gond2025tokenweave}
\bibfield{author}{\bibinfo{person}{Raja Gond}, \bibinfo{person}{Nipun Kwatra},
  {and} \bibinfo{person}{Ramachandran Ramjee}.}
  \bibinfo{year}{2025}\natexlab{}.
\newblock \showarticletitle{TokenWeave: Efficient Compute-Communication Overlap
  for Distributed LLM Inference}.
\newblock \bibinfo{journal}{\emph{arXiv preprint arXiv:2505.11329}}
  (\bibinfo{year}{2025}).
\newblock


\bibitem[Gopal(2024)]%
        {Suchi_AIgold}
\bibfield{author}{\bibinfo{person}{Suchi Gopal}.}
  \bibinfo{year}{2024}\natexlab{}.
\newblock \bibinfo{title}{The AI Gold Rush: Can Utilities Keep Up with the
  Energy Demand?}
\newblock
\urldef\tempurl%
\url{https://floodlightglobal.com/the-ai-gold-rush-can-utilities-keep-up-with-the-energy-demand/}
\showURL{%
\tempurl}


\bibitem[Gu and Kong(2020)]%
        {gu2020fully}
\bibfield{author}{\bibinfo{person}{Jiatao Gu} {and} \bibinfo{person}{Xiang
  Kong}.} \bibinfo{year}{2020}\natexlab{}.
\newblock \showarticletitle{Fully non-autoregressive neural machine
  translation: Tricks of the trade}.
\newblock \bibinfo{journal}{\emph{arXiv preprint arXiv:2012.15833}}
  (\bibinfo{year}{2020}).
\newblock


\bibitem[Gu et~al\mbox{.}(2024)]%
        {gu2024minillm}
\bibfield{author}{\bibinfo{person}{Yuxian Gu}, \bibinfo{person}{Li Dong},
  \bibinfo{person}{Furu Wei}, {and} \bibinfo{person}{Minlie Huang}.}
  \bibinfo{year}{2024}\natexlab{}.
\newblock \showarticletitle{Mini{LLM}: Knowledge Distillation of Large Language
  Models}. In \bibinfo{booktitle}{\emph{Proc. Int. Conf. Learn. Represent.
  (ICLR)}}.
\newblock


\bibitem[Guo et~al\mbox{.}(2023)]%
        {10.1145/3579371.3589038}
\bibfield{author}{\bibinfo{person}{Cong Guo}, \bibinfo{person}{Jiaming Tang},
  \bibinfo{person}{Weiming Hu}, \bibinfo{person}{Jingwen Leng},
  \bibinfo{person}{Chen Zhang}, {et~al\mbox{.}}}
  \bibinfo{year}{2023}\natexlab{}.
\newblock \showarticletitle{OliVe: Accelerating Large Language Models via
  Hardware-friendly Outlier-Victim Pair Quantization}. In
  \bibinfo{booktitle}{\emph{Proc. Annual Int. Symp. Comput. Arch.}}
\newblock


\bibitem[Guo et~al\mbox{.}(2024)]%
        {guo2024deepseek}
\bibfield{author}{\bibinfo{person}{Daya Guo}, \bibinfo{person}{Qihao Zhu},
  \bibinfo{person}{Dejian Yang}, \bibinfo{person}{Zhenda Xie}, {et~al\mbox{.}}}
  \bibinfo{year}{2024}\natexlab{}.
\newblock \showarticletitle{DeepSeek-Coder: When the Large Language Model Meets
  Programming--The Rise of Code Intelligence}.
\newblock \bibinfo{journal}{\emph{arXiv preprint arXiv:2401.14196}}
  (\bibinfo{year}{2024}).
\newblock


\bibitem[Guo et~al\mbox{.}(2019)]%
        {Guo_Tan_He_Qin_Xu_Liu_2019}
\bibfield{author}{\bibinfo{person}{Junliang Guo}, \bibinfo{person}{Xu Tan},
  \bibinfo{person}{Di He}, \bibinfo{person}{Tao Qin}, \bibinfo{person}{Linli
  Xu}, {and} \bibinfo{person}{Tie-Yan Liu}.} \bibinfo{year}{2019}\natexlab{}.
\newblock \showarticletitle{Non-Autoregressive Neural Machine Translation with
  Enhanced Decoder Input}.
\newblock \bibinfo{journal}{\emph{Proc. AAAI}} \bibinfo{volume}{33},
  \bibinfo{number}{01} (\bibinfo{year}{2019}), \bibinfo{pages}{3723--3730}.
\newblock


\bibitem[Hao et~al\mbox{.}(2024)]%
        {10.1145/3662006.3662067}
\bibfield{author}{\bibinfo{person}{Zixu Hao} {et~al\mbox{.}}}
  \bibinfo{year}{2024}\natexlab{}.
\newblock \showarticletitle{Hybrid SLM and LLM for Edge-Cloud Collaborative
  Inference}. In \bibinfo{booktitle}{\emph{Proc. EdgeFM}}.
  \bibinfo{pages}{36–41}.
\newblock


\bibitem[He et~al\mbox{.}(2024)]%
        {10591707}
\bibfield{author}{\bibinfo{person}{Ying He} {et~al\mbox{.}}}
  \bibinfo{year}{2024}\natexlab{}.
\newblock \showarticletitle{Large Language Models (LLMs) Inference Offloading
  and Resource Allocation in Cloud-Edge Computing: An Active Inference
  Approach}.
\newblock \bibinfo{journal}{\emph{IEEE Trans. Mobile Comput.}}
  \bibinfo{volume}{23}, \bibinfo{number}{12} (\bibinfo{year}{2024}),
  \bibinfo{pages}{11253--11264}.
\newblock


\bibitem[Henderson et~al\mbox{.}(2008)]%
        {henderson2008network}
\bibfield{author}{\bibinfo{person}{Thomas~R Henderson},
  \bibinfo{person}{Mathieu Lacage}, \bibinfo{person}{George~F Riley},
  \bibinfo{person}{Craig Dowell}, {and} \bibinfo{person}{Joseph Kopena}.}
  \bibinfo{year}{2008}\natexlab{}.
\newblock \showarticletitle{Network simulations with the ns-3 simulator}.
\newblock \bibinfo{journal}{\emph{SIGCOMM demonstration}} \bibinfo{volume}{14},
  \bibinfo{number}{14} (\bibinfo{year}{2008}), \bibinfo{pages}{527}.
\newblock


\bibitem[Hendrycks et~al\mbox{.}(2021)]%
        {hendrycks2020measuring}
\bibfield{author}{\bibinfo{person}{Dan Hendrycks}, \bibinfo{person}{Collin
  Burns}, \bibinfo{person}{Steven Basart}, \bibinfo{person}{Andy Zou},
  \bibinfo{person}{Mantas Mazeika}, \bibinfo{person}{Dawn Song}, {and}
  \bibinfo{person}{Jacob Steinhardt}.} \bibinfo{year}{2021}\natexlab{}.
\newblock \showarticletitle{Measuring Massive Multitask Language
  Understanding}. In \bibinfo{booktitle}{\emph{Proc. Int. Conf. Learn.
  Represent. (ICLR)}}.
\newblock


\bibitem[Hermann et~al\mbox{.}(2015)]%
        {NIPS2015_afdec700}
\bibfield{author}{\bibinfo{person}{Karl~Moritz Hermann}, \bibinfo{person}{Tomas
  Kocisky}, \bibinfo{person}{Edward Grefenstette}, \bibinfo{person}{Lasse
  Espeholt}, {et~al\mbox{.}}} \bibinfo{year}{2015}\natexlab{}.
\newblock \showarticletitle{Teaching Machines to Read and Comprehend}. In
  \bibinfo{booktitle}{\emph{Proc. Adv. Neural Infor. Process. Syst.
  (NeurIPS)}}, Vol.~\bibinfo{volume}{28}.
\newblock


\bibitem[Ho et~al\mbox{.}(2023)]%
        {ho-etal-2023-large}
\bibfield{author}{\bibinfo{person}{Namgyu Ho} {et~al\mbox{.}}}
  \bibinfo{year}{2023}\natexlab{}.
\newblock \showarticletitle{Large Language Models Are Reasoning Teachers}. In
  \bibinfo{booktitle}{\emph{Proc. ACL}}. \bibinfo{pages}{14852--14882}.
\newblock


\bibitem[Holmes et~al\mbox{.}(2024)]%
        {holmes2024deepspeed}
\bibfield{author}{\bibinfo{person}{Connor Holmes}, \bibinfo{person}{Masahiro
  Tanaka}, \bibinfo{person}{Michael Wyatt}, \bibinfo{person}{Ammar~Ahmad Awan},
  {et~al\mbox{.}}} \bibinfo{year}{2024}\natexlab{}.
\newblock \showarticletitle{Deepspeed-fastgen: High-throughput text generation
  for llms via mii and deepspeed-inference}.
\newblock \bibinfo{journal}{\emph{arXiv preprint arXiv:2401.08671}}
  (\bibinfo{year}{2024}).
\newblock


\bibitem[Hong et~al\mbox{.}(2025)]%
        {hong2025semi}
\bibfield{author}{\bibinfo{person}{Ke Hong}, \bibinfo{person}{Lufang Chen},
  \bibinfo{person}{Zhong Wang}, \bibinfo{person}{Xiuhong Li}, {et~al\mbox{.}}}
  \bibinfo{year}{2025}\natexlab{}.
\newblock \showarticletitle{semi-pd: Towards efficient llm serving via
  phase-wise disaggregated computation and unified storage}.
\newblock \bibinfo{journal}{\emph{arXiv preprint arXiv:2504.19867}}
  (\bibinfo{year}{2025}).
\newblock


\bibitem[Hooper et~al\mbox{.}(2024)]%
        {NEURIPS2024_028fcbcf}
\bibfield{author}{\bibinfo{person}{Coleman Hooper}, \bibinfo{person}{Sehoon
  Kim}, \bibinfo{person}{Hiva Mohammadzadeh}, {et~al\mbox{.}}}
  \bibinfo{year}{2024}\natexlab{}.
\newblock \showarticletitle{KVQuant: Towards 10 Million Context Length LLM
  Inference with KV Cache Quantization}. In \bibinfo{booktitle}{\emph{Adv.
  Neural Infor. Process. Syst. (NeurIPS)}}, Vol.~\bibinfo{volume}{37}.
  \bibinfo{pages}{1270--1303}.
\newblock


\bibitem[Hsieh et~al\mbox{.}(2024)]%
        {hsieh2024ruler}
\bibfield{author}{\bibinfo{person}{Cheng-Ping Hsieh}, \bibinfo{person}{Simeng
  Sun}, \bibinfo{person}{Samuel Kriman}, \bibinfo{person}{Shantanu Acharya},
  \bibinfo{person}{Dima Rekesh}, \bibinfo{person}{Fei Jia}, {and}
  \bibinfo{person}{Boris Ginsburg}.} \bibinfo{year}{2024}\natexlab{}.
\newblock \showarticletitle{{RULER}: What{\textquoteright}s the Real Context
  Size of Your Long-Context Language Models?}. In
  \bibinfo{booktitle}{\emph{Proc. Conf. Language Modeling}}.
\newblock


\bibitem[Hu et~al\mbox{.}(2022)]%
        {9996638}
\bibfield{author}{\bibinfo{person}{Yang Hu}, \bibinfo{person}{Connor Imes},
  \bibinfo{person}{Xuanang Zhao}, \bibinfo{person}{Souvik Kundu},
  \bibinfo{person}{Peter~A. Beerel}, {et~al\mbox{.}}}
  \bibinfo{year}{2022}\natexlab{}.
\newblock \showarticletitle{PipeEdge: Pipeline Parallelism for Large-Scale
  Model Inference on Heterogeneous Edge Devices}. In
  \bibinfo{booktitle}{\emph{Proc. DSD}}. \bibinfo{pages}{298--307}.
\newblock


\bibitem[Hu et~al\mbox{.}(2021)]%
        {hu2021pipeline}
\bibfield{author}{\bibinfo{person}{Yang Hu}, \bibinfo{person}{Connor Imes},
  \bibinfo{person}{Xuanang Zhao}, \bibinfo{person}{Souvik Kundu},
  \bibinfo{person}{Peter~A Beerel}, \bibinfo{person}{Stephen~P Crago}, {and}
  \bibinfo{person}{John Paul~N Walters}.} \bibinfo{year}{2021}\natexlab{}.
\newblock \showarticletitle{Pipeline parallelism for inference on heterogeneous
  edge computing}.
\newblock \bibinfo{journal}{\emph{arXiv preprint arXiv:2110.14895}}
  (\bibinfo{year}{2021}).
\newblock


\bibitem[Hu et~al\mbox{.}(2025)]%
        {10949701}
\bibfield{author}{\bibinfo{person}{Yitao Hu}, \bibinfo{person}{Xiulong Liu},
  \bibinfo{person}{Guotao Yang}, \bibinfo{person}{Linxuan Li},
  \bibinfo{person}{Kai Zeng}, {et~al\mbox{.}}} \bibinfo{year}{2025}\natexlab{}.
\newblock \showarticletitle{TightLLM: Maximizing Throughput for LLM Inference
  via Adaptive Offloading Policy}.
\newblock \bibinfo{journal}{\emph{IEEE Trans. Comput.}} \bibinfo{volume}{74},
  \bibinfo{number}{7} (\bibinfo{year}{2025}), \bibinfo{pages}{2195--2209}.
\newblock


\bibitem[Hua et~al\mbox{.}(2021)]%
        {9352968}
\bibfield{author}{\bibinfo{person}{Sheng Hua}, \bibinfo{person}{Yong Zhou},
  \bibinfo{person}{Kai Yang}, \bibinfo{person}{Yuanming Shi}, {and}
  \bibinfo{person}{Kunlun Wang}.} \bibinfo{year}{2021}\natexlab{}.
\newblock \showarticletitle{Reconfigurable Intelligent Surface for Green Edge
  Inference}.
\newblock \bibinfo{journal}{\emph{IEEE Trans. Green Commun. and Netw.}}
  \bibinfo{volume}{5}, \bibinfo{number}{2} (\bibinfo{year}{2021}),
  \bibinfo{pages}{964--979}.
\newblock


\bibitem[Huang et~al\mbox{.}(2019)]%
        {8741198}
\bibfield{author}{\bibinfo{person}{Chongwen Huang}, \bibinfo{person}{Alessio
  Zappone}, {et~al\mbox{.}}} \bibinfo{year}{2019}\natexlab{}.
\newblock \showarticletitle{Reconfigurable Intelligent Surfaces for Energy
  Efficiency in Wireless Communication}.
\newblock \bibinfo{journal}{\emph{IEEE Trans. Wireless Commun.}}
  \bibinfo{volume}{18}, \bibinfo{number}{8} (\bibinfo{year}{2019}),
  \bibinfo{pages}{4157--4170}.
\newblock


\bibitem[{Hugging Face}(2021)]%
        {Huggingface_optim}
\bibfield{author}{\bibinfo{person}{{Hugging Face}}.}
  \bibinfo{year}{2021}\natexlab{}.
\newblock \bibinfo{booktitle}{\emph{Optimum}}.
\newblock
\urldef\tempurl%
\url{https://github.com/huggingface/optimum}
\showURL{%
\tempurl}


\bibitem[Hui et~al\mbox{.}(2024)]%
        {hui2024qwen2}
\bibfield{author}{\bibinfo{person}{Binyuan Hui}, \bibinfo{person}{Jian Yang},
  \bibinfo{person}{Zeyu Cui}, \bibinfo{person}{Jiaxi Yang},
  \bibinfo{person}{Dayiheng Liu}, \bibinfo{person}{Lei Zhang},
  \bibinfo{person}{Tianyu Liu}, \bibinfo{person}{Jiajun Zhang},
  \bibinfo{person}{Bowen Yu}, \bibinfo{person}{Keming Lu}, {et~al\mbox{.}}}
  \bibinfo{year}{2024}\natexlab{}.
\newblock \showarticletitle{Qwen2. 5-coder technical report}.
\newblock \bibinfo{journal}{\emph{arXiv preprint arXiv:2409.12186}}
  (\bibinfo{year}{2024}).
\newblock


\bibitem[Jacob et~al\mbox{.}(2018)]%
        {Jacob_2018_CVPR}
\bibfield{author}{\bibinfo{person}{Benoit Jacob}, \bibinfo{person}{Skirmantas
  Kligys}, {et~al\mbox{.}}} \bibinfo{year}{2018}\natexlab{}.
\newblock \showarticletitle{Quantization and Training of Neural Networks for
  Efficient Integer-Arithmetic-Only Inference}. In
  \bibinfo{booktitle}{\emph{Proc. IEEE Conf. Comput. Vis. and Pattern Recog.
  (CVPR)}}.
\newblock


\bibitem[Jiang et~al\mbox{.}(2023b)]%
        {jiang2023mistral7b}
\bibfield{author}{\bibinfo{person}{Albert~Q. Jiang}, \bibinfo{person}{Alexandre
  Sablayrolles}, {et~al\mbox{.}}} \bibinfo{year}{2023}\natexlab{b}.
\newblock \showarticletitle{Mistral 7B}.
\newblock \bibinfo{journal}{\emph{arXiv preprint arXiv:2310.06825}}
  (\bibinfo{year}{2023}).
\newblock


\bibitem[Jiang et~al\mbox{.}(2024a)]%
        {jiang2024mixtral}
\bibfield{author}{\bibinfo{person}{Albert~Q Jiang}, \bibinfo{person}{Alexandre
  Sablayrolles}, {et~al\mbox{.}}} \bibinfo{year}{2024}\natexlab{a}.
\newblock \showarticletitle{Mixtral of experts}.
\newblock \bibinfo{journal}{\emph{arXiv preprint arXiv:2401.04088}}
  (\bibinfo{year}{2024}).
\newblock


\bibitem[Jiang et~al\mbox{.}(2024c)]%
        {jiang2024neo}
\bibfield{author}{\bibinfo{person}{Xuanlin Jiang}, \bibinfo{person}{Yang Zhou},
  \bibinfo{person}{Shiyi Cao}, \bibinfo{person}{Ion Stoica}, {and}
  \bibinfo{person}{Minlan Yu}.} \bibinfo{year}{2024}\natexlab{c}.
\newblock \showarticletitle{Neo: Saving gpu memory crisis with cpu offloading
  for online llm inference}.
\newblock \bibinfo{journal}{\emph{arXiv preprint arXiv:2411.01142}}
  (\bibinfo{year}{2024}).
\newblock


\bibitem[Jiang et~al\mbox{.}(2023a)]%
        {jiang-etal-2023-lion}
\bibfield{author}{\bibinfo{person}{Yuxin Jiang}, \bibinfo{person}{Chunkit
  Chan}, \bibinfo{person}{Mingyang Chen}, {and} \bibinfo{person}{Wei Wang}.}
  \bibinfo{year}{2023}\natexlab{a}.
\newblock \showarticletitle{Lion: Adversarial Distillation of Proprietary Large
  Language Models}. In \bibinfo{booktitle}{\emph{Proc. Conf. Empirical Methods
  in Natural Language Process.}} \bibinfo{pages}{3134--3154}.
\newblock


\bibitem[Jiang et~al\mbox{.}(2024b)]%
        {pmlr-v235-jiang24f}
\bibfield{author}{\bibinfo{person}{Youhe Jiang}, \bibinfo{person}{Ran Yan},
  \bibinfo{person}{Xiaozhe Yao}, \bibinfo{person}{Yang Zhou}, {et~al\mbox{.}}}
  \bibinfo{year}{2024}\natexlab{b}.
\newblock \showarticletitle{{H}ex{G}en: Generative Inference of Large Language
  Model over Heterogeneous Environment}. In \bibinfo{booktitle}{\emph{Proc.
  Int. Conf. Machine Learn. (ICML)}}, Vol.~\bibinfo{volume}{235}.
  \bibinfo{pages}{21946--21961}.
\newblock


\bibitem[Kafetzis et~al\mbox{.}(2025)]%
        {kafetzis2025large}
\bibfield{author}{\bibinfo{person}{Dimitrios Kafetzis}, \bibinfo{person}{Ramin
  Khalili}, {and} \bibinfo{person}{Iordanis Koutsopoulos}.}
  \bibinfo{year}{2025}\natexlab{}.
\newblock \showarticletitle{Large Language Model partitioning for low-latency
  inference at the edge}.
\newblock \bibinfo{journal}{\emph{arXiv preprint arXiv:2505.02533}}
  (\bibinfo{year}{2025}).
\newblock


\bibitem[Kaplan et~al\mbox{.}(2020)]%
        {kaplan2020scaling}
\bibfield{author}{\bibinfo{person}{Jared Kaplan} {et~al\mbox{.}}}
  \bibinfo{year}{2020}\natexlab{}.
\newblock \showarticletitle{Scaling laws for neural language models}.
\newblock \bibinfo{journal}{\emph{arXiv preprint arXiv:2001.08361}}
  (\bibinfo{year}{2020}).
\newblock


\bibitem[Khoshnoodi et~al\mbox{.}(2024)]%
        {khoshnoodi2024comprehensive}
\bibfield{author}{\bibinfo{person}{Mahsa Khoshnoodi}, \bibinfo{person}{Vinija
  Jain}, \bibinfo{person}{Mingye Gao}, \bibinfo{person}{Malavika Srikanth},
  {and} \bibinfo{person}{Aman Chadha}.} \bibinfo{year}{2024}\natexlab{}.
\newblock \showarticletitle{A comprehensive survey of accelerated generation
  techniques in large language models}.
\newblock \bibinfo{journal}{\emph{arXiv preprint arXiv:2405.13019}}
  (\bibinfo{year}{2024}).
\newblock


\bibitem[Krishnamoorthi(2018)]%
        {krishnamoorthi2018quantizing}
\bibfield{author}{\bibinfo{person}{Raghuraman Krishnamoorthi}.}
  \bibinfo{year}{2018}\natexlab{}.
\newblock \showarticletitle{Quantizing deep convolutional networks for
  efficient inference: A whitepaper}.
\newblock \bibinfo{journal}{\emph{arXiv preprint arXiv:1806.08342}}
  (\bibinfo{year}{2018}).
\newblock


\bibitem[Kurti\'{c} et~al\mbox{.}(2023)]%
        {NEURIPS2023_ced46a50}
\bibfield{author}{\bibinfo{person}{Eldar Kurti\'{c}}, \bibinfo{person}{Elias
  Frantar}, {and} \bibinfo{person}{Dan Alistarh}.}
  \bibinfo{year}{2023}\natexlab{}.
\newblock \showarticletitle{ZipLM: Inference-Aware Structured Pruning of
  Language Models}. In \bibinfo{booktitle}{\emph{Proc. Adv. Neural Infor.
  Process. Syst. (NeurIPS)}}, Vol.~\bibinfo{volume}{36}.
  \bibinfo{pages}{65597--65617}.
\newblock


\bibitem[Kwon et~al\mbox{.}(2023)]%
        {10.1145/3600006.3613165}
\bibfield{author}{\bibinfo{person}{Woosuk Kwon}, \bibinfo{person}{Zhuohan Li},
  \bibinfo{person}{Siyuan Zhuang}, \bibinfo{person}{Ying Sheng},
  \bibinfo{person}{Lianmin Zheng}, {et~al\mbox{.}}}
  \bibinfo{year}{2023}\natexlab{}.
\newblock \showarticletitle{Efficient Memory Management for Large Language
  Model Serving with PagedAttention}. In \bibinfo{booktitle}{\emph{Proc.
  SOSP}}. \bibinfo{pages}{611–--626}.
\newblock
\showISBNx{9798400702297}


\bibitem[Lan et~al\mbox{.}(2019)]%
        {lan2019albert}
\bibfield{author}{\bibinfo{person}{Zhenzhong Lan}, \bibinfo{person}{Mingda
  Chen}, \bibinfo{person}{Sebastian Goodman}, \bibinfo{person}{Kevin Gimpel},
  \bibinfo{person}{Piyush Sharma}, {and} \bibinfo{person}{Radu Soricut}.}
  \bibinfo{year}{2019}\natexlab{}.
\newblock \showarticletitle{Albert: A lite bert for self-supervised learning of
  language representations}.
\newblock \bibinfo{journal}{\emph{arXiv preprint arXiv:1909.11942}}
  (\bibinfo{year}{2019}).
\newblock


\bibitem[Lantz et~al\mbox{.}(2010)]%
        {10.1145/1868447.1868466}
\bibfield{author}{\bibinfo{person}{Bob Lantz}, \bibinfo{person}{Brandon
  Heller}, {and} \bibinfo{person}{Nick McKeown}.}
  \bibinfo{year}{2010}\natexlab{}.
\newblock \showarticletitle{A network in a laptop: rapid prototyping for
  software-defined networks}. In \bibinfo{booktitle}{\emph{Proc. ACM SIGCOMM
  Workshop on Hot Topics in Netw.}}
\newblock


\bibitem[Larsson et~al\mbox{.}(2014)]%
        {6736761}
\bibfield{author}{\bibinfo{person}{Erik~G. Larsson}, \bibinfo{person}{Ove
  Edfors}, \bibinfo{person}{Fredrik Tufvesson}, {and}
  \bibinfo{person}{Thomas~L. Marzetta}.} \bibinfo{year}{2014}\natexlab{}.
\newblock \showarticletitle{Massive MIMO for next generation wireless systems}.
\newblock \bibinfo{journal}{\emph{IEEE Commun. Mag.}} \bibinfo{volume}{52},
  \bibinfo{number}{2} (\bibinfo{year}{2014}), \bibinfo{pages}{186--195}.
\newblock


\bibitem[Lee et~al\mbox{.}(2018)]%
        {lee2018deterministic}
\bibfield{author}{\bibinfo{person}{Jason Lee}, \bibinfo{person}{Elman
  Mansimov}, {and} \bibinfo{person}{Kyunghyun Cho}.}
  \bibinfo{year}{2018}\natexlab{}.
\newblock \showarticletitle{Deterministic non-autoregressive neural sequence
  modeling by iterative refinement}.
\newblock \bibinfo{journal}{\emph{arXiv preprint arXiv:1802.06901}}
  (\bibinfo{year}{2018}).
\newblock


\bibitem[Li et~al\mbox{.}(2024a)]%
        {10938426}
\bibfield{author}{\bibinfo{person}{Baolin Li}, \bibinfo{person}{Yankai Jiang},
  \bibinfo{person}{Vijay Gadepally}, {and} \bibinfo{person}{Devesh Tiwari}.}
  \bibinfo{year}{2024}\natexlab{a}.
\newblock \showarticletitle{LLM Inference Serving: Survey of Recent Advances
  and Opportunities}. In \bibinfo{booktitle}{\emph{Proc. IEEE High Perf.
  Extreme Comput. Conf. (HPEC)}}. \bibinfo{pages}{1--8}.
\newblock


\bibitem[Li et~al\mbox{.}(2024b)]%
        {li-etal-2024-sprout}
\bibfield{author}{\bibinfo{person}{Baolin Li}, \bibinfo{person}{Yankai Jiang},
  \bibinfo{person}{Vijay Gadepally}, {and} \bibinfo{person}{Devesh Tiwari}.}
  \bibinfo{year}{2024}\natexlab{b}.
\newblock \showarticletitle{Sprout: Green Generative {AI} with Carbon-Efficient
  {LLM} Inference}. In \bibinfo{booktitle}{\emph{Proc. Conf. Empirical Methods
  in Natural Language Process.}} \bibinfo{pages}{21799--21813}.
\newblock


\bibitem[Li et~al\mbox{.}(2024c)]%
        {li2024survey}
\bibfield{author}{\bibinfo{person}{Haoyang Li}, \bibinfo{person}{Yiming Li},
  \bibinfo{person}{Anxin Tian}, \bibinfo{person}{Tianhao Tang}, {and}
  \bibinfo{person}{othres}.} \bibinfo{year}{2024}\natexlab{c}.
\newblock \showarticletitle{A survey on large language model acceleration based
  on kv cache management}.
\newblock \bibinfo{journal}{\emph{arXiv preprint arXiv:2412.19442}}
  (\bibinfo{year}{2024}).
\newblock


\bibitem[Li et~al\mbox{.}(2020)]%
        {li2020cascadebert}
\bibfield{author}{\bibinfo{person}{Lei Li}, \bibinfo{person}{Yankai Lin},
  \bibinfo{person}{Deli Chen}, \bibinfo{person}{Shuhuai Ren},
  \bibinfo{person}{Peng Li}, \bibinfo{person}{Jie Zhou}, {and}
  \bibinfo{person}{Xu Sun}.} \bibinfo{year}{2020}\natexlab{}.
\newblock \showarticletitle{Cascadebert: Accelerating inference of pre-trained
  language models via calibrated complete models cascade}.
\newblock \bibinfo{journal}{\emph{arXiv preprint arXiv:2012.14682}}
  (\bibinfo{year}{2020}).
\newblock


\bibitem[Liang et~al\mbox{.}(2023)]%
        {pmlr-v202-liang23j}
\bibfield{author}{\bibinfo{person}{Chen Liang}, \bibinfo{person}{Simiao Zuo},
  \bibinfo{person}{Qingru Zhang}, {et~al\mbox{.}}}
  \bibinfo{year}{2023}\natexlab{}.
\newblock \showarticletitle{Less is More: Task-aware Layer-wise Distillation
  for Language Model Compression}. In \bibinfo{booktitle}{\emph{Proc. Int.
  Conf. Mach. Learn. (ICML)}}, Vol.~\bibinfo{volume}{202}.
  \bibinfo{pages}{20852--20867}.
\newblock


\bibitem[Lin et~al\mbox{.}(2024)]%
        {MLSYS2024_42a452cb}
\bibfield{author}{\bibinfo{person}{Ji Lin}, \bibinfo{person}{Jiaming Tang},
  \bibinfo{person}{Haotian Tang}, \bibinfo{person}{Shang Yang}, {and}
  \bibinfo{person}{othres}.} \bibinfo{year}{2024}\natexlab{}.
\newblock \showarticletitle{{AWQ}: Activation-aware Weight Quantization for
  On-Device LLM Compression and Acceleration}. In
  \bibinfo{booktitle}{\emph{Proc. Mach. Learn. Syst. (MLSys)}},
  Vol.~\bibinfo{volume}{6}. \bibinfo{pages}{87--100}.
\newblock


\bibitem[Lin et~al\mbox{.}(2022)]%
        {lin-etal-2022-truthfulqa}
\bibfield{author}{\bibinfo{person}{Stephanie Lin}, \bibinfo{person}{Jacob
  Hilton}, {and} \bibinfo{person}{Owain Evans}.}
  \bibinfo{year}{2022}\natexlab{}.
\newblock \showarticletitle{{T}ruthful{QA}: Measuring How Models Mimic Human
  Falsehoods}. In \bibinfo{booktitle}{\emph{Proc. ACL}}.
  \bibinfo{pages}{3214--3252}.
\newblock


\bibitem[Liu et~al\mbox{.}(2024a)]%
        {liu2024deepseek}
\bibfield{author}{\bibinfo{person}{Aixin Liu}, \bibinfo{person}{Bei Feng},
  \bibinfo{person}{Bing Xue}, \bibinfo{person}{Bingxuan Wang},
  \bibinfo{person}{Bochao Wu}, \bibinfo{person}{Chengda Lu},
  \bibinfo{person}{Chenggang Zhao}, \bibinfo{person}{Chengqi Deng},
  \bibinfo{person}{Chenyu Zhang}, \bibinfo{person}{Chong Ruan},
  {et~al\mbox{.}}} \bibinfo{year}{2024}\natexlab{a}.
\newblock \showarticletitle{Deepseek-v3 technical report}.
\newblock \bibinfo{journal}{\emph{arXiv preprint arXiv:2412.19437}}
  (\bibinfo{year}{2024}).
\newblock


\bibitem[Liu et~al\mbox{.}(2024b)]%
        {Liu_2024_CVPR}
\bibfield{author}{\bibinfo{person}{Haotian Liu}, \bibinfo{person}{Chunyuan Li},
  \bibinfo{person}{Yuheng Li}, {and} \bibinfo{person}{Yong~Jae Lee}.}
  \bibinfo{year}{2024}\natexlab{b}.
\newblock \showarticletitle{Improved Baselines with Visual Instruction Tuning}.
  In \bibinfo{booktitle}{\emph{Proc. IEEE CVPR}}.
  \bibinfo{pages}{26296--26306}.
\newblock


\bibitem[Liu et~al\mbox{.}(2019)]%
        {liu2019roberta}
\bibfield{author}{\bibinfo{person}{Yinhan Liu}, \bibinfo{person}{Myle Ott},
  \bibinfo{person}{Naman Goyal}, \bibinfo{person}{Jingfei Du},
  \bibinfo{person}{Mandar Joshi}, \bibinfo{person}{Danqi Chen},
  \bibinfo{person}{Omer Levy}, {et~al\mbox{.}}}
  \bibinfo{year}{2019}\natexlab{}.
\newblock \showarticletitle{Roberta: A robustly optimized bert pretraining
  approach}.
\newblock \bibinfo{journal}{\emph{arXiv preprint arXiv:1907.11692}}
  (\bibinfo{year}{2019}).
\newblock


\bibitem[Liu et~al\mbox{.}(2023)]%
        {liu2023llm}
\bibfield{author}{\bibinfo{person}{Zechun Liu}, \bibinfo{person}{Barlas Oguz},
  \bibinfo{person}{Changsheng Zhao}, \bibinfo{person}{Ernie Chang},
  \bibinfo{person}{Pierre Stock}, {et~al\mbox{.}}}
  \bibinfo{year}{2023}\natexlab{}.
\newblock \showarticletitle{Llm-qat: Data-free quantization aware training for
  large language models}.
\newblock \bibinfo{journal}{\emph{arXiv preprint arXiv:2305.17888}}
  (\bibinfo{year}{2023}).
\newblock


\bibitem[Longpre et~al\mbox{.}(2023)]%
        {pmlr-v202-longpre23a}
\bibfield{author}{\bibinfo{person}{Shayne Longpre}, \bibinfo{person}{Le Hou},
  \bibinfo{person}{Tu Vu}, \bibinfo{person}{Albert Webson}, {et~al\mbox{.}}}
  \bibinfo{year}{2023}\natexlab{}.
\newblock \showarticletitle{The Flan Collection: Designing Data and Methods for
  Effective Instruction Tuning}. In \bibinfo{booktitle}{\emph{Proc. Int. Conf.
  on Mach. Learn. (ICML)}}, Vol.~\bibinfo{volume}{202}.
  \bibinfo{pages}{22631--22648}.
\newblock


\bibitem[Lu et~al\mbox{.}(2014)]%
        {6798744}
\bibfield{author}{\bibinfo{person}{Lu Lu}, \bibinfo{person}{Geoffrey~Ye Li},
  \bibinfo{person}{A.~Lee Swindlehurst}, \bibinfo{person}{Alexei Ashikhmin},
  {and} \bibinfo{person}{Rui Zhang}.} \bibinfo{year}{2014}\natexlab{}.
\newblock \showarticletitle{An Overview of Massive MIMO: Benefits and
  Challenges}.
\newblock \bibinfo{journal}{\emph{IEEE J. Sel. Topics in Signal Process.}}
  \bibinfo{volume}{8}, \bibinfo{number}{5} (\bibinfo{year}{2014}),
  \bibinfo{pages}{742--758}.
\newblock


\bibitem[Ma et~al\mbox{.}(2023)]%
        {NEURIPS2023_44956951}
\bibfield{author}{\bibinfo{person}{Xinyin Ma}, \bibinfo{person}{Gongfan Fang},
  {and} \bibinfo{person}{Xinchao Wang}.} \bibinfo{year}{2023}\natexlab{}.
\newblock \showarticletitle{LLM-Pruner: On the Structural Pruning of Large
  Language Models}. In \bibinfo{booktitle}{\emph{Proc. Adv. Neural Infor.
  Process. Syst. (NeurIPS)}}, Vol.~\bibinfo{volume}{36}.
  \bibinfo{pages}{21702--21720}.
\newblock


\bibitem[Matsubara et~al\mbox{.}(2022)]%
        {10.1145/3527155}
\bibfield{author}{\bibinfo{person}{Yoshitomo Matsubara}, \bibinfo{person}{Marco
  Levorato}, {and} \bibinfo{person}{Francesco Restuccia}.}
  \bibinfo{year}{2022}\natexlab{}.
\newblock \showarticletitle{Split Computing and Early Exiting for Deep Learning
  Applications: Survey and Research Challenges}.
\newblock \bibinfo{journal}{\emph{ACM Comput. Surv.}} \bibinfo{volume}{55},
  \bibinfo{number}{5} (\bibinfo{year}{2022}).
\newblock


\bibitem[Mattson et~al\mbox{.}(2020)]%
        {9001257}
\bibfield{author}{\bibinfo{person}{Peter Mattson},
  \bibinfo{person}{Vijay~Janapa Reddi}, \bibinfo{person}{Christine Cheng},
  \bibinfo{person}{Cody Coleman}, {et~al\mbox{.}}}
  \bibinfo{year}{2020}\natexlab{}.
\newblock \showarticletitle{MLPerf: An Industry Standard Benchmark Suite for
  Machine Learning Performance}.
\newblock \bibinfo{journal}{\emph{IEEE Micro}} \bibinfo{volume}{40},
  \bibinfo{number}{2} (\bibinfo{year}{2020}), \bibinfo{pages}{8--16}.
\newblock


\bibitem[Mei et~al\mbox{.}(2025)]%
        {10.1145/3669940.3707215}
\bibfield{author}{\bibinfo{person}{Yixuan Mei}, \bibinfo{person}{Yonghao
  Zhuang}, \bibinfo{person}{Xupeng Miao}, \bibinfo{person}{Juncheng Yang},
  \bibinfo{person}{Zhihao Jia}, {and} \bibinfo{person}{Rashmi Vinayak}.}
  \bibinfo{year}{2025}\natexlab{}.
\newblock \showarticletitle{Helix: Serving Large Language Models over
  Heterogeneous GPUs and Network via Max-Flow}. In
  \bibinfo{booktitle}{\emph{Proc. ASPLOS}}, Vol.~\bibinfo{volume}{1}.
  \bibinfo{pages}{586–602}.
\newblock


\bibitem[Miao et~al\mbox{.}(2025)]%
        {10.1145/3754448}
\bibfield{author}{\bibinfo{person}{Xupeng Miao}, \bibinfo{person}{Gabriele
  Oliaro}, \bibinfo{person}{Zhihao Zhang}, \bibinfo{person}{Xinhao Cheng},
  \bibinfo{person}{Hongyi Jin}, {et~al\mbox{.}}}
  \bibinfo{year}{2025}\natexlab{}.
\newblock \showarticletitle{Towards Efficient Generative Large Language Model
  Serving: A Survey from Algorithms to Systems}.
\newblock \bibinfo{journal}{\emph{ACM Comput. Surv.}} \bibinfo{volume}{58},
  \bibinfo{number}{1} (\bibinfo{year}{2025}), \bibinfo{pages}{1--37}.
\newblock


\bibitem[Miao et~al\mbox{.}(2024)]%
        {10.1145/3620666.3651335}
\bibfield{author}{\bibinfo{person}{Xupeng Miao}, \bibinfo{person}{Gabriele
  Oliaro}, \bibinfo{person}{Zhihao Zhang}, {and} \bibinfo{person}{othres}.}
  \bibinfo{year}{2024}\natexlab{}.
\newblock \showarticletitle{SpecInfer: Accelerating Large Language Model
  Serving with Tree-based Speculative Inference and Verification}. In
  \bibinfo{booktitle}{\emph{Proc. ASPLOS}}. \bibinfo{pages}{932–949}.
\newblock


\bibitem[Mitchell(2024)]%
        {Debates_Melanie}
\bibfield{author}{\bibinfo{person}{Melanie Mitchell}.}
  \bibinfo{year}{2024}\natexlab{}.
\newblock \showarticletitle{Debates on the nature of artificial general
  intelligence}.
\newblock \bibinfo{journal}{\emph{Science}} \bibinfo{volume}{383},
  \bibinfo{number}{6689} (\bibinfo{year}{2024}), \bibinfo{pages}{eado7069}.
\newblock


\bibitem[Mudvari et~al\mbox{.}(2024)]%
        {mudvari2024splitllm}
\bibfield{author}{\bibinfo{person}{Akrit Mudvari}, \bibinfo{person}{Yuang
  Jiang}, {and} \bibinfo{person}{Leandros Tassiulas}.}
  \bibinfo{year}{2024}\natexlab{}.
\newblock \showarticletitle{Splitllm: Collaborative inference of llms for model
  placement and throughput optimization}.
\newblock \bibinfo{journal}{\emph{arXiv preprint arXiv:2410.10759}}
  (\bibinfo{year}{2024}).
\newblock


\bibitem[Nadeem et~al\mbox{.}(2021)]%
        {nadeem-etal-2021-stereoset}
\bibfield{author}{\bibinfo{person}{Moin Nadeem}, \bibinfo{person}{Anna Bethke},
  {and} \bibinfo{person}{Siva Reddy}.} \bibinfo{year}{2021}\natexlab{}.
\newblock \showarticletitle{{S}tereo{S}et: Measuring stereotypical bias in
  pretrained language models}. In \bibinfo{booktitle}{\emph{Proc. ACL}},
  \bibfield{editor}{\bibinfo{person}{Chengqing Zong}, \bibinfo{person}{Fei
  Xia}, \bibinfo{person}{Wenjie Li}, {and} \bibinfo{person}{Roberto Navigli}}
  (Eds.). \bibinfo{pages}{5356--5371}.
\newblock


\bibitem[Narayan and othres(2018)]%
        {narayan-etal-2018-dont}
\bibfield{author}{\bibinfo{person}{Shashi Narayan} {and}
  \bibinfo{person}{othres}.} \bibinfo{year}{2018}\natexlab{}.
\newblock \showarticletitle{Don{'}t Give Me the Details, Just the Summary!
  Topic-Aware Convolutional Neural Networks for Extreme Summarization}. In
  \bibinfo{booktitle}{\emph{Conf. Empirical Methods in Natural Language
  Process.}} \bibinfo{pages}{1797--1807}.
\newblock


\bibitem[Nguyen et~al\mbox{.}(2022)]%
        {9760729}
\bibfield{author}{\bibinfo{person}{Van-Dinh Nguyen}, \bibinfo{person}{Symeon
  Chatzinotas}, {et~al\mbox{.}}} \bibinfo{year}{2022}\natexlab{}.
\newblock \showarticletitle{FedFog: Network-Aware Optimization of Federated
  Learning Over Wireless Fog-Cloud Systems}.
\newblock \bibinfo{journal}{\emph{IEEE Trans. Wireless Commun.}}
  \bibinfo{volume}{21}, \bibinfo{number}{10} (\bibinfo{year}{2022}),
  \bibinfo{pages}{8581--8599}.
\newblock


\bibitem[{NVIDIA Corporation}(2021)]%
        {nvidia_fast_transformer}
\bibfield{author}{\bibinfo{person}{{NVIDIA Corporation}}.}
  \bibinfo{year}{2021}\natexlab{}.
\newblock \bibinfo{booktitle}{\emph{FasterTransformer}}.
\newblock
\urldef\tempurl%
\url{https://github.com/NVIDIA/FasterTransformer}
\showURL{%
\tempurl}


\bibitem[{NVIDIA Corporation}(2023)]%
        {nvidia_tensorrt_llm}
\bibfield{author}{\bibinfo{person}{{NVIDIA Corporation}}.}
  \bibinfo{year}{2023}\natexlab{}.
\newblock \bibinfo{booktitle}{\emph{TensorRT-LLM: A TensorRT Toolbox for
  Optimized Large Language Model Inference}}.
\newblock
\urldef\tempurl%
\url{https://github.com/NVIDIA/TensorRT-LLM}
\showURL{%
\tempurl}


\bibitem[{ONNX Runtime}(2025)]%
        {onnx_runtime_mobile}
\bibfield{author}{\bibinfo{person}{{ONNX Runtime}}.}
  \bibinfo{year}{2025}\natexlab{}.
\newblock \bibinfo{title}{Get started with ONNX Runtime Mobile}.
\newblock
  \bibinfo{howpublished}{\url{https://onnxruntime.ai/docs/get-started/with-mobile.html}}.
\newblock


\bibitem[Parashar et~al\mbox{.}(2019)]%
        {8695666}
\bibfield{author}{\bibinfo{person}{Angshuman Parashar},
  \bibinfo{person}{Priyanka Raina}, \bibinfo{person}{Yakun~Sophia Shao},
  \bibinfo{person}{Yu-Hsin Chen}, {et~al\mbox{.}}}
  \bibinfo{year}{2019}\natexlab{}.
\newblock \showarticletitle{Timeloop: A Systematic Approach to DNN Accelerator
  Evaluation}. In \bibinfo{booktitle}{\emph{Proc. IEEE ISPASS}}.
  \bibinfo{pages}{304--315}.
\newblock


\bibitem[Park et~al\mbox{.}(2025)]%
        {park2025survey}
\bibfield{author}{\bibinfo{person}{Sihyeong Park}, \bibinfo{person}{Sungryeol
  Jeon}, \bibinfo{person}{Chaelyn Lee}, \bibinfo{person}{Seokhun Jeon},
  {et~al\mbox{.}}} \bibinfo{year}{2025}\natexlab{}.
\newblock \showarticletitle{A Survey on Inference Engines for Large Language
  Models: Perspectives on Optimization and Efficiency}.
\newblock \bibinfo{journal}{\emph{arXiv preprint arXiv:2505.01658}}
  (\bibinfo{year}{2025}).
\newblock


\bibitem[Patel et~al\mbox{.}(2024)]%
        {10609649}
\bibfield{author}{\bibinfo{person}{Pratyush Patel}, \bibinfo{person}{Esha
  Choukse}, \bibinfo{person}{Chaojie Zhang}, \bibinfo{person}{Aashaka Shah},
  {et~al\mbox{.}}} \bibinfo{year}{2024}\natexlab{}.
\newblock \showarticletitle{Splitwise: Efficient Generative LLM Inference Using
  Phase Splitting}. In \bibinfo{booktitle}{\emph{Proc. ACM/IEEE Annual Int.
  Symp. Comput. Arch. (ISCA)}}. \bibinfo{pages}{118--132}.
\newblock


\bibitem[Picano et~al\mbox{.}(2025)]%
        {10966456}
\bibfield{author}{\bibinfo{person}{Benedetta Picano},
  \bibinfo{person}{Dinh~Thai Hoang}, {and} \bibinfo{person}{Diep~N. Nguyen}.}
  \bibinfo{year}{2025}\natexlab{}.
\newblock \showarticletitle{A Matching Game for LLM Layer Deployment in
  Heterogeneous Edge Networks}.
\newblock \bibinfo{journal}{\emph{IEEE Open J. the Commun. Soc.}}
  \bibinfo{volume}{6} (\bibinfo{year}{2025}), \bibinfo{pages}{3795--3805}.
\newblock


\bibitem[Piccialli et~al\mbox{.}(2025)]%
        {piccialli2025federated}
\bibfield{author}{\bibinfo{person}{Francesco Piccialli},
  \bibinfo{person}{Diletta Chiaro}, \bibinfo{person}{Pian Qi},
  \bibinfo{person}{Valerio Bellandi}, {and} \bibinfo{person}{Ernesto Damiani}.}
  \bibinfo{year}{2025}\natexlab{}.
\newblock \showarticletitle{Federated and edge learning for large language
  models}.
\newblock \bibinfo{journal}{\emph{Information fusion}}  \bibinfo{volume}{117}
  (\bibinfo{year}{2025}), \bibinfo{pages}{102840}.
\newblock


\bibitem[Porian et~al\mbox{.}(2024)]%
        {NEURIPS2024_b6341525}
\bibfield{author}{\bibinfo{person}{Tomer Porian}, \bibinfo{person}{Mitchell
  Wortsman}, \bibinfo{person}{Jenia Jitsev}, {et~al\mbox{.}}}
  \bibinfo{year}{2024}\natexlab{}.
\newblock \showarticletitle{Resolving Discrepancies in Compute-Optimal Scaling
  of Language Models}. In \bibinfo{booktitle}{\emph{Proc. Adv. Neural Infor.
  Process. Syst. (NeurIPS)}}, Vol.~\bibinfo{volume}{37}.
  \bibinfo{pages}{100535--100570}.
\newblock


\bibitem[Prabhakar et~al\mbox{.}(2024)]%
        {NEURIPS2024_0f4d1fc0}
\bibfield{author}{\bibinfo{person}{Rohan~Baskar Prabhakar},
  \bibinfo{person}{Hengrui Zhang}, {and} \bibinfo{person}{David Wentzlaff}.}
  \bibinfo{year}{2024}\natexlab{}.
\newblock \showarticletitle{Kraken: Inherently Parallel Transformers For
  Efficient Multi-Device Inference}. In \bibinfo{booktitle}{\emph{Proc. Adv.
  Neural Infor. Process. Syst. (NeurIPS)}}, Vol.~\bibinfo{volume}{37}.
  \bibinfo{pages}{7957--7980}.
\newblock


\bibitem[Protection(2018)]%
        {protection2018general}
\bibfield{author}{\bibinfo{person}{Formerly~Data Protection}.}
  \bibinfo{year}{2018}\natexlab{}.
\newblock \showarticletitle{General data protection regulation (GDPR)}.
\newblock \bibinfo{journal}{\emph{Intersoft Consulting, Accessed in October}}
  \bibinfo{volume}{24}, \bibinfo{number}{1} (\bibinfo{year}{2018}).
\newblock


\bibitem[Qu et~al\mbox{.}(2025)]%
        {10835069}
\bibfield{author}{\bibinfo{person}{Guanqiao Qu}, \bibinfo{person}{Qiyuan Chen},
  \bibinfo{person}{Wei Wei}, \bibinfo{person}{Zheng Lin},
  \bibinfo{person}{Xianhao Chen}, {and} \bibinfo{person}{Kaibin Huang}.}
  \bibinfo{year}{2025}\natexlab{}.
\newblock \showarticletitle{Mobile Edge Intelligence for Large Language Models:
  A Contemporary Survey}.
\newblock \bibinfo{journal}{\emph{IEEE Commun. Surveys Tuts.}}
  (\bibinfo{year}{2025}), \bibinfo{pages}{1--1}.
\newblock


\bibitem[{Qualcomm Innovation Center, Inc.}(2024)]%
        {Qualcomm_ai_engine}
\bibfield{author}{\bibinfo{person}{{Qualcomm Innovation Center, Inc.}}}
  \bibinfo{year}{2024}\natexlab{}.
\newblock \bibinfo{booktitle}{\emph{QAI AppBuilder: Quick AI Application
  Builder}}.
\newblock
\urldef\tempurl%
\url{https://github.com/quic/ai-engine-direct-helper}
\showURL{%
\tempurl}


\bibitem[{Qualcomm Technologies, Inc.}(2024)]%
        {qualcomm_llama3_snapdragon_2024}
\bibfield{author}{\bibinfo{person}{{Qualcomm Technologies, Inc.}}}
  \bibinfo{year}{2024}\natexlab{}.
\newblock \bibinfo{title}{Qualcomm Enables {Meta Llama 3} to Run on Devices
  Powered by Snapdragon}.
\newblock
  \bibinfo{howpublished}{\url{https://www.qualcomm.com/news/releases/2024/04/qualcomm-enables-meta-llama-3-to-run-on-devices-powered-by-snapd}}.
\newblock


\bibitem[Raffel et~al\mbox{.}(2020)]%
        {JMLR:v21:20-074}
\bibfield{author}{\bibinfo{person}{Colin Raffel}, \bibinfo{person}{Noam
  Shazeer}, \bibinfo{person}{Adam Roberts}, \bibinfo{person}{Katherine Lee},
  {et~al\mbox{.}}} \bibinfo{year}{2020}\natexlab{}.
\newblock \showarticletitle{Exploring the Limits of Transfer Learning with a
  Unified Text-to-Text Transformer}.
\newblock \bibinfo{journal}{\emph{J. Mach. Learn. Res. (JMLR)}}
  \bibinfo{volume}{21}, \bibinfo{number}{140} (\bibinfo{year}{2020}),
  \bibinfo{pages}{1--67}.
\newblock


\bibitem[Rajbhandari et~al\mbox{.}(2020)]%
        {9355301}
\bibfield{author}{\bibinfo{person}{Samyam Rajbhandari}, \bibinfo{person}{Jeff
  Rasley}, \bibinfo{person}{Olatunji Ruwase}, {and} \bibinfo{person}{Yuxiong
  He}.} \bibinfo{year}{2020}\natexlab{}.
\newblock \showarticletitle{ZeRO: Memory optimizations Toward Training Trillion
  Parameter Models}. In \bibinfo{booktitle}{\emph{Proc. Int. Conf. High Perfor.
  Comput., Netw., Storage and Analysis}}. \bibinfo{pages}{1--16}.
\newblock


\bibitem[Ranaweera et~al\mbox{.}(2021)]%
        {9364272}
\bibfield{author}{\bibinfo{person}{Pasika Ranaweera},
  \bibinfo{person}{Anca~Delia Jurcut}, {and} \bibinfo{person}{Madhusanka
  Liyanage}.} \bibinfo{year}{2021}\natexlab{}.
\newblock \showarticletitle{Survey on Multi-Access Edge Computing Security and
  Privacy}.
\newblock \bibinfo{journal}{\emph{IEEE Commun. Surveys Tuts.}}
  \bibinfo{volume}{23}, \bibinfo{number}{2} (\bibinfo{year}{2021}),
  \bibinfo{pages}{1078--1124}.
\newblock


\bibitem[Rehman et~al\mbox{.}(2025)]%
        {rehman2025claf}
\bibfield{author}{\bibinfo{person}{Abdul Rehman}, \bibinfo{person}{Kamran~Ahmad
  Awan}, \bibinfo{person}{Mahmood~Ul Hassan}, {et~al\mbox{.}}}
  \bibinfo{year}{2025}\natexlab{}.
\newblock \showarticletitle{CLAF-IoT: Context-Aware LLMs-Enhanced
  Authentication Framework for Internet of Things}.
\newblock \bibinfo{journal}{\emph{IEEE Int. Things J.}} (\bibinfo{year}{2025}).
\newblock


\bibitem[Ren et~al\mbox{.}(2025)]%
        {11161045}
\bibfield{author}{\bibinfo{person}{Jiaqi Ren}, \bibinfo{person}{Chao Wang},
  \bibinfo{person}{Yihan Zhong}, \bibinfo{person}{Shaohua Cao},
  \bibinfo{person}{Danyang Zheng}, {and} \bibinfo{person}{Xiaojun Cao}.}
  \bibinfo{year}{2025}\natexlab{}.
\newblock \showarticletitle{Towards Expert Models Deployment Cost Optimization
  in Edge Computing Networks}. In \bibinfo{booktitle}{\emph{Proc. IEEE Int.
  Conf. Commun. (ICC)}}. \bibinfo{pages}{838--843}.
\newblock


\bibitem[Ren et~al\mbox{.}(2019)]%
        {NEURIPS2019_f63f65b5}
\bibfield{author}{\bibinfo{person}{Yi Ren}, \bibinfo{person}{Yangjun Ruan},
  \bibinfo{person}{Xu Tan}, \bibinfo{person}{Tao Qin}, \bibinfo{person}{Sheng
  Zhao}, \bibinfo{person}{Zhou Zhao}, {and} \bibinfo{person}{Tie-Yan Liu}.}
  \bibinfo{year}{2019}\natexlab{}.
\newblock \showarticletitle{FastSpeech: Fast, Robust and Controllable Text to
  Speech}. In \bibinfo{booktitle}{\emph{Proc. Adv. Neural Infor. Process. Syst.
  (NeurIPS)}}, Vol.~\bibinfo{volume}{32}.
\newblock


\bibitem[Sahu et~al\mbox{.}(2023)]%
        {sahu-etal-2023-promptmix}
\bibfield{author}{\bibinfo{person}{Gaurav Sahu}, \bibinfo{person}{Olga
  Vechtomova}, {et~al\mbox{.}}} \bibinfo{year}{2023}\natexlab{}.
\newblock \showarticletitle{{P}rompt{M}ix: A Class Boundary Augmentation Method
  for Large Language Model Distillation}. In \bibinfo{booktitle}{\emph{Proc.
  Conf. Empirical Methods in Natural Language Process.}}
  \bibinfo{pages}{5316--5327}.
\newblock


\bibitem[Schuster et~al\mbox{.}(2022)]%
        {NEURIPS2022_6fac9e31}
\bibfield{author}{\bibinfo{person}{Tal Schuster}, \bibinfo{person}{Adam Fisch},
  \bibinfo{person}{Jai Gupta}, \bibinfo{person}{Mostafa Dehghani},
  \bibinfo{person}{Dara Bahri}, \bibinfo{person}{Vinh Tran},
  \bibinfo{person}{Yi Tay}, {and} \bibinfo{person}{Donald Metzler}.}
  \bibinfo{year}{2022}\natexlab{}.
\newblock \showarticletitle{Confident Adaptive Language Modeling}. In
  \bibinfo{booktitle}{\emph{Proc. Adv. Neural Infor. Process. Syst.
  (NeurIPS)}}, Vol.~\bibinfo{volume}{35}. \bibinfo{pages}{17456--17472}.
\newblock


\bibitem[Sharma et~al\mbox{.}(2024)]%
        {sharma2023truth}
\bibfield{author}{\bibinfo{person}{Pratyusha Sharma},
  \bibinfo{person}{Jordan~T. Ash}, {and} \bibinfo{person}{Dipendra Misra}.}
  \bibinfo{year}{2024}\natexlab{}.
\newblock \showarticletitle{The Truth is in There: Improving Reasoning in
  Language Models with Layer-Selective Rank Reduction}. In
  \bibinfo{booktitle}{\emph{Proc. Int. Conf. Learn. Represent. (ICLR)}}.
\newblock


\bibitem[She et~al\mbox{.}(2024)]%
        {10682995}
\bibfield{author}{\bibinfo{person}{Yechao She}, \bibinfo{person}{Tuo Shi},
  \bibinfo{person}{Jianping Wang}, {and} \bibinfo{person}{Bin Liu}.}
  \bibinfo{year}{2024}\natexlab{}.
\newblock \showarticletitle{Dynamic Batching and Early-Exiting for Accurate and
  Timely Edge Inference}. In \bibinfo{booktitle}{\emph{Proc. IEEE Veh. Technol.
  Conf. (VTC-Spring)}}. \bibinfo{pages}{1--6}.
\newblock


\bibitem[Shen et~al\mbox{.}(2020)]%
        {Shen_Dong_Ye_Ma_Yao_Gholami_Mahoney_Keutzer_2020}
\bibfield{author}{\bibinfo{person}{Sheng Shen}, \bibinfo{person}{Zhen Dong},
  \bibinfo{person}{Jiayu Ye}, \bibinfo{person}{Linjian Ma},
  \bibinfo{person}{Zhewei Yao}, \bibinfo{person}{Amir Gholami},
  \bibinfo{person}{Michael~W. Mahoney}, {and} \bibinfo{person}{Kurt Keutzer}.}
  \bibinfo{year}{2020}\natexlab{}.
\newblock \showarticletitle{Q-BERT: Hessian Based Ultra Low Precision
  Quantization of BERT}.
\newblock \bibinfo{journal}{\emph{Proc. AAAI}} \bibinfo{volume}{34},
  \bibinfo{number}{05} (\bibinfo{year}{2020}), \bibinfo{pages}{8815--8821}.
\newblock


\bibitem[Sheng et~al\mbox{.}(2023)]%
        {pmlr-v202-sheng23a}
\bibfield{author}{\bibinfo{person}{Ying Sheng}, \bibinfo{person}{Lianmin
  Zheng}, \bibinfo{person}{Binhang Yuan}, {et~al\mbox{.}}}
  \bibinfo{year}{2023}\natexlab{}.
\newblock \showarticletitle{{F}lex{G}en: High-Throughput Generative Inference
  of Large Language Models with a Single {GPU}}. In
  \bibinfo{booktitle}{\emph{Proc. Int. Conf. Mach. Learn. (ICML)}},
  Vol.~\bibinfo{volume}{202}. \bibinfo{pages}{31094--31116}.
\newblock


\bibitem[Shoeybi et~al\mbox{.}(2019)]%
        {shoeybi2019megatron}
\bibfield{author}{\bibinfo{person}{Mohammad Shoeybi}, \bibinfo{person}{Mostofa
  Patwary}, \bibinfo{person}{Raul Puri}, {et~al\mbox{.}}}
  \bibinfo{year}{2019}\natexlab{}.
\newblock \showarticletitle{Megatron-LM: Training Multi-Billion Parameter
  Language Models Using Model Parallelism}.
\newblock \bibinfo{journal}{\emph{CoRR}}  \bibinfo{volume}{abs/1909.08053}
  (\bibinfo{year}{2019}).
\newblock


\bibitem[Simoulin and Crabb{\'e}(2021)]%
        {simoulin-crabbe-2021-many}
\bibfield{author}{\bibinfo{person}{Antoine Simoulin} {and}
  \bibinfo{person}{Benoit Crabb{\'e}}.} \bibinfo{year}{2021}\natexlab{}.
\newblock \showarticletitle{How Many Layers and Why? {A}n Analysis of the Model
  Depth in Transformers}. In \bibinfo{booktitle}{\emph{Proc. ACL}}.
  \bibinfo{pages}{221--228}.
\newblock


\bibitem[Song et~al\mbox{.}(2025)]%
        {song2025mixture}
\bibfield{author}{\bibinfo{person}{Qiuchen Song}, \bibinfo{person}{Shusen
  Jing}, \bibinfo{person}{Shuai Zhang}, \bibinfo{person}{Songyang Zhang}, {and}
  \bibinfo{person}{Chuan Huang}.} \bibinfo{year}{2025}\natexlab{}.
\newblock \showarticletitle{Mixture-of-Experts for Distributed Edge Computing
  with Channel-Aware Gating Function}. In \bibinfo{booktitle}{\emph{Proc. IEEE
  Int. Conf. Commun. (ICC)}}.
\newblock


\bibitem[Stojkovic et~al\mbox{.}(2025)]%
        {10946802}
\bibfield{author}{\bibinfo{person}{Jovan Stojkovic}, \bibinfo{person}{Chaojie
  Zhang}, \bibinfo{person}{Íñigo Goiri}, \bibinfo{person}{Josep Torrellas},
  {and} \bibinfo{person}{Esha Choukse}.} \bibinfo{year}{2025}\natexlab{}.
\newblock \showarticletitle{DynamoLLM: Designing LLM Inference Clusters for
  Performance and Energy Efficiency}. In \bibinfo{booktitle}{\emph{Proc. IEEE
  HPCA}}. \bibinfo{pages}{1348--1362}.
\newblock


\bibitem[Sun et~al\mbox{.}(2024)]%
        {sun2023simple}
\bibfield{author}{\bibinfo{person}{Mingjie Sun}, \bibinfo{person}{Zhuang Liu},
  \bibinfo{person}{Anna Bair}, {and} \bibinfo{person}{J~Zico Kolter}.}
  \bibinfo{year}{2024}\natexlab{}.
\newblock \showarticletitle{A Simple and Effective Pruning Approach for Large
  Language Models}. In \bibinfo{booktitle}{\emph{Proc. Int. Conf. Learn.
  Represent. (ICLR)}}.
\newblock


\bibitem[Sun et~al\mbox{.}(2023)]%
        {NEURIPS2023_6034a661}
\bibfield{author}{\bibinfo{person}{Ziteng Sun},
  \bibinfo{person}{Ananda~Theertha Suresh}, \bibinfo{person}{Jae~Hun Ro},
  {et~al\mbox{.}}} \bibinfo{year}{2023}\natexlab{}.
\newblock \showarticletitle{SpecTr: Fast Speculative Decoding via Optimal
  Transport}. In \bibinfo{booktitle}{\emph{Proc. Adv. Neural Infor. Process.
  Syst. (NeurIPS)}}, Vol.~\bibinfo{volume}{36}. \bibinfo{pages}{30222--30242}.
\newblock


\bibitem[Tao et~al\mbox{.}(2023)]%
        {tao-etal-2023-structured}
\bibfield{author}{\bibinfo{person}{Chaofan Tao}, \bibinfo{person}{Lu Hou},
  \bibinfo{person}{Haoli Bai}, \bibinfo{person}{Jiansheng Wei},
  \bibinfo{person}{Xin Jiang}, \bibinfo{person}{Qun Liu}, \bibinfo{person}{Ping
  Luo}, {and} \bibinfo{person}{Ngai Wong}.} \bibinfo{year}{2023}\natexlab{}.
\newblock \showarticletitle{Structured Pruning for Efficient Generative
  Pre-trained Language Models}. In \bibinfo{booktitle}{\emph{Proc. ACL}}.
  \bibinfo{pages}{10880--10895}.
\newblock


\bibitem[Team et~al\mbox{.}(2024)]%
        {team2024gemma}
\bibfield{author}{\bibinfo{person}{Gemma Team}, \bibinfo{person}{Thomas
  Mesnard}, \bibinfo{person}{Cassidy Hardin}, {et~al\mbox{.}}}
  \bibinfo{year}{2024}\natexlab{}.
\newblock \showarticletitle{Gemma: Open models based on gemini research and
  technology}.
\newblock \bibinfo{journal}{\emph{arXiv preprint arXiv:2403.08295}}
  (\bibinfo{year}{2024}).
\newblock


\bibitem[Team(2024)]%
        {team2024qwen2}
\bibfield{author}{\bibinfo{person}{Qwen Team}.}
  \bibinfo{year}{2024}\natexlab{}.
\newblock \showarticletitle{Qwen2 technical report}.
\newblock \bibinfo{journal}{\emph{arXiv preprint arXiv:2407.10671}}
  \bibinfo{volume}{2} (\bibinfo{year}{2024}).
\newblock


\bibitem[Touvron et~al\mbox{.}(2023)]%
        {touvron2023llama}
\bibfield{author}{\bibinfo{person}{Hugo Touvron}, \bibinfo{person}{Thibaut
  Lavril}, \bibinfo{person}{Gautier Izacard}, \bibinfo{person}{Xavier
  Martinet}, {et~al\mbox{.}}} \bibinfo{year}{2023}\natexlab{}.
\newblock \showarticletitle{Llama: Open and efficient foundation language
  models}.
\newblock \bibinfo{journal}{\emph{arXiv preprint arXiv:2302.13971}}
  (\bibinfo{year}{2023}).
\newblock


\bibitem[Tuli and Jha(2023)]%
        {10120981}
\bibfield{author}{\bibinfo{person}{Shikhar Tuli} {and}
  \bibinfo{person}{Niraj~K. Jha}.} \bibinfo{year}{2023}\natexlab{}.
\newblock \showarticletitle{AccelTran: A Sparsity-Aware Accelerator for Dynamic
  Inference With Transformers}.
\newblock \bibinfo{journal}{\emph{IEEE Trans. Computer-Aided Design Integr.
  Circuits Syst.}} \bibinfo{volume}{42}, \bibinfo{number}{11}
  (\bibinfo{year}{2023}), \bibinfo{pages}{4038--4051}.
\newblock


\bibitem[Vaswani et~al\mbox{.}(2017)]%
        {NIPS2017_3f5ee243}
\bibfield{author}{\bibinfo{person}{Ashish Vaswani}, \bibinfo{person}{Noam
  Shazeer}, \bibinfo{person}{Niki Parmar}, \bibinfo{person}{Jakob Uszkoreit},
  \bibinfo{person}{Llion Jones}, \bibinfo{person}{Aidan~N Gomez},
  \bibinfo{person}{\L~ukasz Kaiser}, {and} \bibinfo{person}{Illia Polosukhin}.}
  \bibinfo{year}{2017}\natexlab{}.
\newblock \showarticletitle{Attention is All you Need}. In
  \bibinfo{booktitle}{\emph{Proc. Adv. Neural Infor. Process. Syst.
  (NeurIPS)}}, Vol.~\bibinfo{volume}{30}.
\newblock


\bibitem[Wan et~al\mbox{.}(2024)]%
        {wan2024knowledge}
\bibfield{author}{\bibinfo{person}{Fanqi Wan}, \bibinfo{person}{Xinting Huang},
  \bibinfo{person}{Deng Cai}, \bibinfo{person}{Xiaojun Quan},
  \bibinfo{person}{Wei Bi}, {and} \bibinfo{person}{Shuming Shi}.}
  \bibinfo{year}{2024}\natexlab{}.
\newblock \showarticletitle{Knowledge Fusion of Large Language Models}. In
  \bibinfo{booktitle}{\emph{Proc. Int. Conf. Learn. Rep. (ICLR)}}.
\newblock


\bibitem[Wang et~al\mbox{.}(2024c)]%
        {10.1145/3620666.3651357}
\bibfield{author}{\bibinfo{person}{Haoran Wang}, \bibinfo{person}{Lei Wang},
  \bibinfo{person}{Haobo Xu}, \bibinfo{person}{Ying Wang},
  \bibinfo{person}{Yuming Li}, {and} \bibinfo{person}{Yinhe Han}.}
  \bibinfo{year}{2024}\natexlab{c}.
\newblock \showarticletitle{PrimePar: Efficient Spatial-temporal Tensor
  Partitioning for Large Transformer Model Training}. In
  \bibinfo{booktitle}{\emph{ACM ASPLOS}}. \bibinfo{pages}{801–--817}.
\newblock
\showISBNx{9798400703867}


\bibitem[Wang et~al\mbox{.}(2022a)]%
        {wang-etal-2022-skipbert}
\bibfield{author}{\bibinfo{person}{Jue Wang}, \bibinfo{person}{Ke Chen},
  \bibinfo{person}{Gang Chen}, \bibinfo{person}{Lidan Shou}, {and}
  \bibinfo{person}{Julian McAuley}.} \bibinfo{year}{2022}\natexlab{a}.
\newblock \showarticletitle{{S}kip{BERT}: Efficient Inference with Shallow
  Layer Skipping}. In \bibinfo{booktitle}{\emph{Proc. Annual Meeting of the
  ACL)}}. \bibinfo{pages}{7287--7301}.
\newblock


\bibitem[Wang et~al\mbox{.}(2024b)]%
        {wang2024mixture}
\bibfield{author}{\bibinfo{person}{Junlin Wang}, \bibinfo{person}{Jue Wang},
  \bibinfo{person}{Ben Athiwaratkun}, \bibinfo{person}{Ce Zhang}, {and}
  \bibinfo{person}{James Zou}.} \bibinfo{year}{2024}\natexlab{b}.
\newblock \showarticletitle{Mixture-of-agents enhances large language model
  capabilities}.
\newblock \bibinfo{journal}{\emph{arXiv preprint arXiv:2406.04692}}
  (\bibinfo{year}{2024}).
\newblock


\bibitem[Wang et~al\mbox{.}(2020)]%
        {NEURIPS2020_3f5ee243}
\bibfield{author}{\bibinfo{person}{Wenhui Wang}, \bibinfo{person}{Furu Wei},
  \bibinfo{person}{Li Dong}, {et~al\mbox{.}}} \bibinfo{year}{2020}\natexlab{}.
\newblock \showarticletitle{MiniLM: Deep Self-Attention Distillation for
  Task-Agnostic Compression of Pre-Trained Transformers}. In
  \bibinfo{booktitle}{\emph{Proc. Adv. Neural Infor. Process. Syst.
  (NeurIPS)}}, Vol.~\bibinfo{volume}{33}. \bibinfo{pages}{5776--5788}.
\newblock


\bibitem[Wang et~al\mbox{.}(2022b)]%
        {wang2022self}
\bibfield{author}{\bibinfo{person}{Xuezhi Wang}, \bibinfo{person}{Jason Wei},
  \bibinfo{person}{Dale Schuurmans}, \bibinfo{person}{Quoc Le},
  \bibinfo{person}{Ed Chi}, {et~al\mbox{.}}} \bibinfo{year}{2022}\natexlab{b}.
\newblock \showarticletitle{Self-consistency improves chain of thought
  reasoning in language models}.
\newblock \bibinfo{journal}{\emph{arXiv preprint arXiv:2203.11171}}
  (\bibinfo{year}{2022}).
\newblock


\bibitem[Wang et~al\mbox{.}(2024a)]%
        {wang2024revisiting}
\bibfield{author}{\bibinfo{person}{Zhibin Wang}, \bibinfo{person}{Shipeng Li},
  \bibinfo{person}{Yuhang Zhou}, \bibinfo{person}{Xue Li},
  \bibinfo{person}{Rong Gu}, \bibinfo{person}{Nguyen Cam-Tu},
  \bibinfo{person}{Chen Tian}, {and} \bibinfo{person}{Sheng Zhong}.}
  \bibinfo{year}{2024}\natexlab{a}.
\newblock \showarticletitle{Revisiting slo and goodput metrics in llm serving}.
\newblock \bibinfo{journal}{\emph{arXiv preprint arXiv:2410.14257}}
  (\bibinfo{year}{2024}).
\newblock


\bibitem[Wei et~al\mbox{.}(2022a)]%
        {wei2022finetuned}
\bibfield{author}{\bibinfo{person}{Jason Wei}, \bibinfo{person}{Maarten Bosma},
  \bibinfo{person}{Vincent Zhao}, \bibinfo{person}{Kelvin Guu},
  \bibinfo{person}{Adams~Wei Yu}, \bibinfo{person}{Brian Lester},
  \bibinfo{person}{Nan Du}, \bibinfo{person}{Andrew~M. Dai}, {and}
  \bibinfo{person}{Quoc~V Le}.} \bibinfo{year}{2022}\natexlab{a}.
\newblock \showarticletitle{Finetuned Language Models are Zero-Shot Learners}.
  In \bibinfo{booktitle}{\emph{Proc. Int. Conf. Learn. Represent. (ICLR)}}.
\newblock


\bibitem[Wei et~al\mbox{.}(2022b)]%
        {NEURIPS2022_9d560961}
\bibfield{author}{\bibinfo{person}{Jason Wei}, \bibinfo{person}{Xuezhi Wang},
  \bibinfo{person}{Dale Schuurmans}, {et~al\mbox{.}}}
  \bibinfo{year}{2022}\natexlab{b}.
\newblock \showarticletitle{Chain-of-Thought Prompting Elicits Reasoning in
  Large Language Models}. In \bibinfo{booktitle}{\emph{Advances in Neural
  Information Processing Systems}}, Vol.~\bibinfo{volume}{35}.
  \bibinfo{pages}{24824--24837}.
\newblock


\bibitem[Wei et~al\mbox{.}(2023)]%
        {wei2023outlier}
\bibfield{author}{\bibinfo{person}{Xiuying Wei}, \bibinfo{person}{Yunchen
  Zhang}, \bibinfo{person}{Yuhang Li}, {et~al\mbox{.}}}
  \bibinfo{year}{2023}\natexlab{}.
\newblock \showarticletitle{Outlier suppression+: Accurate quantization of
  large language models by equivalent and optimal shifting and scaling}.
\newblock \bibinfo{journal}{\emph{arXiv preprint arXiv:2304.09145}}
  (\bibinfo{year}{2023}).
\newblock


\bibitem[Wilhelm et~al\mbox{.}(2025)]%
        {10.1145/3721146.3721953}
\bibfield{author}{\bibinfo{person}{Patrick Wilhelm}, \bibinfo{person}{Thorsten
  Wittkopp}, {and} \bibinfo{person}{Odej Kao}.}
  \bibinfo{year}{2025}\natexlab{}.
\newblock \showarticletitle{Beyond Test-Time Compute Strategies: Advocating
  Energy-per-Token in LLM Inference}. In \bibinfo{booktitle}{\emph{Proc.
  Workshop on Machine Learn. and Sys. (EuroMLSys)}}.
  \bibinfo{pages}{208–215}.
\newblock


\bibitem[Wong et~al\mbox{.}(2021)]%
        {9264694}
\bibfield{author}{\bibinfo{person}{Kai-Kit Wong} {et~al\mbox{.}}}
  \bibinfo{year}{2021}\natexlab{}.
\newblock \showarticletitle{Fluid Antenna Systems}.
\newblock \bibinfo{journal}{\emph{IEEE Trans. Wireless Commun.}}
  \bibinfo{volume}{20}, \bibinfo{number}{3} (\bibinfo{year}{2021}),
  \bibinfo{pages}{1950--1962}.
\newblock


\bibitem[Wu et~al\mbox{.}(2023)]%
        {wu2023fast}
\bibfield{author}{\bibinfo{person}{Bingyang Wu}, \bibinfo{person}{Yinmin
  Zhong}, \bibinfo{person}{Zili Zhang}, \bibinfo{person}{Shengyu Liu},
  \bibinfo{person}{Fangyue Liu}, \bibinfo{person}{Yuanhang Sun},
  \bibinfo{person}{Gang Huang}, \bibinfo{person}{Xuanzhe Liu}, {and}
  \bibinfo{person}{Xin Jin}.} \bibinfo{year}{2023}\natexlab{}.
\newblock \showarticletitle{Fast distributed inference serving for large
  language models}.
\newblock \bibinfo{journal}{\emph{arXiv preprint arXiv:2305.05920}}
  (\bibinfo{year}{2023}).
\newblock


\bibitem[Wu et~al\mbox{.}(2025a)]%
        {wu-etal-2025-tokenselect}
\bibfield{author}{\bibinfo{person}{Wei Wu} {et~al\mbox{.}}}
  \bibinfo{year}{2025}\natexlab{a}.
\newblock \showarticletitle{{T}oken{S}elect: Efficient Long-Context Inference
  and Length Extrapolation for {LLM}s via Dynamic Token-Level {KV} Cache
  Selection}. In \bibinfo{booktitle}{\emph{Proc. Conf. Empirical Methods in
  Natural Language Process.}} \bibinfo{pages}{21275--21292}.
\newblock


\bibitem[Wu et~al\mbox{.}(2025b)]%
        {wu2025arrow}
\bibfield{author}{\bibinfo{person}{Yu Wu}, \bibinfo{person}{Tongxuan Liu},
  \bibinfo{person}{Yuting Zeng}, \bibinfo{person}{Siyu Wu}, {et~al\mbox{.}}}
  \bibinfo{year}{2025}\natexlab{b}.
\newblock \showarticletitle{Arrow: Adaptive Scheduling Mechanisms for
  Disaggregated LLM Inference Architecture}.
\newblock \bibinfo{journal}{\emph{arXiv preprint arXiv:2505.11916}}
  (\bibinfo{year}{2025}).
\newblock


\bibitem[Xia et~al\mbox{.}(2024b)]%
        {xia2024unlocking}
\bibfield{author}{\bibinfo{person}{Heming Xia}, \bibinfo{person}{Zhe Yang},
  \bibinfo{person}{Qingxiu Dong}, \bibinfo{person}{Peiyi Wang},
  {et~al\mbox{.}}} \bibinfo{year}{2024}\natexlab{b}.
\newblock \showarticletitle{Unlocking efficiency in large language model
  inference: A comprehensive survey of speculative decoding}.
\newblock \bibinfo{journal}{\emph{arXiv preprint arXiv:2401.07851}}
  (\bibinfo{year}{2024}).
\newblock


\bibitem[Xia et~al\mbox{.}(2023)]%
        {xia2023flash}
\bibfield{author}{\bibinfo{person}{Haojun Xia}, \bibinfo{person}{Zhen Zheng},
  \bibinfo{person}{Yuchao Li}, \bibinfo{person}{Donglin Zhuang},
  {et~al\mbox{.}}} \bibinfo{year}{2023}\natexlab{}.
\newblock \showarticletitle{Flash-llm: Enabling cost-effective and
  highly-efficient large generative model inference with unstructured
  sparsity}.
\newblock \bibinfo{journal}{\emph{arXiv preprint arXiv:2309.10285}}
  (\bibinfo{year}{2023}).
\newblock


\bibitem[Xia et~al\mbox{.}(2024a)]%
        {xia2023sheared}
\bibfield{author}{\bibinfo{person}{Mengzhou Xia}, \bibinfo{person}{Tianyu Gao},
  \bibinfo{person}{Zhiyuan Zeng}, {and} \bibinfo{person}{Danqi Chen}.}
  \bibinfo{year}{2024}\natexlab{a}.
\newblock \showarticletitle{Sheared llama: Accelerating language model
  pre-training via structured pruning}. In \bibinfo{booktitle}{\emph{Proc. Int.
  Conf. Learn. Represent. (ICLR)}}.
\newblock


\bibitem[Xiao et~al\mbox{.}(2023)]%
        {pmlr-v202-xiao23c}
\bibfield{author}{\bibinfo{person}{Guangxuan Xiao}, \bibinfo{person}{Ji Lin},
  \bibinfo{person}{Mickael Seznec}, {et~al\mbox{.}}}
  \bibinfo{year}{2023}\natexlab{}.
\newblock \showarticletitle{{S}mooth{Q}uant: Accurate and Efficient
  Post-Training Quantization for Large Language Models}. In
  \bibinfo{booktitle}{\emph{Proc. Int. Conf. Mach. Learn. (ICML)}},
  Vol.~\bibinfo{volume}{202}. \bibinfo{pages}{38087--38099}.
\newblock


\bibitem[Xie et~al\mbox{.}(2025)]%
        {xie2025novel}
\bibfield{author}{\bibinfo{person}{Zuan Xie}, \bibinfo{person}{Yang Xu},
  \bibinfo{person}{Hongli Xu}, \bibinfo{person}{Yunming Liao}, {and}
  \bibinfo{person}{Zhiwei Yao}.} \bibinfo{year}{2025}\natexlab{}.
\newblock \showarticletitle{A Novel Hat-Shaped Device-Cloud Collaborative
  Inference Framework for Large Language Models}.
\newblock \bibinfo{journal}{\emph{arXiv preprint arXiv:2503.18989}}
  (\bibinfo{year}{2025}).
\newblock


\bibitem[Xin et~al\mbox{.}(2020)]%
        {xin2020deebert}
\bibfield{author}{\bibinfo{person}{Ji Xin}, \bibinfo{person}{Raphael Tang},
  \bibinfo{person}{Jaejun Lee}, \bibinfo{person}{Yaoliang Yu}, {and}
  \bibinfo{person}{Jimmy Lin}.} \bibinfo{year}{2020}\natexlab{}.
\newblock \showarticletitle{DeeBERT: Dynamic early exiting for accelerating
  BERT inference}.
\newblock \bibinfo{journal}{\emph{arXiv preprint arXiv:2004.12993}}
  (\bibinfo{year}{2020}).
\newblock


\bibitem[Xu et~al\mbox{.}(2025b)]%
        {10812936}
\bibfield{author}{\bibinfo{person}{Daliang Xu}, \bibinfo{person}{Wangsong Yin},
  \bibinfo{person}{Hao Zhang}, \bibinfo{person}{Xin Jin}, {et~al\mbox{.}}}
  \bibinfo{year}{2025}\natexlab{b}.
\newblock \showarticletitle{EdgeLLM: Fast On-Device LLM Inference With
  Speculative Decoding}.
\newblock \bibinfo{journal}{\emph{IEEE Trans. Mobile Comput.}}
  \bibinfo{volume}{24}, \bibinfo{number}{4} (\bibinfo{year}{2025}),
  \bibinfo{pages}{3256--3273}.
\newblock


\bibitem[Xu et~al\mbox{.}(2024b)]%
        {xu2024device}
\bibfield{author}{\bibinfo{person}{Jiajun Xu}, \bibinfo{person}{Zhiyuan Li},
  \bibinfo{person}{Wei Chen}, \bibinfo{person}{Qun Wang}, \bibinfo{person}{Xin
  Gao}, \bibinfo{person}{Qi Cai}, {and} \bibinfo{person}{Ziyuan Ling}.}
  \bibinfo{year}{2024}\natexlab{b}.
\newblock \showarticletitle{On-device language models: A comprehensive review}.
\newblock \bibinfo{journal}{\emph{arXiv preprint arXiv:2409.00088}}
  (\bibinfo{year}{2024}).
\newblock


\bibitem[Xu et~al\mbox{.}(2024a)]%
        {10398474}
\bibfield{author}{\bibinfo{person}{Minrui Xu}, \bibinfo{person}{Hongyang Du},
  \bibinfo{person}{Dusit Niyato}, \bibinfo{person}{Jiawen Kang},
  {et~al\mbox{.}}} \bibinfo{year}{2024}\natexlab{a}.
\newblock \showarticletitle{Unleashing the Power of Edge-Cloud Generative AI in
  Mobile Networks: A Survey of AIGC Services}.
\newblock \bibinfo{journal}{\emph{IEEE Commun. Surveys Tuts.}}
  \bibinfo{volume}{26}, \bibinfo{number}{2} (\bibinfo{year}{2024}),
  \bibinfo{pages}{1127--1170}.
\newblock


\bibitem[Xu et~al\mbox{.}(2025a)]%
        {11154975}
\bibfield{author}{\bibinfo{person}{Xinyi Xu}, \bibinfo{person}{Gang Feng},
  \bibinfo{person}{Yijing Liu}, \bibinfo{person}{Shuang Qin},
  \bibinfo{person}{Jian Wang}, {and} \bibinfo{person}{Yunxiang Wang}.}
  \bibinfo{year}{2025}\natexlab{a}.
\newblock \showarticletitle{Joint Inference Offloading and Model Caching for
  Small and Large Language Model Collaboration}.
\newblock \bibinfo{journal}{\emph{IEEE Trans. Mobile Comput.}}
  (\bibinfo{year}{2025}), \bibinfo{pages}{1--16}.
\newblock


\bibitem[Xue et~al\mbox{.}(2021)]%
        {xue-etal-2021-mt5}
\bibfield{author}{\bibinfo{person}{Linting Xue}, \bibinfo{person}{Noah
  Constant}, \bibinfo{person}{Adam Roberts}, \bibinfo{person}{Mihir Kale},
  \bibinfo{person}{Rami Al-Rfou}, \bibinfo{person}{Aditya Siddhant},
  \bibinfo{person}{Aditya Barua}, {and} \bibinfo{person}{Colin Raffel}.}
  \bibinfo{year}{2021}\natexlab{}.
\newblock \showarticletitle{m{T}5: A Massively Multilingual Pre-trained
  Text-to-Text Transformer}. In \bibinfo{booktitle}{\emph{Proc. ACL}}.
  \bibinfo{pages}{483--498}.
\newblock


\bibitem[Xue et~al\mbox{.}(2025)]%
        {11083676}
\bibfield{author}{\bibinfo{person}{Nan Xue}, \bibinfo{person}{Yaping Sun},
  \bibinfo{person}{Zhiyong Chen}, \bibinfo{person}{Meixia Tao},
  {et~al\mbox{.}}} \bibinfo{year}{2025}\natexlab{}.
\newblock \showarticletitle{WDMoE: Wireless Distributed Mixture of Experts for
  Large Language Models}.
\newblock \bibinfo{journal}{\emph{IEEE Trans. Wireless Commun.}}
  (\bibinfo{year}{2025}), \bibinfo{pages}{1--1}.
\newblock


\bibitem[Yang et~al\mbox{.}(2025b)]%
        {yang2025qwen2}
\bibfield{author}{\bibinfo{person}{An Yang}, \bibinfo{person}{Bowen Yu},
  \bibinfo{person}{Chengyuan Li}, \bibinfo{person}{Dayiheng Liu},
  \bibinfo{person}{Fei Huang}, \bibinfo{person}{Haoyan Huang},
  \bibinfo{person}{Jiandong Jiang}, \bibinfo{person}{Jianhong Tu},
  \bibinfo{person}{Jianwei Zhang}, \bibinfo{person}{Jingren Zhou},
  {et~al\mbox{.}}} \bibinfo{year}{2025}\natexlab{b}.
\newblock \showarticletitle{Qwen2. 5-1m technical report}.
\newblock \bibinfo{journal}{\emph{arXiv preprint arXiv:2501.15383}}
  (\bibinfo{year}{2025}).
\newblock


\bibitem[Yang et~al\mbox{.}(2025a)]%
        {11086391}
\bibfield{author}{\bibinfo{person}{Jin Yang}, \bibinfo{person}{Qiong Wu},
  \bibinfo{person}{Zhiying Feng}, \bibinfo{person}{Zhi Zhou},
  \bibinfo{person}{Deke Guo}, {and} \bibinfo{person}{Xu Chen}.}
  \bibinfo{year}{2025}\natexlab{a}.
\newblock \showarticletitle{Quality-of-Service Aware LLM Routing for Edge
  Computing With Multiple Experts}.
\newblock \bibinfo{journal}{\emph{IEEE Trans. Mobile Comput.}}
  \bibinfo{volume}{24}, \bibinfo{number}{12} (\bibinfo{year}{2025}),
  \bibinfo{pages}{13648--13662}.
\newblock


\bibitem[Yang and Gursoy(2025)]%
        {10978053}
\bibfield{author}{\bibinfo{person}{Yang Yang} {and} \bibinfo{person}{M.~Cenk
  Gursoy}.} \bibinfo{year}{2025}\natexlab{}.
\newblock \showarticletitle{Collaborative Inference in RIS-Assisted MEC
  Networks under Computing Backlog Constraints}.
\newblock \bibinfo{journal}{\emph{IEEE Trans. Commun.}} (\bibinfo{year}{2025}),
  \bibinfo{pages}{1--1}.
\newblock


\bibitem[Yao et~al\mbox{.}(2023)]%
        {NEURIPS2023_271db992}
\bibfield{author}{\bibinfo{person}{Shunyu Yao}, \bibinfo{person}{Dian Yu},
  \bibinfo{person}{Jeffrey Zhao}, \bibinfo{person}{Izhak Shafran},
  {et~al\mbox{.}}} \bibinfo{year}{2023}\natexlab{}.
\newblock \showarticletitle{Tree of Thoughts: Deliberate Problem Solving with
  Large Language Models}. In \bibinfo{booktitle}{\emph{Advances in Neural
  Information Processing Systems}}, Vol.~\bibinfo{volume}{36}.
  \bibinfo{pages}{11809--11822}.
\newblock


\bibitem[Yao et~al\mbox{.}(2022b)]%
        {yao2022react}
\bibfield{author}{\bibinfo{person}{Shunyu Yao}, \bibinfo{person}{Jeffrey Zhao},
  \bibinfo{person}{Dian Yu}, \bibinfo{person}{Nan Du}, {et~al\mbox{.}}}
  \bibinfo{year}{2022}\natexlab{b}.
\newblock \showarticletitle{React: Synergizing reasoning and acting in language
  models}. In \bibinfo{booktitle}{\emph{The eleventh international conference
  on learning representations}}.
\newblock


\bibitem[Yao et~al\mbox{.}(2022a)]%
        {NEURIPS2022_adf7fa39}
\bibfield{author}{\bibinfo{person}{Zhewei Yao}, \bibinfo{person}{Reza
  Yazdani~Aminabadi}, \bibinfo{person}{Minjia Zhang}, {et~al\mbox{.}}}
  \bibinfo{year}{2022}\natexlab{a}.
\newblock \showarticletitle{ZeroQuant: Efficient and Affordable Post-Training
  Quantization for Large-Scale Transformers}. In \bibinfo{booktitle}{\emph{Adv.
  Neural Infor. Process. Sys.}}, Vol.~\bibinfo{volume}{35}.
  \bibinfo{pages}{27168--27183}.
\newblock


\bibitem[Ye et~al\mbox{.}(2025b)]%
        {11044734}
\bibfield{author}{\bibinfo{person}{Shengyuan Ye}, \bibinfo{person}{Bei Ouyang},
  \bibinfo{person}{Liekang Zeng}, {et~al\mbox{.}}}
  \bibinfo{year}{2025}\natexlab{b}.
\newblock \showarticletitle{Jupiter: Fast and Resource-Efficient Collaborative
  Inference of Generative LLMs on Edge Devices}. In
  \bibinfo{booktitle}{\emph{Proc. IEEE Conf. Comput. Commun. (INFOCOM)}}.
  \bibinfo{pages}{1--10}.
\newblock


\bibitem[Ye et~al\mbox{.}(2024)]%
        {10.1145/3636534.3649363}
\bibfield{author}{\bibinfo{person}{Shengyuan Ye}, \bibinfo{person}{Liekang
  Zeng}, \bibinfo{person}{Xiaowen Chu}, {et~al\mbox{.}}}
  \bibinfo{year}{2024}\natexlab{}.
\newblock \showarticletitle{Asteroid: Resource-Efficient Hybrid Pipeline
  Parallelism for Collaborative DNN Training on Heterogeneous Edge Devices}. In
  \bibinfo{booktitle}{\emph{Proc. ACM MobiCom}}. \bibinfo{pages}{312–--326}.
\newblock
\showISBNx{9798400704895}


\bibitem[Ye et~al\mbox{.}(2025a)]%
        {ye2025flashinfer}
\bibfield{author}{\bibinfo{person}{Zihao Ye}, \bibinfo{person}{Lequn Chen},
  \bibinfo{person}{Ruihang Lai}, \bibinfo{person}{Wuwei Lin}, {et~al\mbox{.}}}
  \bibinfo{year}{2025}\natexlab{a}.
\newblock \showarticletitle{FlashInfer: Efficient and Customizable Attention
  Engine for {LLM} Inference Serving}. In \bibinfo{booktitle}{\emph{Proc.
  Annual Conf. Machine Learn. and Sys. (MLSys)}}.
\newblock


\bibitem[Yin et~al\mbox{.}(2024)]%
        {10.5555/3692070.3694428}
\bibfield{author}{\bibinfo{person}{Lu Yin}, \bibinfo{person}{You Wu},
  \bibinfo{person}{Zhenyu Zhang}, {et~al\mbox{.}}}
  \bibinfo{year}{2024}\natexlab{}.
\newblock \showarticletitle{Outlier weighed layerwise sparsity (OWL): a missing
  secret sauce for pruning LLMs to high sparsity}. In
  \bibinfo{booktitle}{\emph{Proc. Int. Conf. Machine Learn. (ICML)}}.
  \bibinfo{pages}{57101--57115}.
\newblock


\bibitem[Yu et~al\mbox{.}(2022)]%
        {280922}
\bibfield{author}{\bibinfo{person}{Gyeong-In Yu}, \bibinfo{person}{Joo~Seong
  Jeong}, \bibinfo{person}{Geon-Woo Kim}, \bibinfo{person}{Soojeong Kim}, {and}
  \bibinfo{person}{Byung-Gon Chun}.} \bibinfo{year}{2022}\natexlab{}.
\newblock \showarticletitle{Orca: A Distributed Serving System for
  {Transformer-Based} Generative Models}. In \bibinfo{booktitle}{\emph{Proc.
  USENIX OSDI}}. \bibinfo{pages}{521--538}.
\newblock
\showISBNx{978-1-939133-28-1}


\bibitem[Yuan et~al\mbox{.}(2023)]%
        {yuan2023asvd}
\bibfield{author}{\bibinfo{person}{Zhihang Yuan}, \bibinfo{person}{Yuzhang
  Shang}, \bibinfo{person}{Yue Song}, \bibinfo{person}{Qiang Wu},
  \bibinfo{person}{Yan Yan}, {and} \bibinfo{person}{Guangyu Sun}.}
  \bibinfo{year}{2023}\natexlab{}.
\newblock \showarticletitle{Asvd: Activation-aware singular value decomposition
  for compressing large language models}.
\newblock \bibinfo{journal}{\emph{arXiv preprint arXiv:2312.05821}}
  (\bibinfo{year}{2023}).
\newblock


\bibitem[Zafrir et~al\mbox{.}(2019)]%
        {9463531}
\bibfield{author}{\bibinfo{person}{Ofir Zafrir}, \bibinfo{person}{Guy
  Boudoukh}, {et~al\mbox{.}}} \bibinfo{year}{2019}\natexlab{}.
\newblock \showarticletitle{Q8BERT: Quantized 8Bit BERT}. In
  \bibinfo{booktitle}{\emph{Proc. EMC2-NIPS}}. \bibinfo{pages}{36--39}.
\newblock


\bibitem[Zellers et~al\mbox{.}(2019)]%
        {zellers-etal-2019-hellaswag}
\bibfield{author}{\bibinfo{person}{Rowan Zellers}, \bibinfo{person}{Ari
  Holtzman}, \bibinfo{person}{Yonatan Bisk}, \bibinfo{person}{Ali Farhadi},
  {and} \bibinfo{person}{Yejin Choi}.} \bibinfo{year}{2019}\natexlab{}.
\newblock \showarticletitle{{H}ella{S}wag: Can a Machine Really Finish Your
  Sentence?}. In \bibinfo{booktitle}{\emph{Proc. ACL}}.
  \bibinfo{pages}{4791--4800}.
\newblock


\bibitem[Zeng et~al\mbox{.}(2024)]%
        {Zeng_Hong_Dai_Zhuang_Chen_2024}
\bibfield{author}{\bibinfo{person}{Ziqian Zeng}, \bibinfo{person}{Yihuai Hong},
  \bibinfo{person}{Hongliang Dai}, {et~al\mbox{.}}}
  \bibinfo{year}{2024}\natexlab{}.
\newblock \showarticletitle{ConsistentEE: A Consistent and Hardness-Guided
  Early Exiting Method for Accelerating Language Models Inference}.
\newblock \bibinfo{journal}{\emph{Proc. AAAI}} \bibinfo{volume}{38},
  \bibinfo{number}{17} (\bibinfo{date}{Mar.} \bibinfo{year}{2024}),
  \bibinfo{pages}{19506--19514}.
\newblock


\bibitem[Zhang et~al\mbox{.}(2024b)]%
        {10382642}
\bibfield{author}{\bibinfo{person}{Deyu Zhang}, \bibinfo{person}{Yunzhen Luo},
  \bibinfo{person}{Yaobo Wang}, \bibinfo{person}{Xiaoyan Kui}, {and}
  \bibinfo{person}{Ju Ren}.} \bibinfo{year}{2024}\natexlab{b}.
\newblock \showarticletitle{BatOpt: Optimizing GPU-Based Deep Learning
  Inference Using Dynamic Batch Processing}.
\newblock \bibinfo{journal}{\emph{IEEE Trans. Cloud Comput.}}
  \bibinfo{volume}{12}, \bibinfo{number}{1} (\bibinfo{year}{2024}),
  \bibinfo{pages}{174--185}.
\newblock


\bibitem[Zhang et~al\mbox{.}(2025a)]%
        {zhang2025communication}
\bibfield{author}{\bibinfo{person}{Kai Zhang}, \bibinfo{person}{Hengtao He},
  \bibinfo{person}{Shenghui Song}, \bibinfo{person}{Jun Zhang}, {and}
  \bibinfo{person}{Khaled~B Letaief}.} \bibinfo{year}{2025}\natexlab{a}.
\newblock \showarticletitle{Distributed on-device llm inference with
  over-the-air computation}. In \bibinfo{booktitle}{\emph{Proc. IEEE Int. Conf.
  Commun. (ICC)}}.
\newblock


\bibitem[Zhang et~al\mbox{.}(2025c)]%
        {10818760}
\bibfield{author}{\bibinfo{person}{Mingjin Zhang}, \bibinfo{person}{Xiaoming
  Shen}, \bibinfo{person}{Jiannong Cao}, \bibinfo{person}{Zeyang Cui}, {and}
  \bibinfo{person}{Shan Jiang}.} \bibinfo{year}{2025}\natexlab{c}.
\newblock \showarticletitle{EdgeShard: Efficient LLM Inference via
  Collaborative Edge Computing}.
\newblock \bibinfo{journal}{\emph{IEEE Int. Things J.}} \bibinfo{volume}{12},
  \bibinfo{number}{10} (\bibinfo{year}{2025}), \bibinfo{pages}{13119--13131}.
\newblock


\bibitem[Zhang et~al\mbox{.}(2024a)]%
        {zhang2024tinyllama}
\bibfield{author}{\bibinfo{person}{Peiyuan Zhang} {et~al\mbox{.}}}
  \bibinfo{year}{2024}\natexlab{a}.
\newblock \showarticletitle{Tinyllama: An open-source small language model}.
\newblock \bibinfo{journal}{\emph{arXiv preprint arXiv:2401.02385}}
  (\bibinfo{year}{2024}).
\newblock


\bibitem[Zhang et~al\mbox{.}(2022)]%
        {zhang2022opt}
\bibfield{author}{\bibinfo{person}{Susan Zhang}, \bibinfo{person}{Stephen
  Roller}, \bibinfo{person}{Naman Goyal}, \bibinfo{person}{Mikel Artetxe},
  {et~al\mbox{.}}} \bibinfo{year}{2022}\natexlab{}.
\newblock \showarticletitle{Opt: Open pre-trained transformer language models}.
\newblock \bibinfo{journal}{\emph{arXiv preprint arXiv:2205.01068}}
  (\bibinfo{year}{2022}).
\newblock


\bibitem[Zhang et~al\mbox{.}(2020)]%
        {zhang-etal-2020-ternarybert}
\bibfield{author}{\bibinfo{person}{Wei Zhang}, \bibinfo{person}{Lu Hou},
  \bibinfo{person}{Yichun Yin}, \bibinfo{person}{Lifeng Shang},
  \bibinfo{person}{Xiao Chen}, \bibinfo{person}{Xin Jiang}, {and}
  \bibinfo{person}{Qun Liu}.} \bibinfo{year}{2020}\natexlab{}.
\newblock \showarticletitle{{T}ernary{BERT}: Distillation-aware Ultra-low Bit
  {BERT}}. In \bibinfo{booktitle}{\emph{Proc. Conf. Empirical Methods in
  Natural Language Process. (EMNLP)}}. \bibinfo{pages}{509--521}.
\newblock


\bibitem[Zhang et~al\mbox{.}(2025b)]%
        {10759588}
\bibfield{author}{\bibinfo{person}{Xinyuan Zhang}, \bibinfo{person}{Jiangtian
  Nie}, {et~al\mbox{.}}} \bibinfo{year}{2025}\natexlab{b}.
\newblock \showarticletitle{Beyond the Cloud: Edge Inference for Generative
  Large Language Models in Wireless Networks}.
\newblock \bibinfo{journal}{\emph{IEEE Trans. Wireless Commun.}}
  \bibinfo{volume}{24}, \bibinfo{number}{1} (\bibinfo{year}{2025}),
  \bibinfo{pages}{643--658}.
\newblock


\bibitem[Zhao et~al\mbox{.}(2024a)]%
        {zhao-etal-2024-lingualinked}
\bibfield{author}{\bibinfo{person}{Junchen Zhao}, \bibinfo{person}{Yurun Song},
  \bibinfo{person}{Simeng Liu}, \bibinfo{person}{Ian~G. Harris}, {and}
  \bibinfo{person}{Sangeetha Abdu~Jyothi}.} \bibinfo{year}{2024}\natexlab{a}.
\newblock \showarticletitle{{L}ingua{L}inked: Distributed Large Language Model
  Inference on Mobile Devices}. In \bibinfo{booktitle}{\emph{Proc. ACL)}}.
  \bibinfo{pages}{160--171}.
\newblock


\bibitem[Zhao et~al\mbox{.}(2024b)]%
        {10.1145/3627535.3638480}
\bibfield{author}{\bibinfo{person}{Juntao Zhao}, \bibinfo{person}{Borui Wan},
  \bibinfo{person}{Chuan Wu}, \bibinfo{person}{Yanghua Peng}, {and}
  \bibinfo{person}{Haibin Lin}.} \bibinfo{year}{2024}\natexlab{b}.
\newblock \showarticletitle{POSTER: LLM-PQ:Serving LLM on Heterogeneous
  Clusters with Phase-Aware Partition and Adaptive Quantization}. In
  \bibinfo{booktitle}{\emph{Proc. ACM PPoPP}}. \bibinfo{pages}{460–--462}.
\newblock
\showISBNx{9798400704352}


\bibitem[Zheng and Yang(2025)]%
        {zheng2025communication}
\bibfield{author}{\bibinfo{person}{Ce Zheng} {and} \bibinfo{person}{Tingting
  Yang}.} \bibinfo{year}{2025}\natexlab{}.
\newblock \showarticletitle{Communication-Efficient Collaborative LLM Inference
  via Distributed Speculative Decoding}. In \bibinfo{booktitle}{\emph{Proc.
  Int. Conf. Wireless Commun. and Signal Process. (WCSP)}}.
\newblock


\bibitem[Zheng et~al\mbox{.}(2025b)]%
        {10832517}
\bibfield{author}{\bibinfo{person}{Guangyuan Zheng}, \bibinfo{person}{Miaowen
  Wen}, {et~al\mbox{.}}} \bibinfo{year}{2025}\natexlab{b}.
\newblock \showarticletitle{Computation-Aware Offloading for DNN Inference
  Tasks in Semantic Communication Assisted MEC Systems}.
\newblock \bibinfo{journal}{\emph{IEEE Trans. Wireless Commun.}}
  \bibinfo{volume}{24}, \bibinfo{number}{4} (\bibinfo{year}{2025}),
  \bibinfo{pages}{2693--2706}.
\newblock


\bibitem[Zheng et~al\mbox{.}(2022)]%
        {280874}
\bibfield{author}{\bibinfo{person}{Lianmin Zheng}, \bibinfo{person}{Zhuohan
  Li}, \bibinfo{person}{Hao Zhang}, \bibinfo{person}{Yonghao Zhuang},
  \bibinfo{person}{Zhifeng Chen}, {et~al\mbox{.}}}
  \bibinfo{year}{2022}\natexlab{}.
\newblock \showarticletitle{Alpa: Automating Inter- and {Intra-Operator}
  Parallelism for Distributed Deep Learning}. In
  \bibinfo{booktitle}{\emph{Proc. USENIX OSDI}}. \bibinfo{pages}{559--578}.
\newblock
\showISBNx{978-1-939133-28-1}


\bibitem[Zheng et~al\mbox{.}(2025a)]%
        {10.1145/3719664}
\bibfield{author}{\bibinfo{person}{Yue Zheng}, \bibinfo{person}{Yuhao Chen},
  \bibinfo{person}{Bin Qian}, \bibinfo{person}{Xiufang Shi},
  \bibinfo{person}{Yuanchao Shu}, {and} \bibinfo{person}{Jiming Chen}.}
  \bibinfo{year}{2025}\natexlab{a}.
\newblock \showarticletitle{A Review on Edge Large Language Models: Design,
  Execution, and Applications}.
\newblock \bibinfo{journal}{\emph{ACM Comput. Surv.}} \bibinfo{volume}{57},
  \bibinfo{number}{8} (\bibinfo{year}{2025}), \bibinfo{pages}{1--35}.
\newblock


\bibitem[Zhong et~al\mbox{.}(2024)]%
        {298687}
\bibfield{author}{\bibinfo{person}{Yinmin Zhong}, \bibinfo{person}{Shengyu
  Liu}, \bibinfo{person}{Junda Chen}, \bibinfo{person}{Jianbo Hu},
  {et~al\mbox{.}}} \bibinfo{year}{2024}\natexlab{}.
\newblock \showarticletitle{{DistServe}: Disaggregating Prefill and Decoding
  for Goodput-optimized Large Language Model Serving}. In
  \bibinfo{booktitle}{\emph{Proc. USENIX OSDI}}. \bibinfo{pages}{193--210}.
\newblock
\showISBNx{978-1-939133-40-3}


\bibitem[Zhou et~al\mbox{.}(2024)]%
        {zhou2024dynamickv}
\bibfield{author}{\bibinfo{person}{Xiabin Zhou}, \bibinfo{person}{Wenbin Wang},
  \bibinfo{person}{Minyan Zeng}, \bibinfo{person}{Jiaxian Guo},
  \bibinfo{person}{Xuebo Liu}, \bibinfo{person}{Li Shen}, \bibinfo{person}{Min
  Zhang}, {and} \bibinfo{person}{Liang Ding}.} \bibinfo{year}{2024}\natexlab{}.
\newblock \showarticletitle{Dynamickv: Task-aware adaptive kv cache compression
  for long context llms}.
\newblock \bibinfo{journal}{\emph{arXiv preprint arXiv:2412.14838}}
  (\bibinfo{year}{2024}).
\newblock


\bibitem[Zhu et~al\mbox{.}(2025a)]%
        {zhu2025efficient}
\bibfield{author}{\bibinfo{person}{Bingjie Zhu}, \bibinfo{person}{Zhixiong
  Chen}, \bibinfo{person}{Liqiang Zhao}, \bibinfo{person}{Hyundong Shin}, {and}
  \bibinfo{person}{Arumugam Nallanathan}.} \bibinfo{year}{2025}\natexlab{a}.
\newblock \showarticletitle{Efficient LLM Inference over Heterogeneous Edge
  Networks with Speculative Decoding}.
\newblock \bibinfo{journal}{\emph{arXiv preprint arXiv:2510.11331}}
  (\bibinfo{year}{2025}).
\newblock


\bibitem[Zhu et~al\mbox{.}(2025b)]%
        {11161094}
\bibfield{author}{\bibinfo{person}{Bingjie Zhu}, \bibinfo{person}{Zhixiong
  Chen}, \bibinfo{person}{Liqiang Zhao}, \bibinfo{person}{Hyundong Shin}, {and}
  \bibinfo{person}{Arumugam Nallanathan}.} \bibinfo{year}{2025}\natexlab{b}.
\newblock \showarticletitle{Joint Caching and Inference for Large Language
  Models in Wireless Networks}. In \bibinfo{booktitle}{\emph{Proc. IEEE Int.
  Conf. Commun. (ICC)}}. \bibinfo{pages}{6285--6290}.
\newblock


\bibitem[Zhu et~al\mbox{.}(2023)]%
        {NEURIPS2023_b914a8fc}
\bibfield{author}{\bibinfo{person}{Banghua Zhu}, \bibinfo{person}{Ying Sheng},
  \bibinfo{person}{Lianmin Zheng}, \bibinfo{person}{Clark Barrett},
  {et~al\mbox{.}}} \bibinfo{year}{2023}\natexlab{}.
\newblock \showarticletitle{Towards Optimal Caching and Model Selection for
  Large Model Inference}. In \bibinfo{booktitle}{\emph{Proc. Adv. Neural Infor.
  Process. Syst. (NeurIPS)}}, Vol.~\bibinfo{volume}{36}.
  \bibinfo{pages}{59062--59094}.
\newblock


\end{thebibliography}

\end{document}